\documentclass[a4paper,10pt]{fullarticle}
\usepackage{sciencestuff}
\usepackage{pdflscape}

\title{Trustworthiness of Laser-Induced Breakdown Spectroscopy Predictions via Simulation-based Synthetic Data Augmentation and Multitask Learning}

\author[1,2]{Riccardo Finotello\emailfoot{riccardo.finotello@cea.fr}}
\author[1]{Daniel L'Hermite\emailfoot{daniel.lhermite@cea.fr}}
\author[1]{Celine Quéré\emailfoot{celine.quere@cea.fr}}
\author[1]{Benjamin Rouge\emailfoot{benjamin.rouge54@gmail.com}}
\author[2]{Mohamed Tamaazousti\emailfoot{mohamed.tamaazousti@cea.fr}}
\author[1]{Jean-Baptiste Sirven\emailfoot{jean-baptiste.sirven@cea.fr}}
\affil[1]{%
  Université Paris-Saclay, CEA, \protect\\
  Service d'études analytiques et de réactivité des surfaces (SEARS), \protect\\
  Gif sur Yvette, F-91191, France
}
\affil[2]{%
  Université Paris-Saclay, CEA, \protect\\
  LIST, \protect\\
  Palaiseau, F-91120, France
}

\date{}

\hypersetup{%
  pdfkeywords={deep learning, spectroscopy, libs, multitask, trustworthy AI},
  pdfsubject={trustworthy AI LIBS}
}

\begin{document}

\maketitle

\begin{abstract}
We consider quantitative analyses of spectral data using laser-induced breakdown spectroscopy.
We address the small size of training data available, and the validation of the predictions during inference on unknown data.
For the purpose, we build robust calibration models using deep convolutional multitask learning architectures to predict the concentration of the analyte, alongside additional spectral information as auxiliary outputs.
These secondary predictions can be used to validate the trustworthiness of the model by taking advantage of the mutual dependencies of the parameters of the multitask neural networks.
Due to the experimental lack of training samples, we introduce a simulation-based data augmentation process to synthesise an arbitrary number of spectra, statistically representative of the experimental data.
Given the nature of the deep learning model, no dimensionality reduction or data selection processes are required.
The procedure is an end-to-end pipeline including the process of synthetic data augmentation, the construction of a suitable robust, homoscedastic, deep learning model, and the validation of its predictions.
In the article, we compare the performance of the multitask model with traditional univariate and multivariate analyses, to highlight the separate contributions of each element introduced in the process. 
\end{abstract}

\highlights{we build an end-to-end pipeline to deal with data augmentation, robust multitask deep calibration models and the trustworthiness assessment of \libs predictions.}

\keywords{\libs, calibration, deep learning, data augmentation, trustworthy \ai}

\clearpage

{\small\tableofcontents}

\clearpage

\section{Introduction}

Deep Learning (\dl) has been widely developed with great success in many areas, among which computer vision and natural language processing have become well-known mainstream applications.
The performance of those algorithms in solving complex tasks stimulated their development in many other fields. 
Analytical chemistry is no exception to the rule.
As recently reviewed by Dubus et al., \dl has already proved successful in chromatography, spectroscopy, chemical sensing, imaging applications~\cite{Debus:2021:Deep}.
Historically, some optical spectrometry techniques have particularly benefited from multivariate approaches.
The example of near-infrared spectroscopy shows that they have become unavoidable for decades~\cite{Martens:1990:NIR}.
Their development is more recent in the Laser-Induced Breakdown Spectroscopy (\libs) community, but it has been growing for more than 15 years.
In this field, the possibility to make fast measurements with detectors of hundreds to several millions pixels for \emph{Echelle} spectrometers rapidly leads to a large amount of experimental data, and this is all the truer for hyperspectral imaging applications~\cite{Moncayo:2018:Exploration, Finotello:2022:HyperPCA}.
Data processing algorithms with a variable degree of complexity were successfully applied to \libs datasets for efficient detection of compounds, samples classification, quantitative analysis or chemical mapping.
Basic artificial Neural Networks (\nns) were implemented as soon as the '90s for identification of polymers~\cite{Sattman:1998:Applied}.
Beyond, in the second half of the 2010s, Machine Learning (\ml) techniques first, \dl after, have been more and more developed.
The recent review of Li et al.\ showed that modern concepts and architectures of \nns have already been tested in \libs, and this seems to be a promising direction to explore~\cite{Li:2021:Review}.

In quantitative analysis by spectroscopic methods, the objective is to build a model relating experimental spectra to the concentration of the species of interest.
This is based on a set of known calibration samples, and can be done by a variety of supervised techniques.
A quantitative model is then valid for a given sample matrix and for given experimental conditions.
However, for direct analytical techniques like \libs, which do not require sample preparation, it can happen that an unknown sample has a slightly different matrix compared to calibration ones, or that its measurement conditions are slightly variable.
In this case, it can be expected that the predicted concentration is biased to a certain extent.
Yet, a quantitative model is designed to deliver a concentration but not to estimate to which extent an unknown sample is well represented by calibration ones.
In other words, we do not know how reliable the prediction is.
It is therefore necessary to develop models that are able to estimate, if possible quantitatively, the trustworthiness of a prediction.
Thanks to their ability to perform different tasks in parallel, MultiTask (\mt) \nns are a type of model that could address this issue.
However, a well-known drawback of \nns is that they are easily prone to overfitting of training data, since they require a very large number of spectra to be properly trained.
Experimental, time and cost constraints generally do not enable to acquire enough spectra.
A possibility to overcome this limitation is to generate calibration data.
This can be done either by transforming the initial spectra, or by phenomenological modelling of training spectra, or by hard modelling based on physical equations – in the case of \libs, the equations ruling laser ablation, the plasma formation and emission.

\begin{figure}[t]
    \centering
    \includegraphics[width=\linewidth]{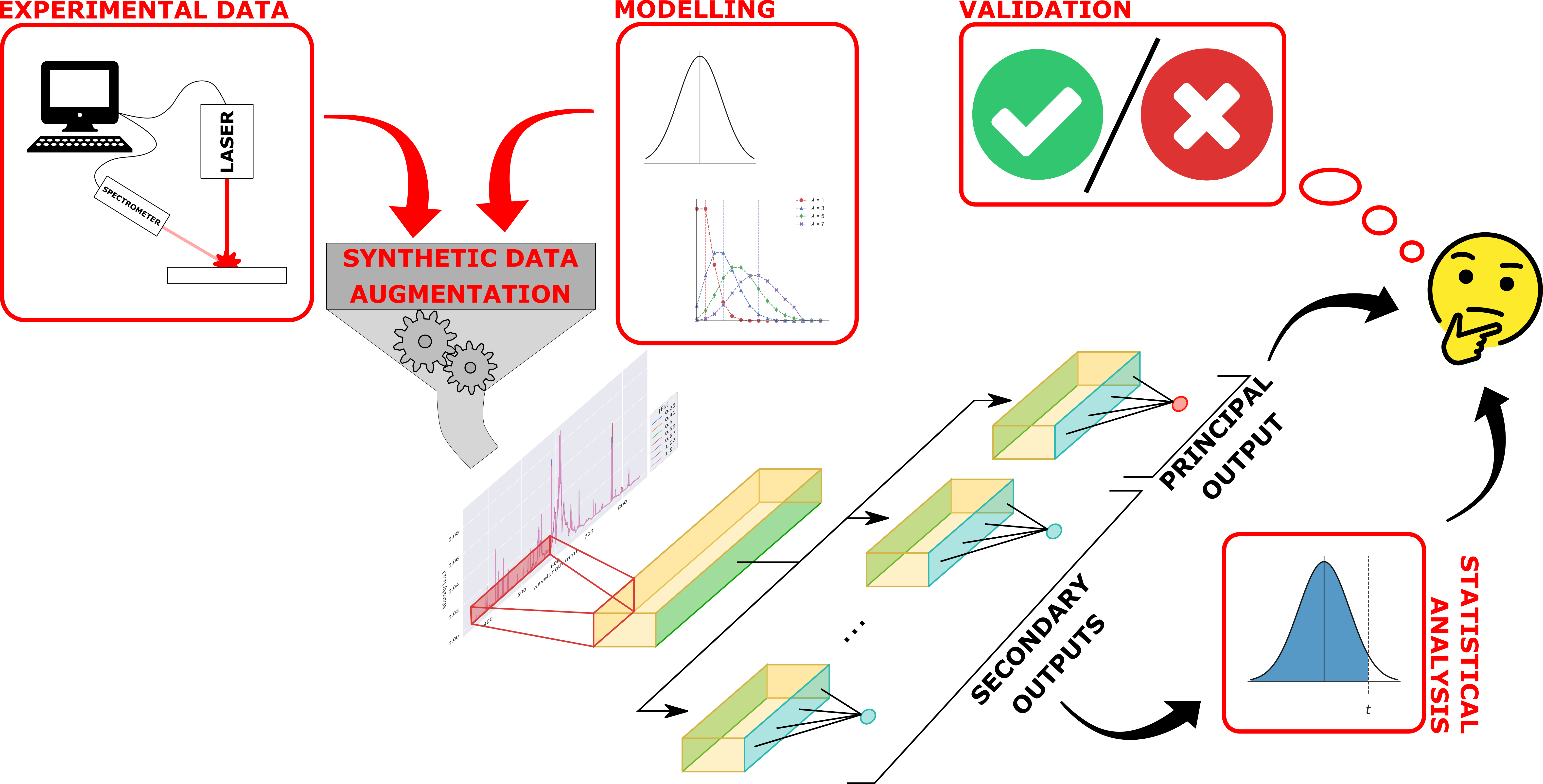}
    \caption{%
        \emph{Schematics of the methodology.}
        Mathematical modelling and experimental data are used as inputs for the data augmentation procedure.
        A robust multitask model is trained on synthetic data to predict a principal variable (e.g.\ the concentration of the analyte) and several secondary spectral quantities, which can be verified on experimental data.
        Given the inter-dependence of the parameters of the network, the statistical analysis of the latter enables the assessment the trustworthiness of the predictions of the principal output.
    }
    \label{fig:graphical}
\end{figure}

The objective of this paper is then to introduce a new approach of quantitative analysis by \libs and \dl, enabling both to generate enough spectra to train a model based on \mt Convolutional Neural Networks (\cnns), and to evaluate the confidence in the model predictions.
In this article, we propose a methodology for calibration models based on three fundamental elements:
\begin{enumerate}
\item simulation-based synthetic data augmentation, namely the generation of an arbitrary number of new spectra from experimental data to train complex \dl architectures;
\item robust deep \mt \cnns, capable of processing entire \libs spectra and increasing the robustness (homoscedasticity) of the model;
\item measure of the trustworthiness of the predictions of the model through the statistical analysis of the inter-dependencies of the parameters and the outputs of the \mt architecture.
\end{enumerate}
The procedure, presented in \Cref{fig:graphical}, aims at providing a way to increase the robustness of the calibration models, and to benchmark the confidence of the Artificial Intelligence (\ai) architecture through its own predictions.

In the article, we first revise the State-Of-The-Art (\sota) of augmentation, \dl and trustworthy \ai in \libs.
We then move to the description of the technique, introducing in detail the three mentioned aspects.
In the following section, we discuss the experimental setup and the details of the implementation.
We then show the results of the analysis, compared to known techniques in the \libs literature, and we provide a separate analysis of the contribution of each element.

\section{Related Works}

Several univariate and multivariate calibration techniques are explored in the literature~\cite{Costa:2020:Calibration}.
The first are widely adopted for their ability to provide interpretable results quickly and with good precision~\cite{MottoRos:2019:Investigation}.
In this scenario, calibration standards are used to build a map between the concentration of a given analyte and the information contained in a single measurable variable, such as the integral intensity of an emission line.
The model is then inverted during inference to use the measured intensity as predictor to predict the concentration of the analyte.
On the other hand, multivariate methods were introduced for their ability to take advantage of more information contained in the input spectra, rather than focusing on a single variable.
Techniques based on principal components and multilinear regression have been widely adopted in \libs~\cite{Hoehse:2012:Multivariate, Dingari:2012:Incorporation, Cisewski:2012:SVM, Yaroshchyk:2012:Comparison}.
As a multivariate technique, \dl has also been explored at length, comparing various types of architectures~\cite{Monch:1997:HighSpeed, MottoRos:2008:Quantitative, Li:2021:Review}.
In particular, \cnns present interesting properties, capable of scanning complex spectral data, with minimal preprocess treatments needed~\cite{Zhang:2019:DeepSpectra}.
It is often coupled to dimensionality reduction algorithms or to input selection criteria~\cite{Hahn:2012:LIBS, Sirven:2006:Qualitative, Takahashi:2017:Quantitative, Luarte:2021:Combining}.
However, the increased computational power witnessed in the last decade, enables avoiding the loss of information in the spectra~\cite{Chen:2020:Convolutional, Zou:2021:Online, Lu:2019:Detection, Chen:2022:DeepLearning}.

\subsection{Data Augmentation}

Though \dl shows potential for various analyses, time and experimental conditions often prevent building large \libs datasets.
The preferred solution seems to be the simplification of the models and an accurate selection of the input data.
Feature engineering and feature selection, for instance through a principal components analysis or \emph{a priori} expertise, have been employed to reduce the size of the input data~\cite{Hahn:2012:LIBS, Takahashi:2017:Quantitative}.
At the same time, the production of purely synthetic data, based on local thermodynamical equilibrium, has been explored in \libs applications~\cite{DAndrea:2015:Hybrid}.
The idea of enriching existing data by means of different representations of the inputs, such as time resolved spectra, was experimented with success~\cite{Narlagiri:2021:Simultaneous}.
However, most of the time, no real augmentation of the training datasets is provided, at least in the same sense as in modern \ml~\cite{Zou:2021:Online}.
More recently, standard data augmentation techniques were used for the classification of \libs maps, though mostly exploiting the spatial information of the two-dimensional maps, rather than the spectral distribution of the signal~\cite{Li:2021:Review}.
Augmentation through synthetic data via the simple addition of random noise to the experimental data is also briefly discussed in the literature, though no implementations seem to be present~\cite{Chen:2022:DeepLearning}.

\subsection{Multitask Learning}

Though explored at length in \ml literature~\cite{Caruana:1993:Multitask, Caruana:1997:Multitask}, recently, \mt learning has seen major developments and a wide range of applications.
In particular, it has been shown to improve the performance of different computer vision models~\cite{Mullapudi:2018:Hydranets}, used, for instance, for multiple object detection in self-driving cars.
At the same time, the possibility to create latent common representations for multiple tasks has been considered for computations in fundamental research in physics and mathematics~\cite{Erbin:2021:Machine, Erbin:2021:Inception, Erbin:2021:Deep}.
Its improved generalisation ability on multiple tasks is used for the creation of versatile agents~\cite{Reed:2022:Generalist}, towards the definition of Artificial General Intelligence~\cite{Shevlin:2019:Limits}.
In \libs applications, univariate and multivariate analyses focus on the prediction of a single output, usually the concentration of the analyte.
However, some examples of multi-output algorithms were recently explored, based on \nns~\cite{Narlagiri:2021:Simultaneous, Li:2020:LIBS} and on Partial Least Squares (\pls) with two outputs, i.e.\ \pls[2]~\cite{Anderson:2012:Clustering}.
In these cases, the response variables are usually the concentrations of multiple elements.
Interestingly, multi-output \nn architectures found successful applications for the computation of plasma parameters from \libs data~\cite{Borges:2014:Fast, Saeidfirozeh:2022:ANN}.
\mt learning was also introduced for the simultaneous predictions of concentrations of analytes and lithology classes: different \nns were used to process the data with different loss functions, using a latent representation of the input, computed by a common backbone architecture~\cite{Chen:2022:Simultaneous}.
This shows the viability of \mt learning as an analysis technique in \libs, capable of providing information of different nature on related tasks, at the same time.

\subsection{Trustworthy AI}

Once the calibration model is defined, it is difficult to assess the correctness of the predictions for unknown samples.
Standard techniques in statistics are usually employed~\cite{Mermet:2006:Limit, Mermet:2010:Calibration}, though they rely on strong assumptions on the type of data analysed, notably their heteroscedasticity~\cite{Grimmett:2020:Probability}.
Moreover, calibration-free methods in \libs have been proven successful in providing robust elemental analyses~\cite{Ciucci:1999:New, Cremers:2013:Handbook, Tognoni:2010:Calibration, Cavalcanti:2013:OnePoint, Chen:2018:TwoStep}, with the possibility to compare the predictions with sensible physical hypotheses.
On the other hand, it has been seen heuristically that traditional models do not systematically generalise to unknown data~\cite{Li:2021:Review, Li:2020:LIBS, ElHaddad:2014:Good}.
For instance, algorithms used for calibration can present different performances at different levels of concentrations, hence the possible need to separately work on different sub-populations of samples~\cite{Anderson:2012:Clustering}.
Random permutations and cross-validation of the samples used in the \nn models have also been proposed to test the robustness of the algorithms, by measuring the differences in the prediction on given samples~\cite{ElHaddad:2014:Good}.
The computation of misclassification probability based on \ml methods has also been considered, in order to assess the validity of classification models~\cite{Morais:2019:Uncertainty}.
Recently, different teams proposed \dl frameworks to understand the learning process of the \ai architectures for spectral data.
These interpretable-\ai methods either analyse the progression of the feature maps in the hidden layers~\cite{Zhao:2021:Interpretable}, or study the importance of the spectral variables leading to the prediction~\cite{Zhang:2020:Understanding}.
This represents indeed a step towards the comprehension of the mechanisms behind \nns, as it enables a semi-quantitative analysis of the hidden layers.
Nevertheless, it does not deal with the confidence of the predictions, or the automatic detection of changes in the distribution of the samples.
As a matter of fact, the validation of the outputs of a model, and the study of their trustworthiness, remains a challenging task, with only a few algorithmic implementations or statistical treatments.

\section{Methodology}

We use a \dl approach based on \mt learning to increase the robustness of the models.
We predict, at the same time, the concentration of the analyte and a set of related spectral quantities.
Given the complexity of the architecture, we introduce a simulation-based data augmentation to increase the number of available spectra for training.
Though formally inspired by traditional data augmentation techniques, we focus on generating new synthetic spectra from a theoretical prior on the experimental data, rather than transforming the experimental spectra.
Finally, we validate the predictions of the model using the additional outputs of the \mt architecture as reference for a statistical analysis.
The procedure is used to reveal possible anomalies in training or modifications of the experimental conditions.

\subsection{Simulation-based Synthetic Data Augmentation}\label{sec:simulation}

For the analysis, we build complex \dl architectures, with a large number of free parameters, compared to traditional models.
In this regime, training large \nns by optimising the bias-variance trade-off may lead to phenomena such as poor stability and bad generalisation in inference on unknown samples.
Many training samples are usually required to train more complex architectures to overcome the issues.
Inspired by usual \dl practices, we thus introduce a data augmentation technique for \libs.

\begin{figure}[t]
    \centering\includegraphics[width=0.66\linewidth]{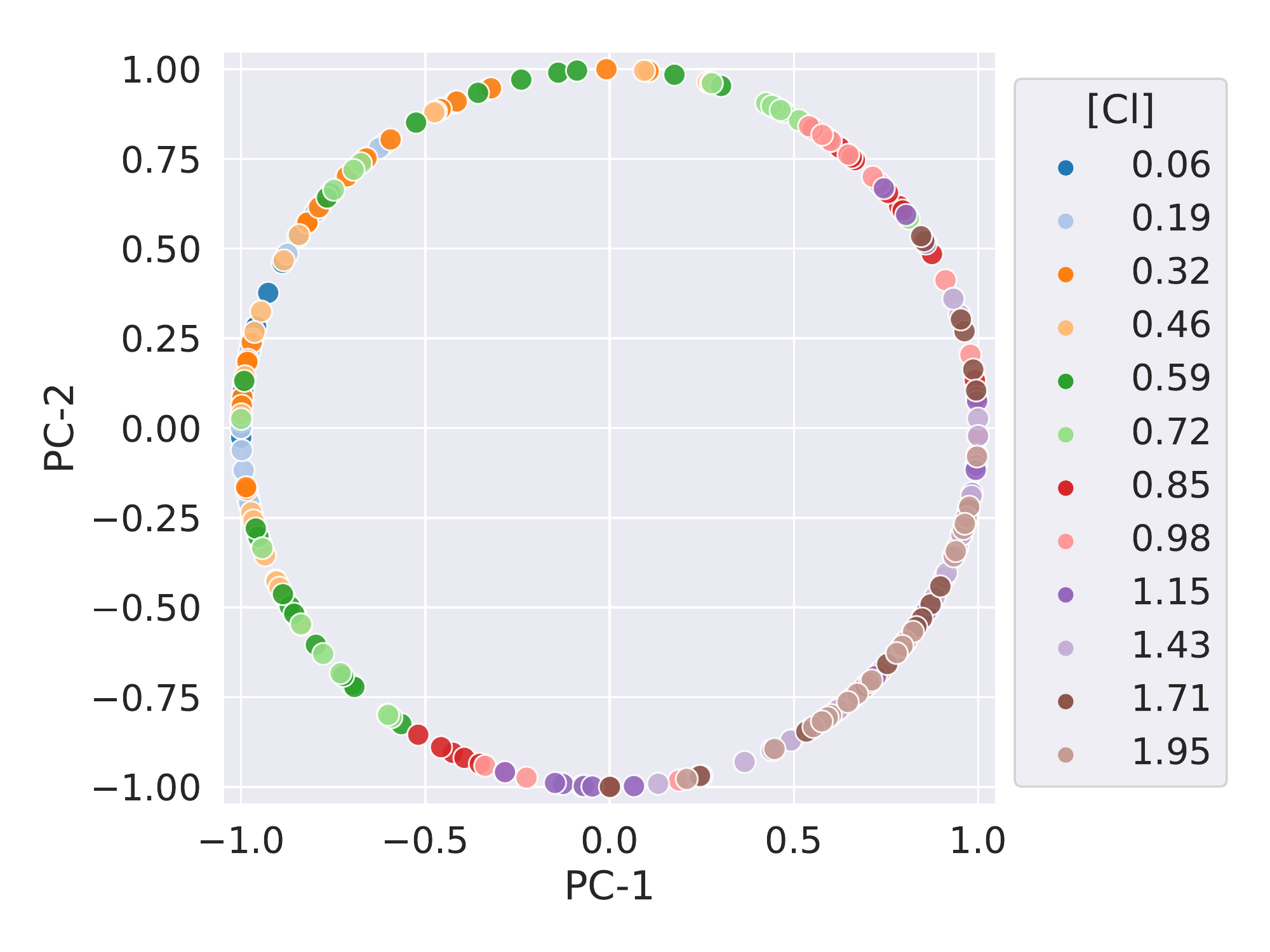}
    \caption{%
        \emph{Sample training distribution at different concentrations.}
        Example of the training distribution of cement samples (described in the text) as principal components in feature space, normalised to the unit sphere.
        Differences in distribution display as rotations.
    }
    \label{fig:train_pca}
\end{figure}

We train the \mt model on entire \libs spectra, without \emph{a priori} data selection.
For the \libs analysis, typical data augmentations used in computer vision, and based on rotations, reflections and crops, may jeopardise the predictive power of the model: the spatial distribution of the wavelength channels, and their neighbouring relations, should not be altered randomly by transformations, in order to retain a physical meaning.
At the same time, given the scarcity of training data, the simple addition of random noise to the experimental spectra may result in a training distribution no longer representative, at least locally, of the use case.
As a reference, \Cref{fig:train_pca} shows the span of the first two principal components of the distribution, normalised to the unit sphere to highlight differences as rotations.
We first proceed to model the distribution of the original spectra, and then to synthetically produce an arbitrary number of spectra.
We show that noise can be added to each individual channel, provided that its global average effect is negligible, in order to make the synthetic distribution more realistic.
For instance, this influences locally the models which use small mini-batches of spectra for training, such as \nns (see \Cref{sec:trad_an} for additional details).
The procedure ensures to enlarge consistently the feature space spanned by the synthetic spectra.
The generated samples are able to cover an extended range of variations inside the original distribution, before the addition of noise.
In turn, we noticed that the model is able to generalise to unknown data more efficiently.

From a technical point of view, we first compute the average experimental spectrum of each sample as input of the data synthesis technique.
This alleviates the effects of possible experimental local signal fluctuations.
It also ensures the statistical consistency of the frequentist analysis.
We then consider each wavelength channel as independently modelled by a random variable following a probability distribution function, whose expected value is represented by the average spectrum.
As a whole, the expected values follow another probability distribution, which reproduces the profile of the emission lines (or bands) in the average spectrum, experimentally available.
That is, we consider the set of $n$ spectra with $p$ wavelength channels $\qty{ \bfx^{(i)} \in \R^p}_{i \in \qty[1,\ n]}$ and the corresponding average spectrum $\barbfx = \qty( \barx_r ) \in \R^p$, where $\barx_r = \Ev{x_r}$.
For each wavelength channel $r = 1,\, 2,\, \dots,\, p$, we fix the expected value of a random variable $y_r \in \R$ such that $\Ev{y_r} = \barx_r$.
Generalising to an arbitrary number $m$ of random variables, we consider the set:
\begin{equation}
    \qty{
        y_r^{(j)} \in \R~\mid~
        \Ev{y_r^{(j)}} = \barx_r,
        \quad
        j = 1,\, 2,\, \dots,\, m
    }.
\end{equation}
An arbitrary number $m \gg n$ of full spectra can be constructed by similarly proceeding for all wavelength channels $r \in \qty[1,\, p]$.

\begin{figure}[t]
    \centering
    \begin{tabular}{@{}c|c@{}}
        {\LARGE $\beta = 0.00$}                                         & {\LARGE $\beta = 0.03$}
        \\
        \includegraphics[width=0.48\linewidth]{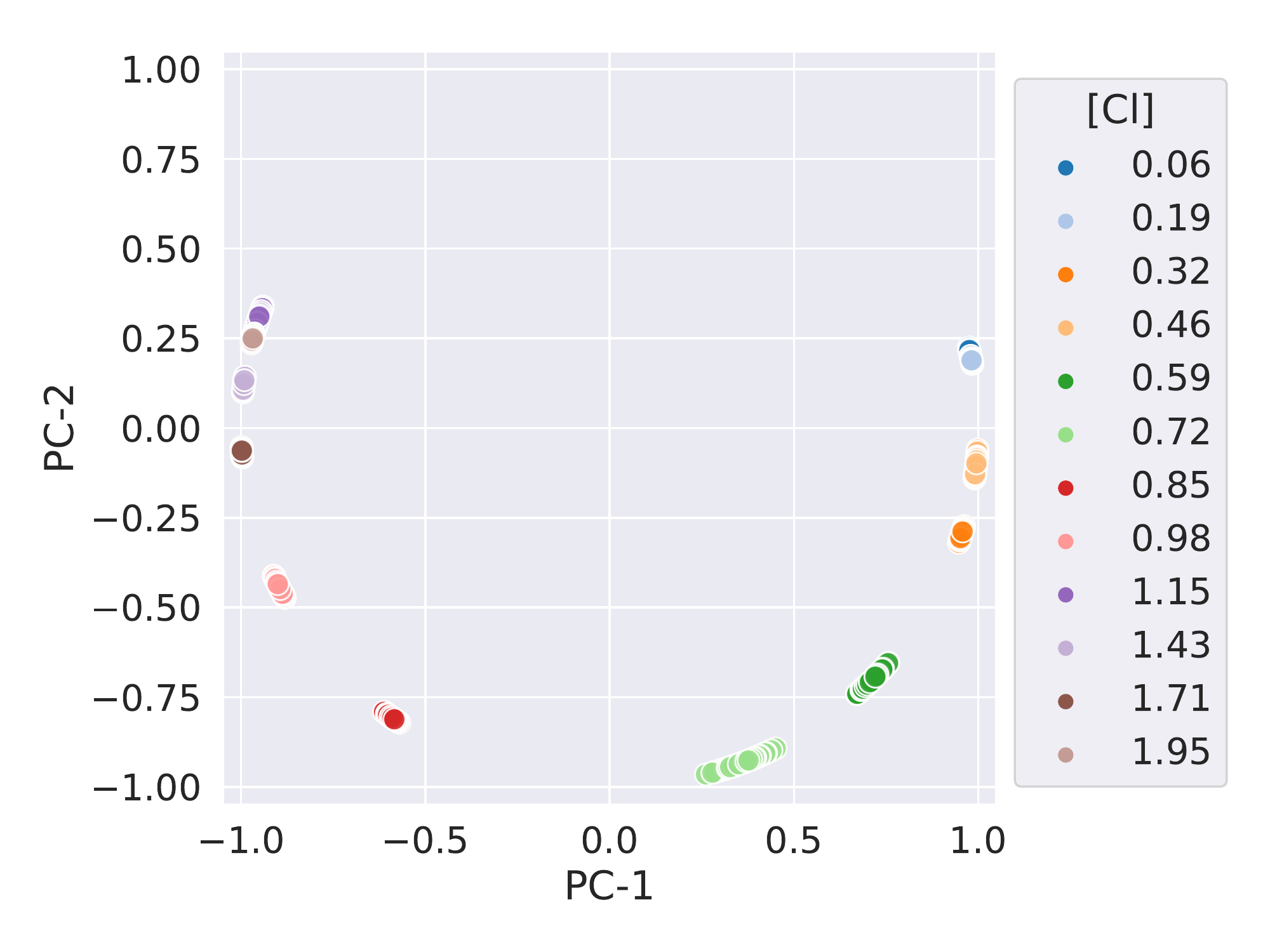} & \includegraphics[width=0.48\linewidth]{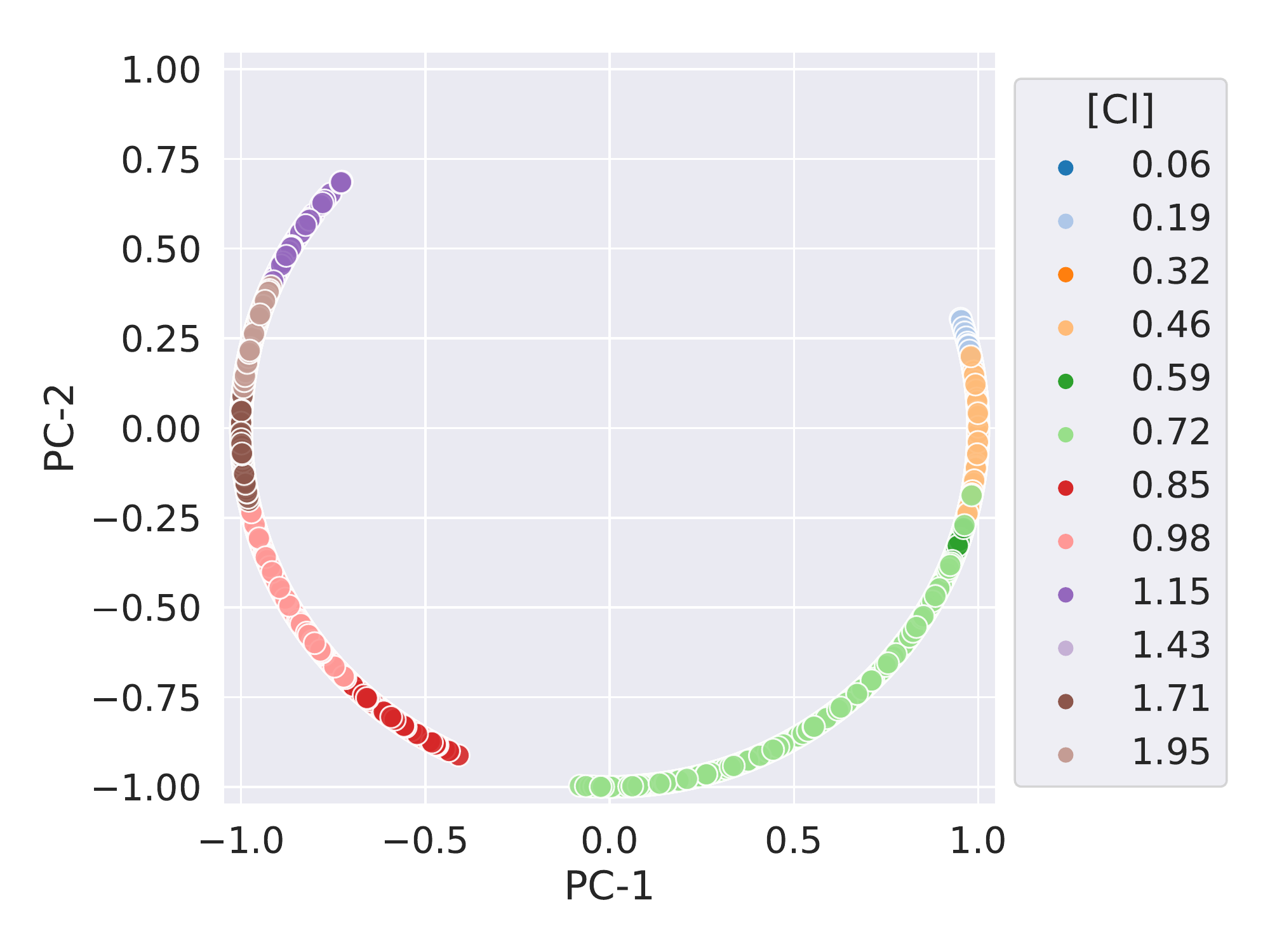}
        \\  \midrule
        {\LARGE $\beta = 0.05$}                                         & {\LARGE $\beta = 0.10$}
        \\
        \includegraphics[width=0.48\linewidth]{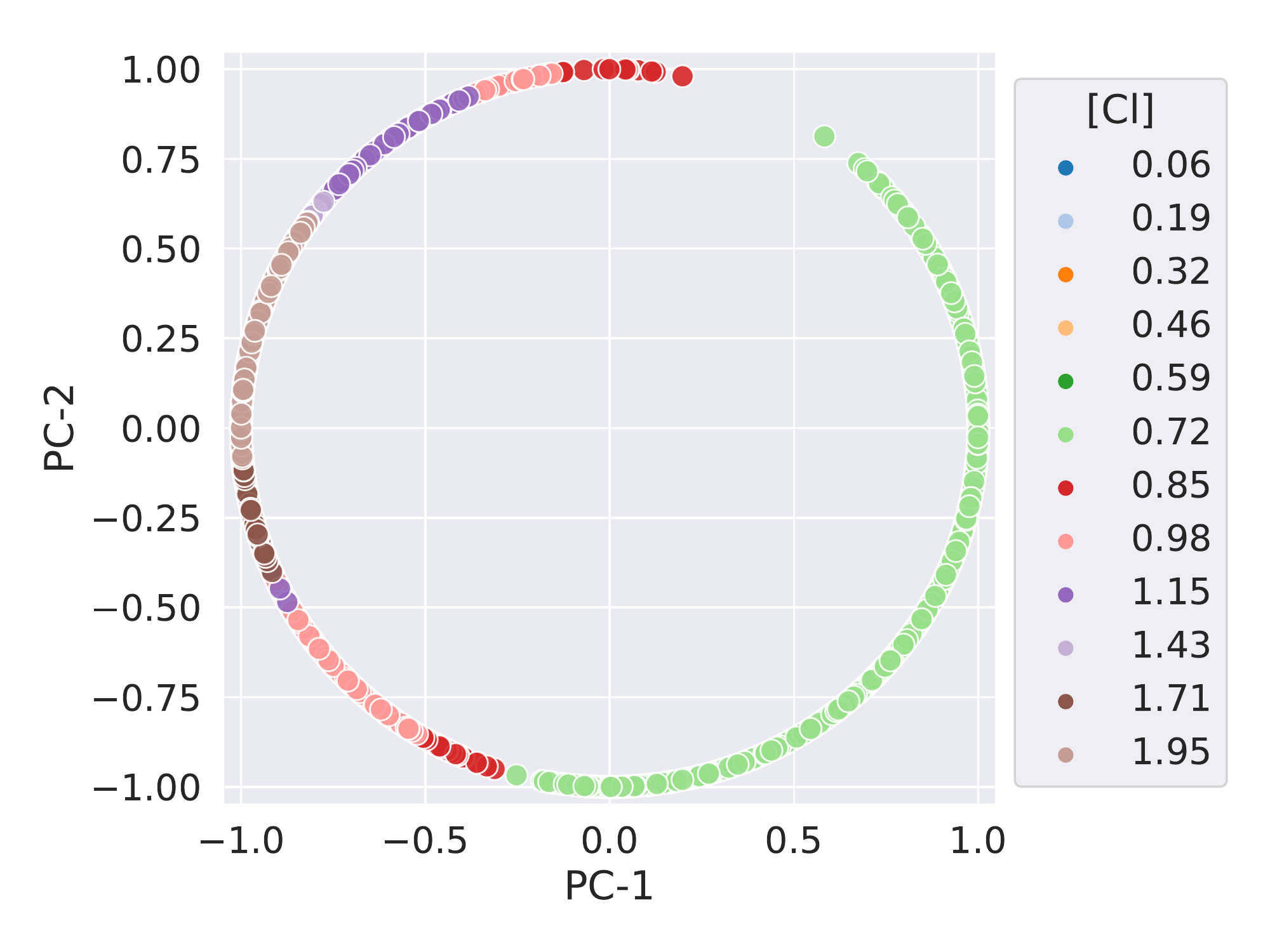} & \includegraphics[width=0.48\linewidth]{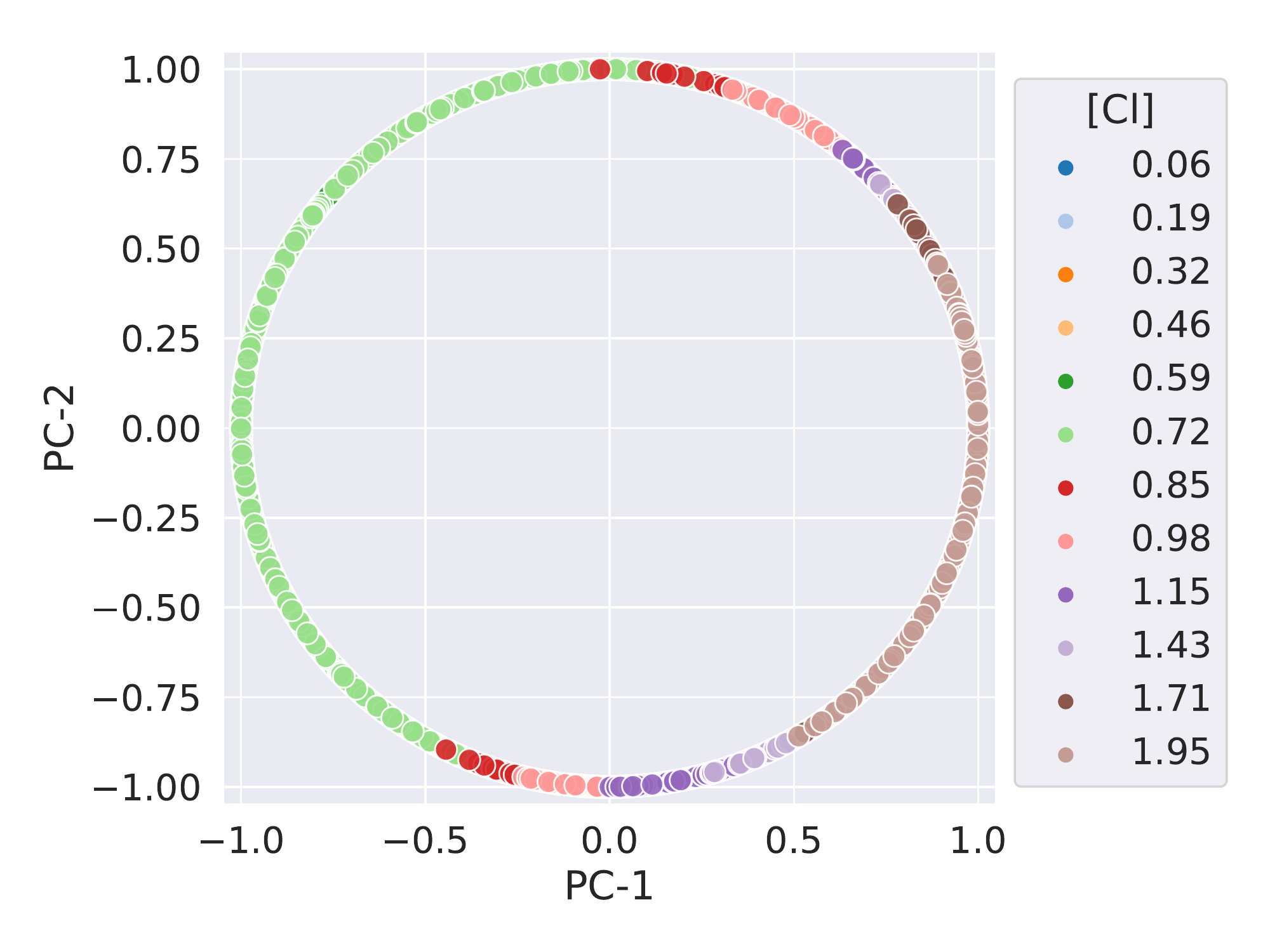}
    \end{tabular}
    \caption{%
        \emph{Qualitative effects of the confounder parameter $\beta$.}
        Synthetic data under the influence of a noise parameter locally span the entire feature space, rather than forming cluster structures.
        The overall distribution is more realistic (see \Cref{fig:train_pca}).
    }
    \label{fig:pca_beta}
\end{figure}

In detail, we model the intensity of a given wavelength channel as a probability distribution
\begin{equation}
    \rP\qty( \lambda, \beta ) = \rP_1\qty( \lambda~|~\beta )\, \rP_2\qty( \beta ),
\end{equation}
where $\lambda$ represents the experimental photon-like contributions to the \libs spectra (either the genuine signal or low-intensity photon noise), while $\beta$ is a confounder variable, modelling external factors.
In this realisation, we achieve good performance by choosing independent variables $\lambda$ and $\beta$ (i.e.\ $\rP_1\qty( \lambda~|~\beta ) = \rP_1\qty( \lambda )$), leading to $\rP\qty( \lambda, \beta ) = \rP_1\qty( \lambda )\, \rP_2\qty( \beta )$, for the synthetic spectra.
In \Cref{fig:pca_beta}, we show the effect of $\beta$ on the distribution of the synthetic spectra.
Qualitatively, the confounder parameter makes the synthetic set locally more realistic, by enlarging the portion of feature space spanned by the spectra.
Specifically, we model the spectral intensity of a wavelength channel $r$ through a variable $y_r$ using a Poisson-like law.
We decompose
\begin{equation}
    y_r = z_r\, p_r,
    \qquad
    z_r \sim \cN\qty( 1, \beta ),
    \qquad
    p_r \sim \cP\qty( \barx_r ),
    \label{eq:synthesis}
\end{equation}
where $\cN\qty( \mu, \sigma )$ is a normal distribution with expected value $\mu$ and variance $\sigma^2$, and $\cP\qty( \lambda )$ is a Poisson distribution with expected value and variance $\lambda$.
As $z_r$ and $p_r$ are independently sampled, we have $\Ev{y_r} = \Ev{z_r} \Ev{p_r} = \barx_r$ for $r = 1,\, 2,\, \dots,\, p$.

\begin{figure}[t]
    \centering
    \begin{tabular}{@{}c|c@{}}
        {\LARGE \textsc{Global Effects}}                                & {\LARGE \textsc{Local Coverage}}
        \\
        (whole spectra)                                                 & (CaCl band at \SI{593.46}{\nano\meter})
        \\
        \includegraphics[width=0.48\linewidth]{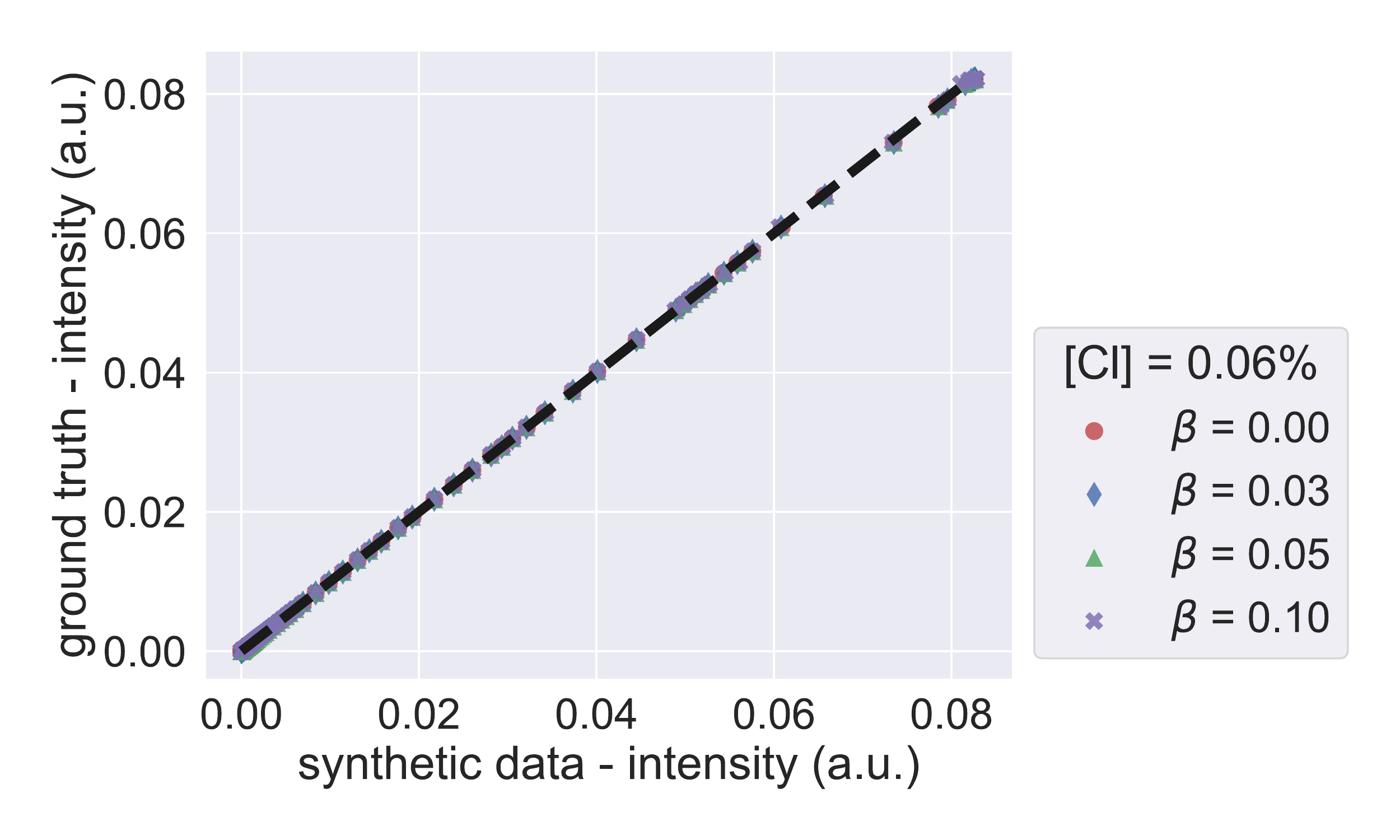}  & \includegraphics[width=0.48\linewidth]{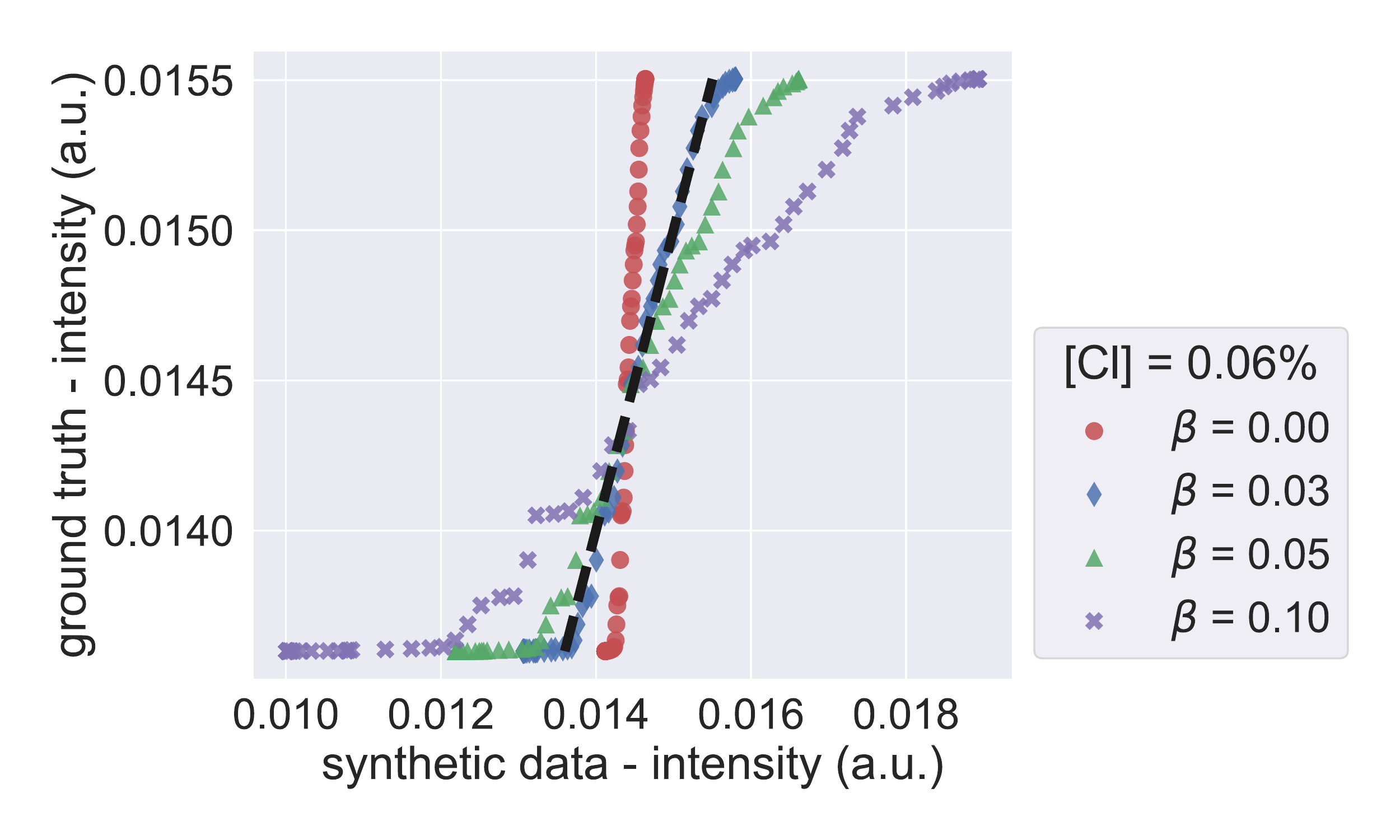}
        \\ \midrule
        \includegraphics[width=0.48\linewidth]{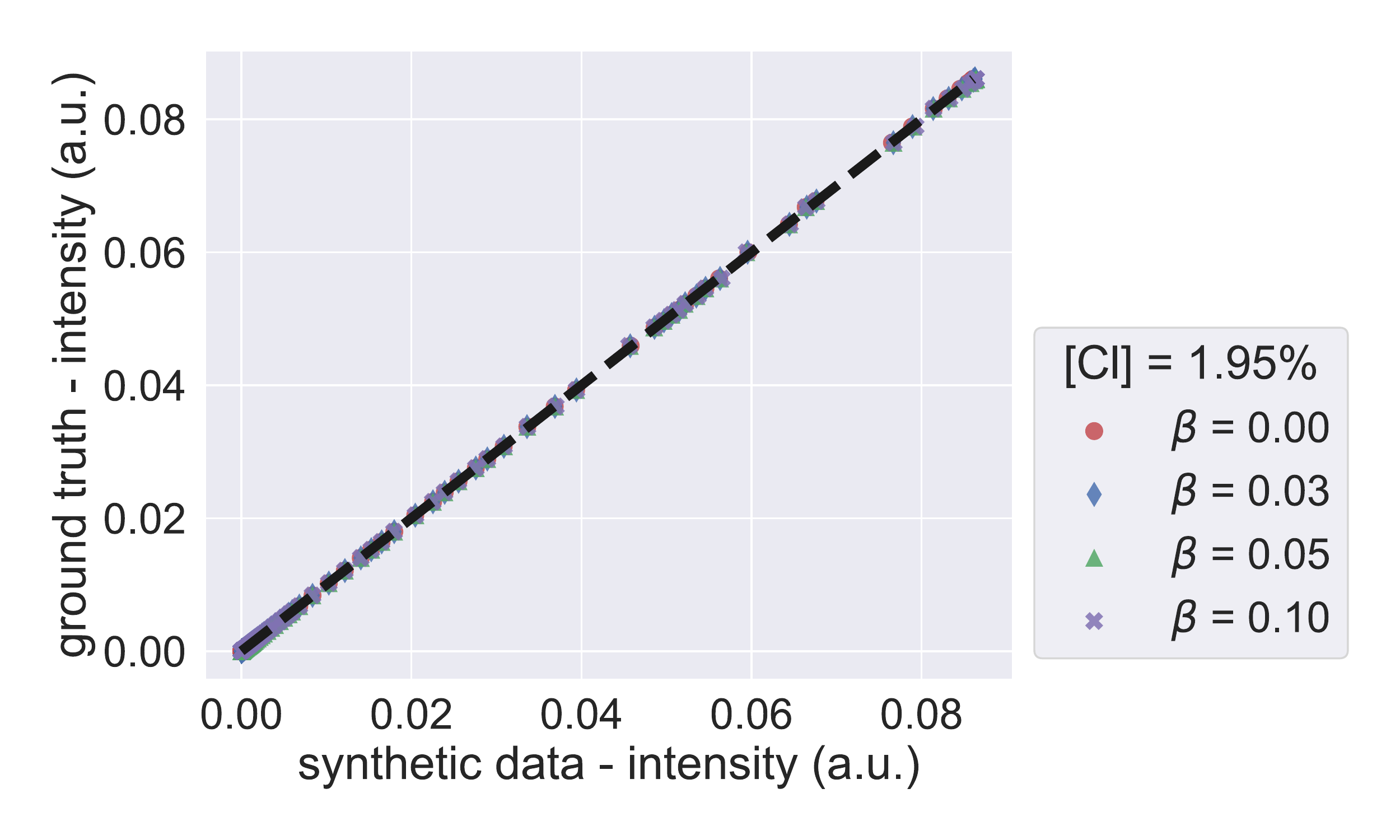} & \includegraphics[width=0.48\linewidth]{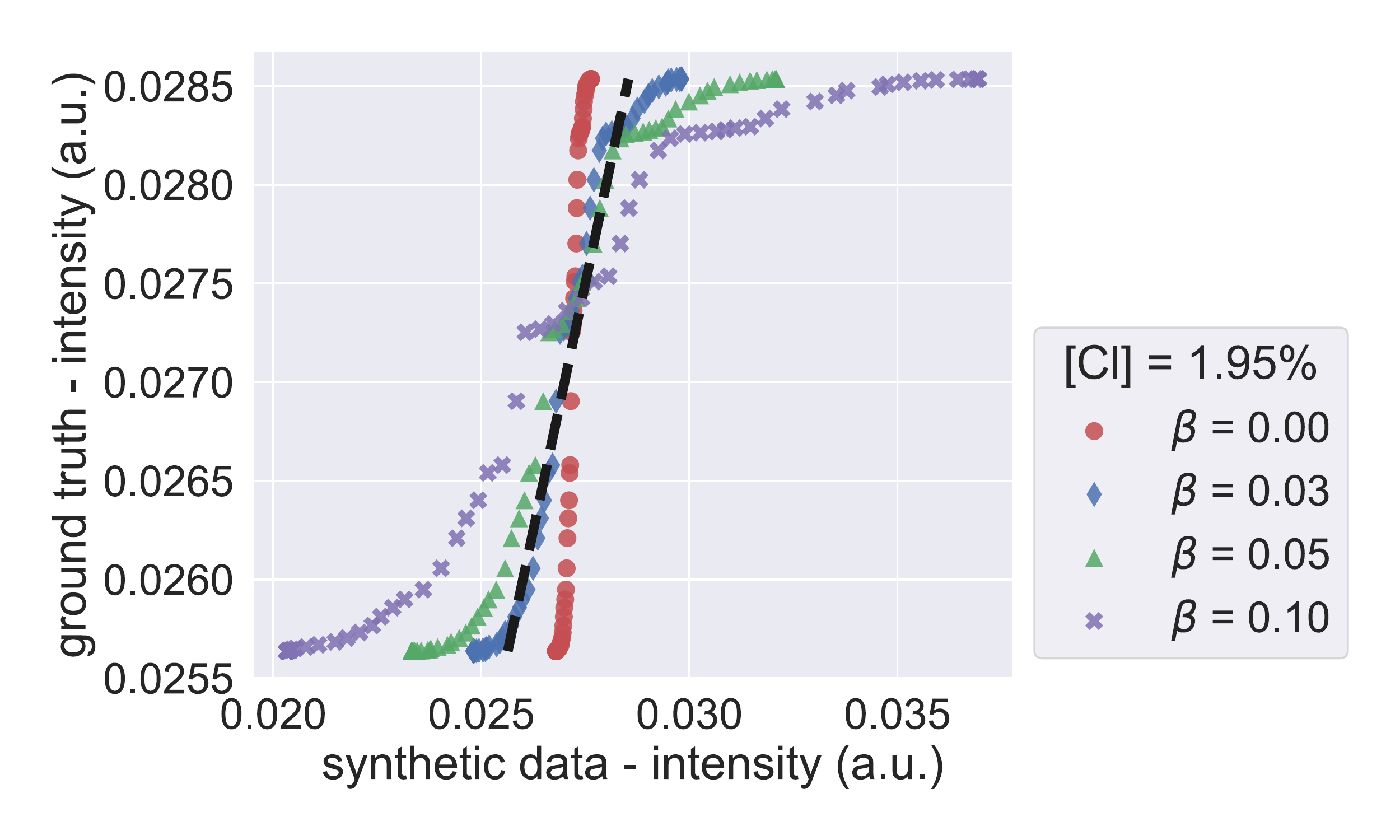}
    \end{tabular}
    \caption{%
        \emph{Global and local quantitative effects of the noise parameter (low concentration of the analyte at the top, higher concentration at the bottom).}
        The confounder parameter $\beta$ does not impact the average (global) distribution of the spectra.
        However, it presents local differences at given wavelengths and for different concentrations of the analyte.
    }
    \label{fig:qqplots}
\end{figure}

We can use the spectra in the training sets to choose appropriate values of $\beta$.
For illustration purposes, in \Cref{fig:qqplots} we present the quantitative effects of different choices of $\beta$ on the modelled distribution using a quantile-quantile diagram of the average spectra to compare the real and the synthetic distributions.
We use a dataset of cement samples, described in \Cref{sec:data}, for the prediction of chlorine concentration.
On average, the generated samples seem to reproduce correctly, on a global scale, the ground truths, especially at low signal intensity, as shown on the left of \Cref{fig:qqplots}: as expected, $\beta$ as a low impact as a global effect.
Moreover, the average spectra of the synthetic samples do not present strong deviations from the original distribution, suggesting that the theoretical model may correctly represent the experimental data.
On the right, we study in detail the statistical coverage of the simulated spectra at a given wavelength, namely the CaCl emission band at \SI{593.46}{\nano\meter}, used later in the analysis to infer the concentration of Cl in the samples.
In the absence of the parameter $\beta$, the statistical model is over-confident at (i.e.\ it does not reproduce the full distribution) low intensity and under-confident at higher intensities (i.e.\ it creates out-of-distribution intensities), while the opposite occurs for higher values of the noise parameter.
This suggests that there is an optimum of the parameter $\beta$, for which the synthetic distribution covers correctly the variance of the training distribution at a local level.
The value of $\beta$ can be chosen deterministically by maximising the coefficient of determination (\rs) between the ground truths and the synthetic quantiles at a given wavelengths of interest, or by weighting the \rs at different relevant wavelengths.
In \Cref{fig:co_coverage,fig:ni_coverage,fig:sn_coverage,fig:zr_coverage} in \Cref{app:figures}, we show the same diagrams for different alloy matrices, introduced in \Cref{sec:data}.
The effect of the noise parameter $\beta$ is different depending on the matrix, and results suggest that it may be advisable to model independently each sample at different concentrations of the analyte, even though the global results do not deviate from the experimental distribution.
However, as shown later in \Cref{sec:results}, the case of the alloy matrices presents peculiarities in the \dl model which mitigate the differences in the theoretical priors, leading to a less intense impact of the choice of $\beta$ with respect to the cement samples.
In the article, we provide the results for several values of $\beta$ for comparison, without focusing on finding the optimal value of the parameter.

\subsection{Multitask Convolutional Neural Networks}\label{sec:mt_cnn}

\begin{figure}[t]
    \centering
    \includegraphics[width=\linewidth]{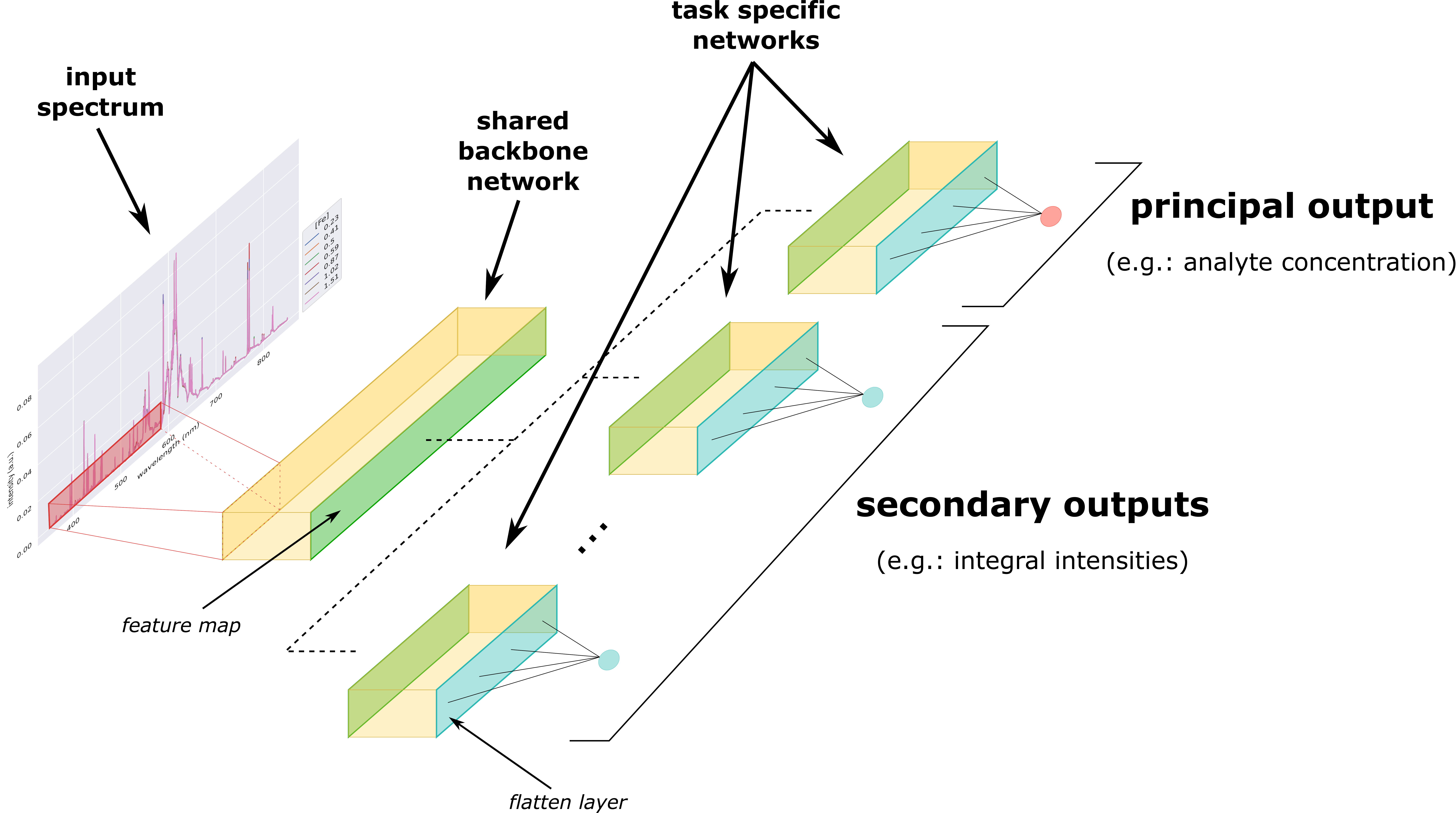}
    \caption{%
        \emph{Structure of the \dl architecture.}
        The model is a hard parameter sharing structure built using a common backbone network and several heads, connected to the shared feature map and computing regression scalars.
    }
    \label{fig:schema}
\end{figure}

By definition, multi-output \nns are a broad class of algorithms, which provide multiple predictions at the same time, using a shared structure of weights, trained simultaneously.
\mt networks are a subset of these architectures, implementing different strategies for sharing the training of their free parameters.
This property gives the networks great versatility, as it is capable of using information on one task to improve its generalisation.
This strategy acts as a regularisation and reduces the overfitting of the training data, as the model supposedly learns new representations, which should generalise well on all tasks~\cite{Caruana:1993:Multitask, Caruana:1997:Multitask}.
However, outputs need to be chosen carefully, such that the every single output could benefit from learning the others~\cite{Baxter:2000:Model, Crawshaw:2020:MultiTask}.

\begin{figure}[t]
    \centering
    \includegraphics[width=\linewidth]{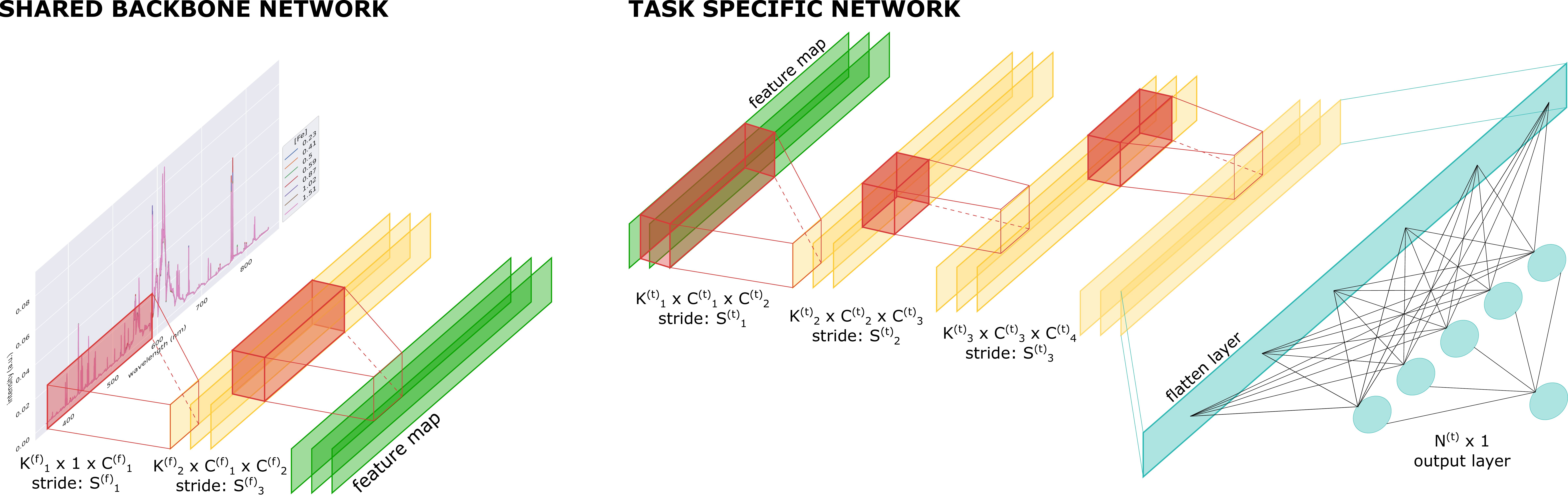}
    \caption{%
        \emph{Structures in the \mt architecture.}
        The shared backbone is a one dimensional \cnn, which processes the spectra and produces a latent vector representation.
        The task-specific networks are \cnns with shallow \fcnns as top layers.
    }
    \label{fig:structures}
\end{figure}

In this analysis, we use a \emph{hard parameter sharing} implementation of \mt learning, with a common set of bottom layers (see the general schematics in \Cref{fig:schema}).
The \mt architecture is made by different sub-structures, shown in \Cref{fig:structures}.
The innermost backbone processes the input spectra and produces a new latent vector representation.
The task-specific heads separately use the latent representation as inputs to compute a scalar regression output.
We use one-dimensional convolutions in the spectral dimension of the data cube as main operation.
This is repeated on different channels, thus adding a depth dimension to the output.
Additional details on the exact implementation of the architecture are in \Cref{sec:details}.

The choice of the outputs is particularly relevant, as their predictions from a common representation should be mutually beneficial.
From a likelihood perspective, let the variables $Y_1$ and $Y_2$ be two outputs of the \mt learning architecture, given an input $X$.
The model aims at maximising, as a function of the parameters, the function:
\begin{equation}
    \rP\qty(Y_1, Y_2 \mid X;\, \qty{A, B, \Omega})
    =
    \frac{1}{2}\,
    \rP\qty(Y_1 \mid Y_2;\, X;\, \qty{A, B, \Omega})\,
    \rP\qty(Y_2 \mid X;\, \qty{A, B, \Omega})
    +
    \qty(1 \leftrightarrow 2),
    \label{eq:likelihood}
\end{equation}
where $A$ and $B$ are the task-oriented sets of parameters, and $\Omega$ represents the shared parameters.
The right-hand side of \eqref{eq:likelihood} is thus symmetric in the outputs: all sets of parameters are used to ensure the optimal prediction of $Y_1$ and $Y_2$.
In turn, maximising the likelihood of $Y_2$, helps to maximise of the likelihood of $Y_1$, and viceversa.
The procedure can be generalised to an arbitrary number of outputs by summing over all permutations in the right-hand side of the identity.
Therefore, uncorrelated tasks would work against each other, resulting in worse performance.
In what follows, we choose to use the concentration of the analyte as the principal prediction of the network, and the integral intensities of the associated emission lines or bands as secondary outputs of the network.
In turn, this helps to stabilise the model and increase its robustness during inference.
It also provides a set of secondary results used to validate the performance of the model and to detect anomalies, as described in \Cref{sec:validation}.
Notice that the prediction of the maximal intensity of the emission lines may result in an easy task for the network.
As a consequence, the network may be imbalanced with respect to the prediction of the concentration of the analyte, which requires non-trivial computations.
For this reason, we use integral intensities in the effort to complicate the secondary tasks.
We also implement a specific training strategy, described in \Cref{sec:arch_mt}.

\subsection{Trustworthiness of the Model Via Validation of the Predictions}\label{sec:trust}

The evaluation of the performance of any model is always possible in the training and development sets, by comparing the predictions of the model with the corresponding known ground truths.
However, the validation of the predictions on new data and the detection of anomalies remain complicated issues.
In this implementation, given the mutual dependence of the multiple outputs of the \mt architecture, we rely on the set of secondary outputs of the model to assess its performance on unknown data.
As seen previously, good predictions on sub-tasks related to the principal output concur to improve its stability and overall results.
Conversely, a bad performance on the secondary tasks leads to poor predictions on the principal output.
This is especially true if the sample matrix significantly changes in the inference set or following a modification in the experimental conditions.
For this reason, the set of sub-tasks chosen in this analysis is experimentally measurable on unknown samples.
In fact, the integral intensities of emission lines or bands can be extracted from experimental spectra at any given time, even though the concentration of the analyte remains unknown.
In turn, this can be used to assess the trustworthiness of the predictions of the model.

When dealing with new experimental data, we compute the predictions of the model on a single spectrum basis, independently.
We then average the results per sample, to smooth the influence of defects on the surface.
This procedure allows us to use standard frequentist statistical tests to compare the results to ground truths.
The Mean Absolute Percentage Error (\mape) gives a measure of the deviation of the data and the performance of the model.
For the $i$-th secondary output of the network and a sample $s$, we compute the \mape $M^{(s)}_i$ of the predicted intensities $\bfI^{(s)}_i \in \R^n$ and the corresponding ground truths $\hbfI^{(s)}_i \in \R^n$, with $n$ samples:
\begin{equation}
    M^{(s)}_i \defeq \Ev{\abs{ \frac{\bfI^{(s)}_i - \hbfI^{(s)}_i}{\hbfI^{(s)}_i} }}.
\end{equation}
This offers a first estimate of the error made by the model, independent of the variance of its predictions: it describes the trueness of the model in predicting a given quantity.
We use this value as a first step in the analysis of the trustworthiness of the predictions.
For instance, a larger \mape on the secondary outputs is an indication of worse performance of the model or anomalies, which could reflect on the main prediction.
In order to estimate a soft threshold for the quantity, we use the validation set $V$.
We compute a confidence interval around the \mape $M^{(s)}_i$ of a sample $s \in V$ for the $i$-th secondary output:
\begin{equation}
    \qty[
        M^{(s)}_i - \hatt^{\,n}_{1-\alpha} \frac{\sigma_i^{(s)}}{\sqrt{n}},
        ~
        M^{(s)}_i + \hatt^{\,n}_{1-\alpha} \frac{\sigma_i^{(s)}}{\sqrt{n}}
    ],
    \label{eq:conf}
\end{equation}
where
\begin{equation}
    \qty( \sigma_i^{(s)} )^2
    =
    \Ev{\abs{ \frac{\bfI^{(s)}_i - \hbfI^{(s)}_i}{\hbfI^{(s)}_i} }^2} - \qty( M^{(s)}_i )^2
    =
    \Var{\abs{ \frac{\bfI^{(s)}_i - \hbfI^{(s)}_i}{\hbfI^{(s)}_i} }},
\end{equation}
and $\hatt^{\,\nu}_{1-\alpha}$ is the value of the \emph{Student}'s $t$ variable for $\nu$ degrees of freedom (the number of spectra for each sample), at a confidence level $1 - \alpha$.

A second analysis is based on a Student test on the predicted intensity, given its nature of repeated measurement on the sample.
Supposing that the ground truth value of a sample $s$ for the $i$-th average intensity has a sample variance $\sigma_i^2$, we can compute the random variable
\begin{equation}
    t_i^{(s)}
    \defeq
    \frac{\abs{ \Ev{\bfI^{(s)}_i} - \Ev{\hbfI^{(s)}_i} }}{\sqrt{\Sigma_i^2 + \sigma_i^2}},
\end{equation}
where $\Sigma_i^2$ is the sample variance of the predictions.
By including the dependence on the variance, this test gives a statistical measure of the ability of the \mt model to generalise to unknown data.
In order to discriminate possible anomalies, we can use a standard approach by choosing a threshold value $\hatt^{\,n}_{1-\frac{\alpha}{2}}$ of a two-tailed $t$-test at confidence level $1 - \alpha$, with $n$ degrees of freedom, such that the probability $\rP\qty(t_i^{(s)} \ge \hatt^{\,n}_{1-\frac{\alpha}{2}}) = \alpha$.
Unknown samples can be compared to validation standards according to the compatibility of their principal outputs.
This way, we recover a probabilistic interpretation of the result in terms of confidence: models can be compared based on their performance at different values of $\alpha$ on the secondary outputs.

\begin{figure}[t]
    \centering
    \includegraphics[width=\linewidth]{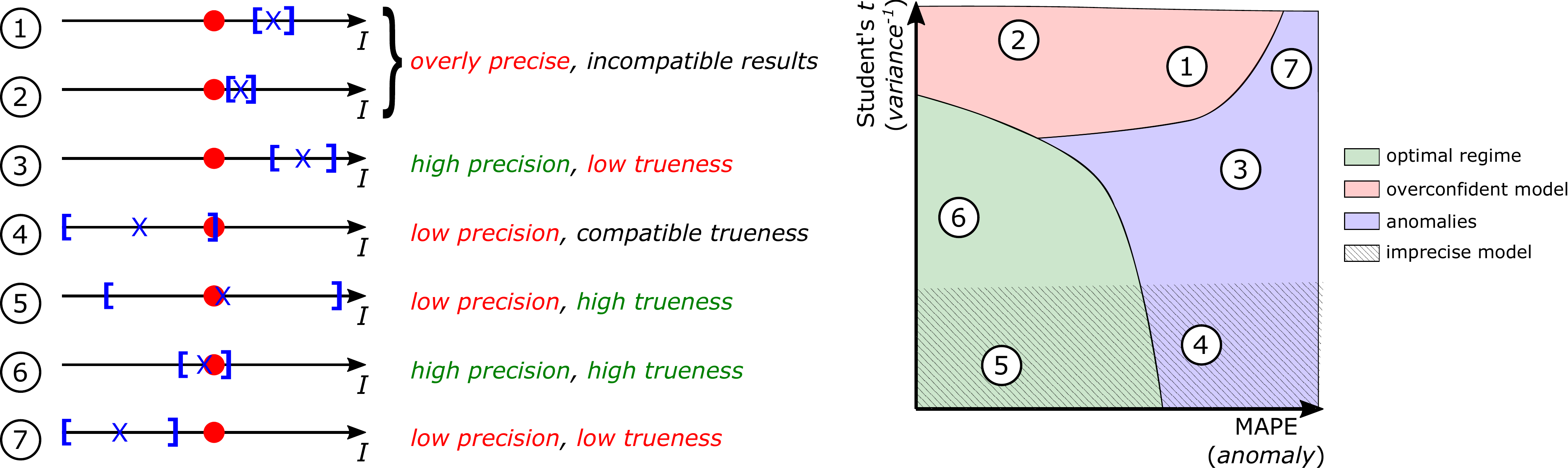}
    \caption{%
        \emph{Interpretation of the trustworthiness assessment results.}
        \mape and Student's $t$ are used as measure of the confidence of the predictions, with possible outcomes summarised on the left of the figure.
        Interpretations of these results are presented schematically on the right.
    }
    \label{fig:validation_cases}
\end{figure}

The values of \mape and of the Student’s $t$ variable can be used together to evaluate the trustworthiness of the model and characterise its predictions.
For instance, large \mape and low $t$ values may be an indication of the model ability to extrapolate the results for out-of-distribution samples or an anomaly, though with high variance.
In other words, the prediction is compatible with the ground truth, though not precise.
Conversely, small values of \mape and large $t$ values can be interpreted as anomalies in training, leading to an artificially good prediction, whose variance is too small to explain the variability of the data.
Hence, in this case, the prediction is close to the ground truth, though not compatible with the real value.
We graphically summarise these interpretations in the plane in \Cref{fig:validation_cases}.
Though the confidence level of the principal output is not easily computed from the confidence of the secondary outputs, this measure gives an implicit feedback on the main output.
Given the dependencies of the \mt parameters in \Cref{sec:mt_cnn}, the information determines whether the prediction of the concentration of the analyte is trustworthy.

\section{Implementation and Experimental Details}\label{sec:details}

In this section, we describe experimental methods.
We also present the details of the \mt model based on \cnns used in the analysis.
We then highlight the differences with the univariate Linear Regression (\lr) and multivariate algorithms such as Multi-Linear Regression (\mlr), Partial Least Squares (\pls[1]), and Fully Connected Neural Networks (\fcnns).

\subsection{Experimental Setup}\label{sec:data}

We compare the predictive ability of different algorithms on two types of datasets.
We consider 19 cement samples, whose elemental compositions are reported in \Cref{tab:cement_samples} in \Cref{app:matrices}, and on 6 alloy matrices with 4 to 6 samples each, summarised in \Cref{tab:alloys_part1,tab:alloys_part2} in \Cref{app:matrices}.
The first were built in the framework of an interlaboratory comparison in \num{2021}~\cite{Volker:2022:Interlaboratory} by the \emph{Bundesanstalt f\"{u}r Materialforschung und -pr\"{u}fung} (BAM) in Berlin, Germany.
All measurements were carried out in air, at room temperature.

\begin{figure}[t]
    \centering
    \includegraphics[width=0.49\linewidth]{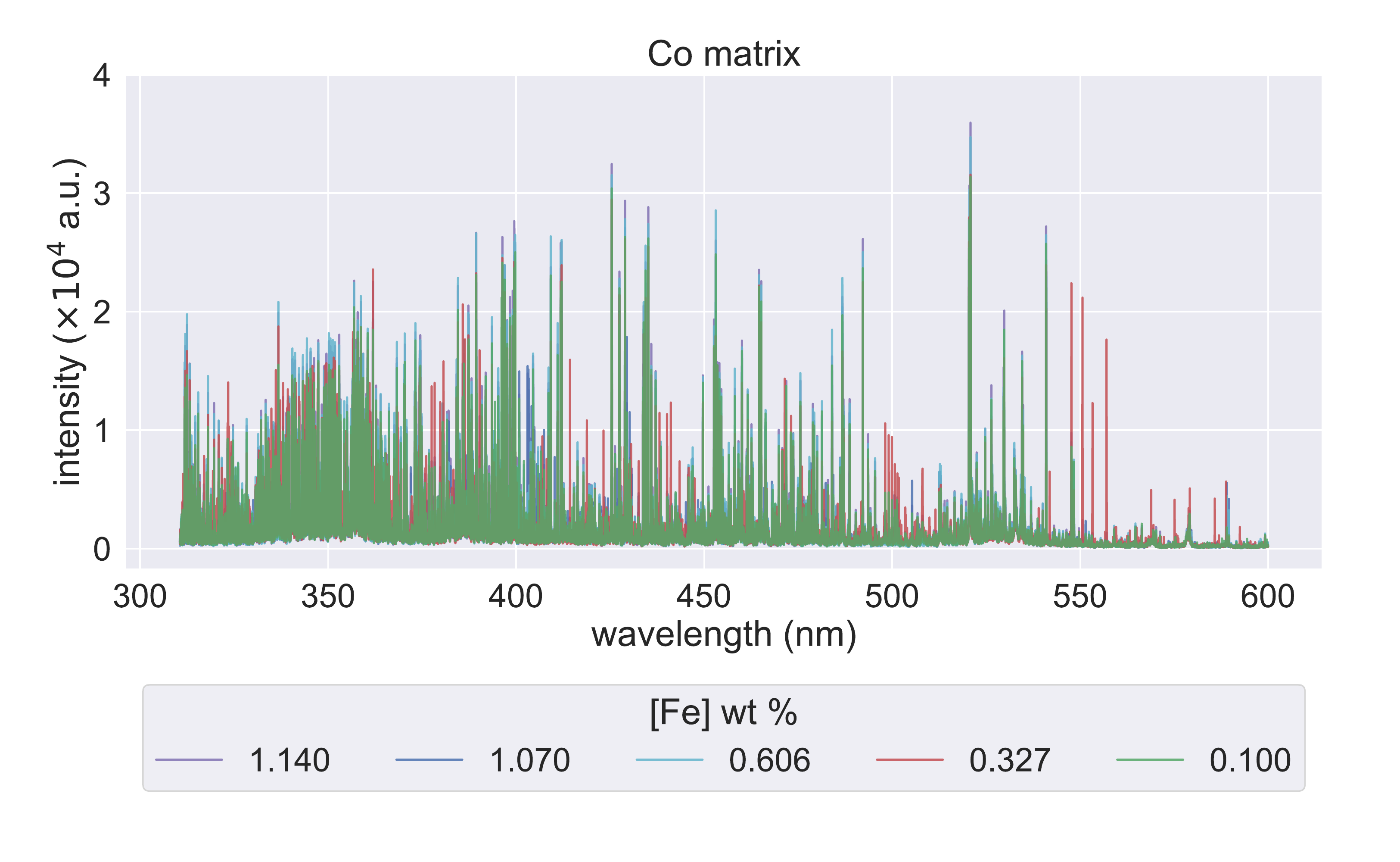}
    \hfill
    \includegraphics[width=0.49\linewidth]{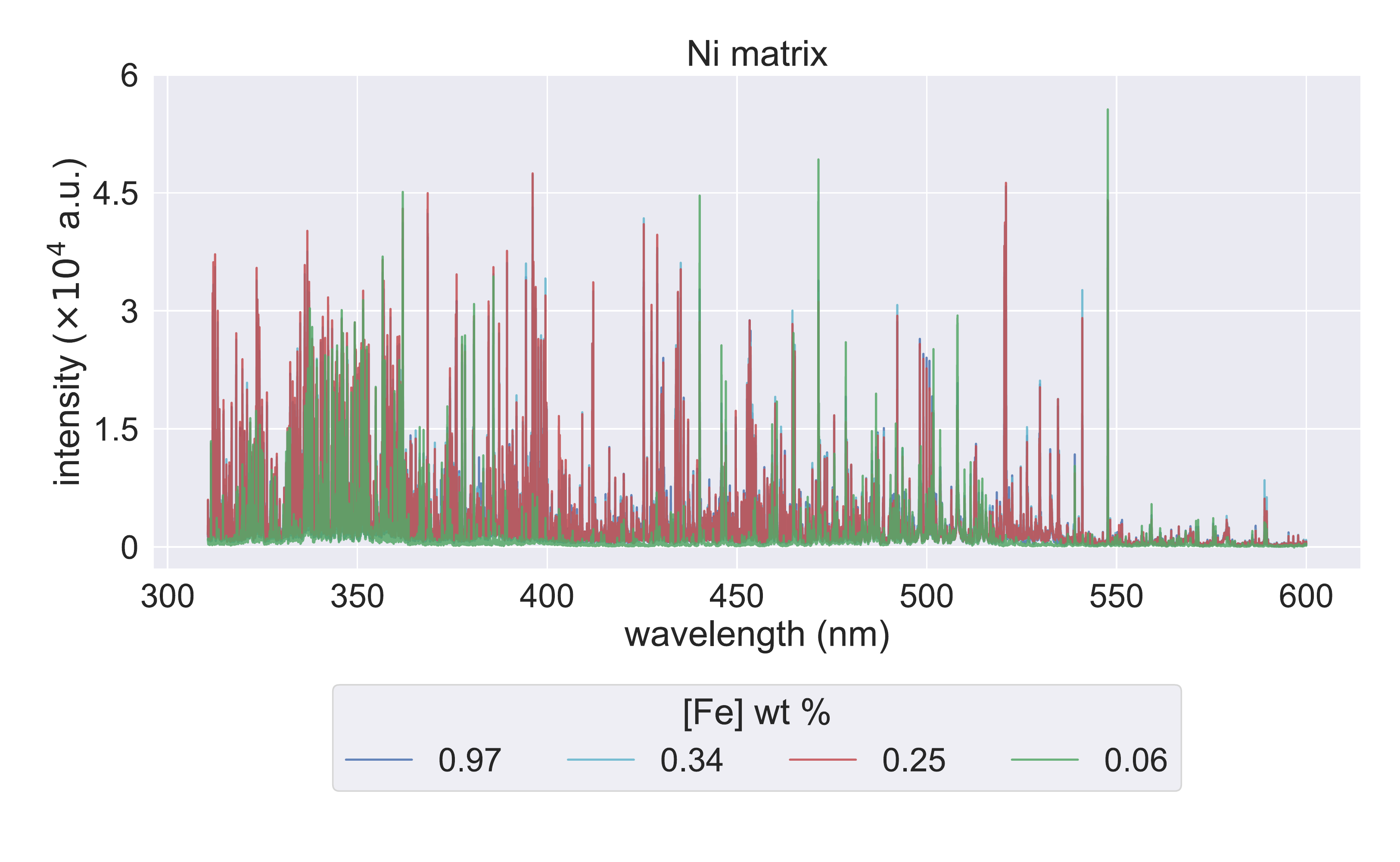}
    \\
    \includegraphics[width=0.49\linewidth]{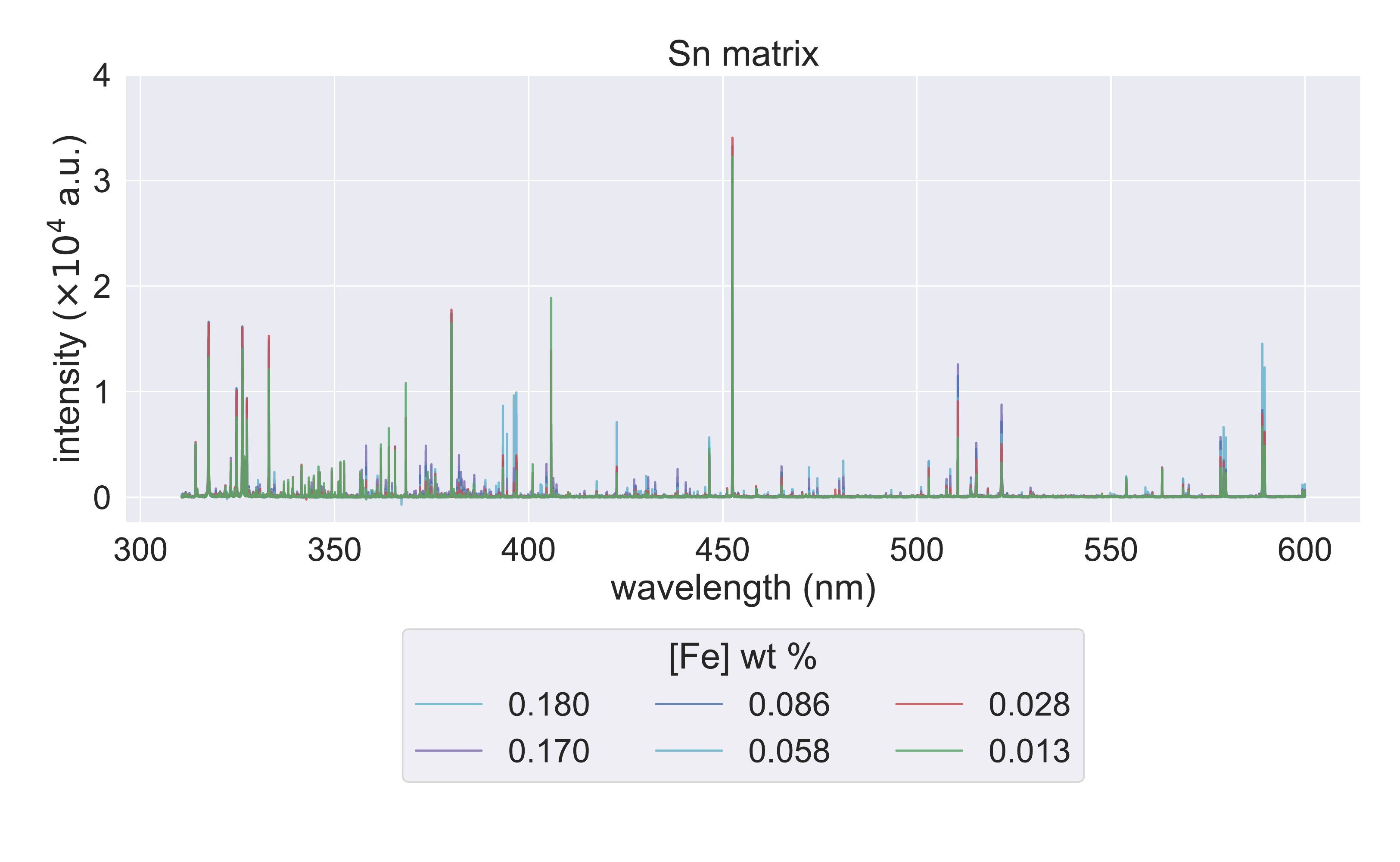}
    \hfill
    \includegraphics[width=0.49\linewidth]{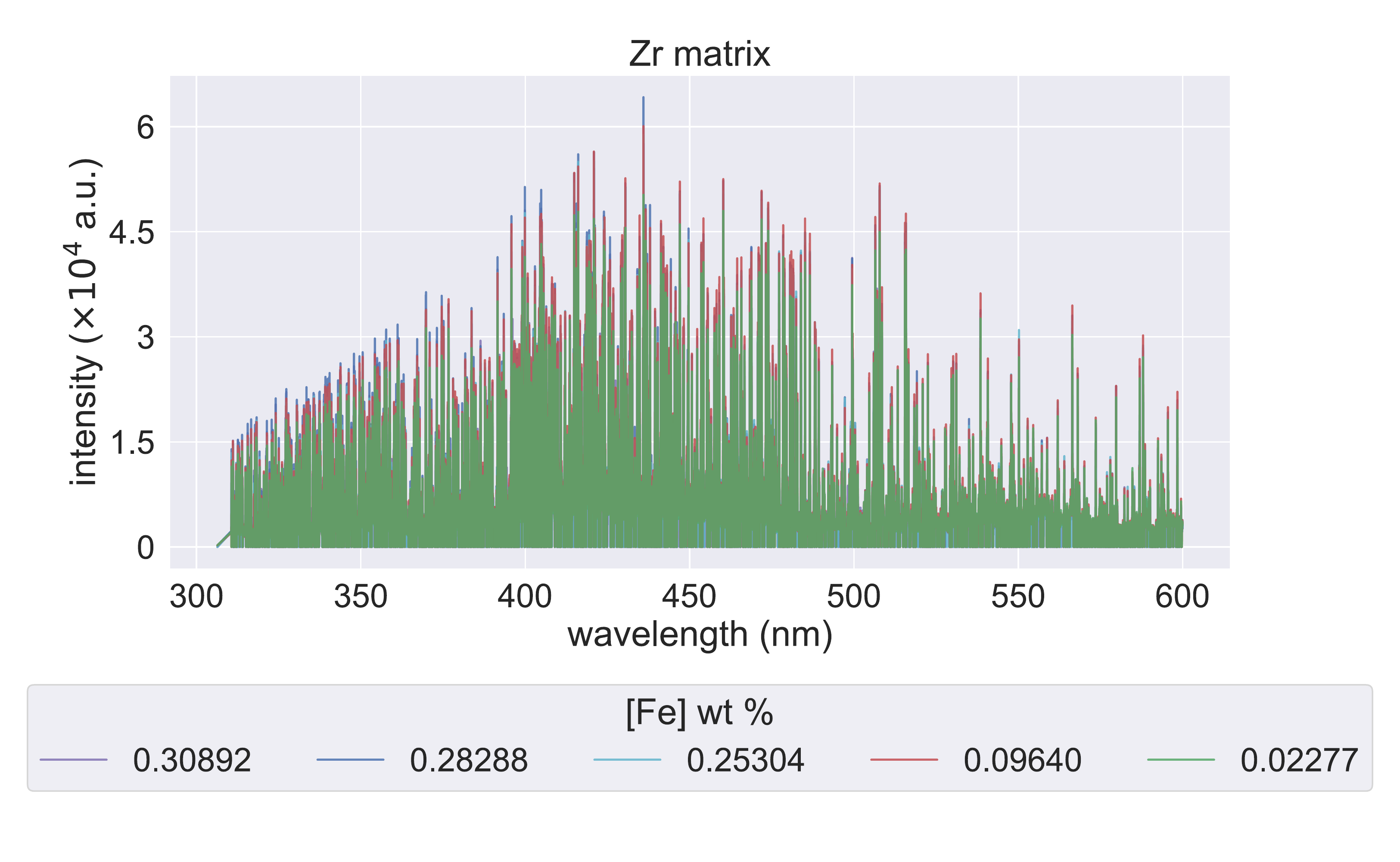}
    \caption{%
        \emph{Average spectra of the selected alloy matrices.}
        This selection allows displaying different degrees of noise and spectral interference at different concentration levels of the analyte.
    }
    \label{fig:alloys_avg}
\end{figure}

Data on alloys were collected using a Nd:YAG laser (\emph{Quantel Ultra}) operating at a wavelength of \SI{266}{\nano\meter}, \SI{6}{\milli\joule} pulse energy, and \SI{4}{\nano\second} pulse length.
The plasma emission was analysed with a \emph{LTB Aryelle 400} spectrometer with a fixed aperture of $\SI{50}{\micro\meter} \times \SI{50}{\micro\meter}$, equipped with an \emph{Andor DH740} ICCD, in the range \SIrange{310}{613}{\nano\meter} (resolving power: $\lambda / \Delta\lambda \simeq \num{1.8e4}$).
We used a gate delay of \SI{1}{\micro\second} and a gate width of \SI{0.5}{\micro\second}.
For each sample, 25 spectra were collected, accumulating 20 laser shots for per crater.
The irradiance on the surface of the samples was \SI{76}{\giga\watt\per\square\centi\meter}.
A selection of average spectra is shown in \Cref{fig:alloys_avg} (see \Cref{fig:full_alloy_avg} in \Cref{app:figures} for the other matrices).

\begin{figure}[t]
    \centering
    \includegraphics[width=0.75\linewidth]{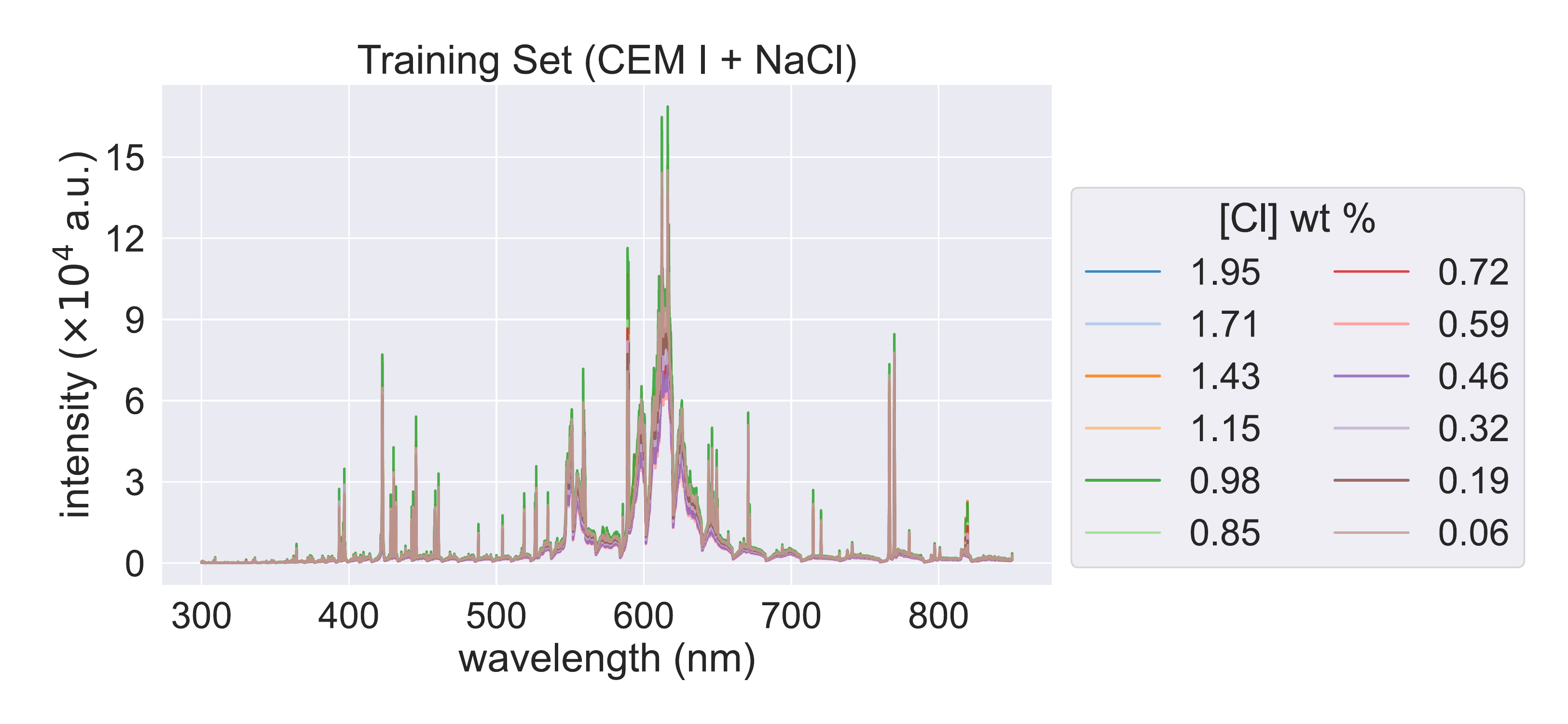}
    \\
    \includegraphics[width=0.75\linewidth]{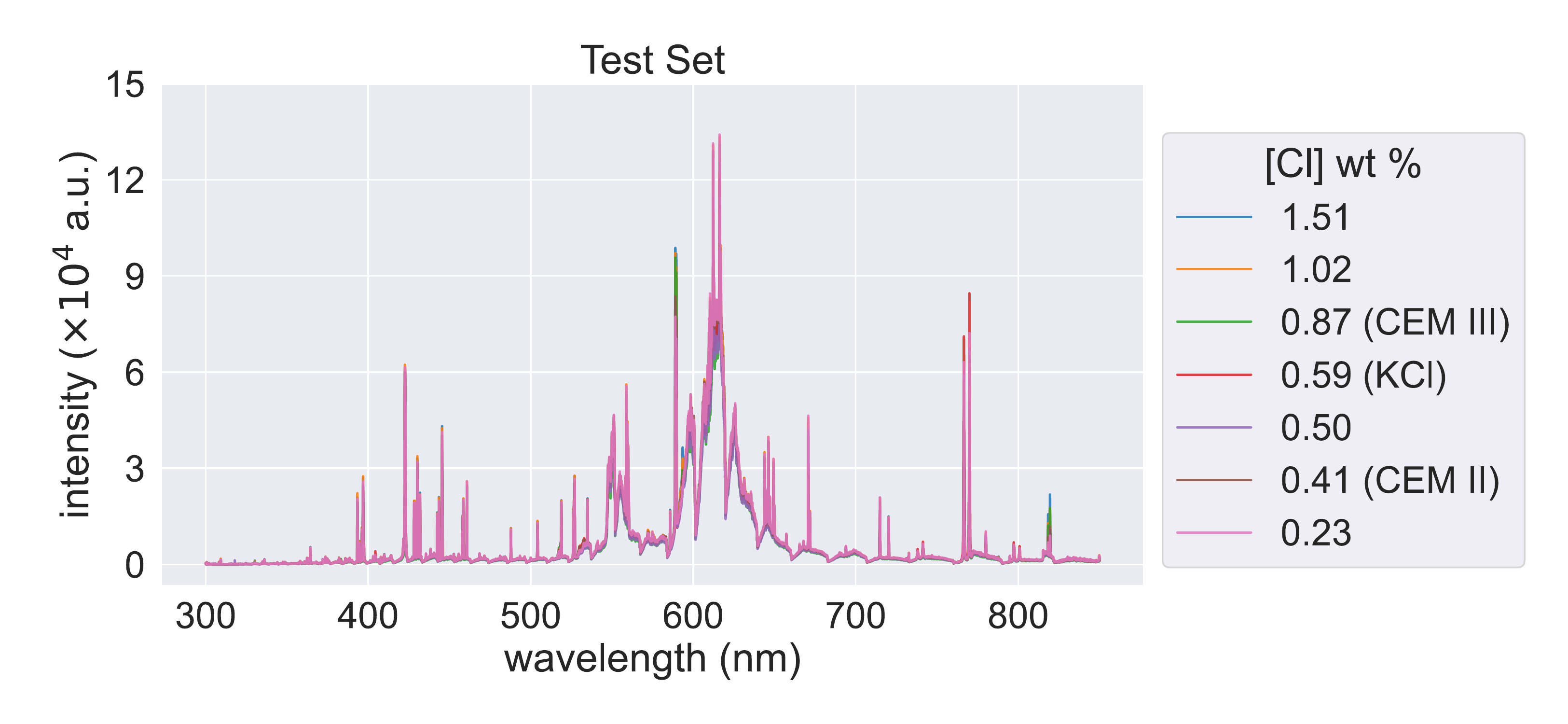}
    \caption{%
        \emph{Average spectra of the cement samples.}
        Spectra are separated into training and test sets.
        Samples were fabricated using CEM I, with the addition of NaCl, unless otherwise stated.
    }
    \label{fig:cl_avg}
\end{figure}

Cement samples, whose average spectra are shown in \Cref{fig:cl_avg}, were probed using a Nd:YAG laser (\emph{Quantel Brio}) at a wavelength of \SI{1064}{\nano\meter}, \SI{15}{\milli\joule} pulse energy and \SI{4}{\nano\second} pulse length.
For each sample, 25 spectra were collected, accumulating 40 laser shots (5 pre-ablation shots).
We used a \emph{Mechelle} spectrometer with an \emph{Andor iStar} ICCD and a fixed aperture of $\SI{50}{\micro\meter} \times \SI{50}{\micro\meter}$, \SI{10}{\micro\second} gate delay, and \SI{100}{\micro\second} gate width in the range \SIrange{200}{975}{\nano\meter} (resolving power $\lambda / \Delta\lambda \simeq \num{4e3}$ measured at \SI{589.60}{\nano\meter} on a Na peak).
The irradiance on the sample surface was \SI{190}{\giga\watt\per\square\centi\meter}.

\subsection{Secondary Outputs}\label{sec:secondary}

As the prediction of a peak intensity may result in simple secondary tasks for the \nns, we use integral intensities on a neighbourhood of wavelength channels to complicate the computation.
Other strategies are also employed to solve the possible issue, though they are discussed in~\Cref{sec:arch_mt}, when we introduce the architecture and the training procedure.
Notice that these secondary outputs of the \mt architecture are considered to be the inputs in the \mlr and \fcnns for this analysis.

For the analysis of the alloy matrices, we use the intensities of the most intense persistent lines of Fe in the spectral range considered, integrated over 10 wavelength channels.
Specifically, we choose 8 lines, reported in \Cref{tab:fe_lines} in \Cref{app:details}.
For the cement samples, we consider two molecular bands centred at \SI{593.46}{\nano\meter} and \SI{617.74}{\nano\meter}, and integrated over 14 channels in the intervals \SIrange{593.20}{593.64}{\nano\meter} and \SIrange{617.47}{617.93}{\nano\meter}.
As a general guideline, choosing a large number of secondary outputs helps the convergence of the network, and preserves a good generalisation performance of the model.
Moreover, since the \cnn architecture of the \mt model is sufficiently versatile, there is no need for an accurate selection of the emission lines.

\subsection{Train and Test Sets Selection}

In the case of alloy matrices, given the small number of samples available, \SI{30}{\percent} of the spectra for each sample is retained as independent test set, while the rest is used as training set for the baseline models, and as input of the data augmentation for the \mt architecture.
This ensures, on average, an in-sample inference for the algorithms (the selection of test samples comes from the same samples in the training set).
Moreover, it provides the means to verify whether the augmentation technique correctly enhances the training distribution.
Notice that this does not automatically translate into a simpler task for the model: spectra in the test set may still differ from the training distribution, due to random and local fluctuations, thus they may represent out-of-distribution data on a spectrum basis.

Differently, cement matrices are separated into training and test sets on a sample basis: 12 matrices are considered as calibration set, while 7 samples are used for inference.
In the test set, we insert specifically samples which present a different matrix (type of cement) or manufacturing procedure (different salts added in the mixture) in order to check the out-of-distribution generalisation ability of the algorithms (see \Cref{tab:cement_samples} in \Cref{app:matrices} for details), and their ability to recognise possible anomalies.

For \lr and \mlr, no additional validation sets have been considered, as no free parameters are present in the algorithms: for these baselines, we do not introduce lasso or ridge regularisations, thus no hyperparameter optimisation procedures are necessary.
For \pls[1], we perform a cross-validation procedure (5-fold, chosen to optimise the number of spectra in each permutation of the data and the computation wall time) to determine the best number of components to retain.
However, in the case of \nns, both \fcnn and \mt \cnn, several hyperparameters (number of hidden layers, number of units, size of the convolutional kernels, etc.) are considered.
Given the time consuming operations involved, we use a single holdout validation set made of \SI{20}{\percent} of the spectra contained in the training set, selected using a stratified strategy to preserve the fraction of spectra for each sample.
Notice that, in the case of the data-augmented training, we provide the results for different sizes of the training set, namely \numlist{100;1000;5000;10000} synthetic spectra for each experimental sample.
In this case, the validation set is chosen from the synthetic data: in order to avoid any data leakage, we avoid using the experimental spectra when optimising the model.
Results always refer to the independent experimental test set.

\subsection{Preprocessing}

We first remove peripheral zones by reducing the spectra of the alloy matrices to the interval \SIrange{310}{600}{\nano\meter}, and the cement spectra in the range \SIrange{300}{850}{\nano\meter}.
This sets the number of wavelength channels in the range \num{67829} to \num{68438} in the first case, and to \num{16917} in the second.

Outliers are then removed from the training set, either experimental or synthetic: in our analysis, at a given wavelength, outliers are spectra which present an intensity outside the interval between the 5\textsuperscript{th} and 95\textsuperscript{th} percentile of the values.
The goal is to build performing calibration models without using extreme configurations.
In order to test the generalisation ability of the models, we retain the outliers in the test set.
For the alloy samples, we focus on the Fe line at \SI{373.49}{\nano\meter} (the most intense persistent line), while, for the cement matrices, we consider the molecular band of CaCl at \SI{593.46}{\nano\meter}.

Finally, spectra are normalised using the integral intensity at a given wavelength.
Since the procedure is performed independently on each spectrum, we can also safely normalise the spectra in the test sets.
For the alloy matrices, we consider the integrated intensity of the most intense emission line of the matrix itself over an interval of 10 wavelength channels (emission lines and corresponding integration intervals are reported in \Cref{tab:normal} in \Cref{app:details}).
The cement samples have been normalised to the intensity of the CaO molecular band at \SI{615.03}{\nano\meter}, integrated over 20 wavelength channels in the interval \SIrange{614.66}{615.38}{\nano\meter}.

\subsection{Evaluation Metrics}

In the framework of the analysis, different evaluation metrics are used to assess the performance of the models on their predictions.
In order to obtain comparable results, we adopt the same metrics and strategy for the univariate and multivariate analyses.
We use common regression metrics such as the Mean Squared Error (\mse, also expressed in the same units as the predictions by its square root, the Root Mean Squared Error, or \rmse), the Mean Absolute Error (\mae), and the \mape to score the results of the algorithms.
We provide the results of the algorithms on the independent test set, made of experimental spectra, unless otherwise specified.

\subsection{Training and Optimisation Strategies}\label{sec:training}

Depending on the algorithms used for the analysis, we use different training techniques and representations of the input data.
We also define different optimisation strategies, in order to fix the optimal hyperparameters of the models, used for inference.

\subsubsection{Input Representation}

In the case of \lr, we consider the integral intensity of selected wavelength channels as inputs of the model.
Specifically, we consider the integral intensity of the Fe emission line at \SI{373.49}{\nano\meter}, integrated over 10 wavelength channels, for the alloy matrices.
For the cement samples, we use the CaCl molecular band at \SI{593.46}{\nano\meter}, integrated over 14 channels.
As already stated at the end of \Cref{sec:secondary}, for the \mlr and the \fcnn, we select several atomic Fe emission lines as inputs, in the case of the alloy matrices.
We use the CaCl molecular bands at \SI{593.46}{\nano\meter} and \SI{617.74}{\nano\meter} for the cement samples.
In both cases, they are the same variables used as secondary outputs by the \mt model.
On the other hand, for \pls[1] and the \mt network, we use the entire spectra as input.
Finally, we choose a quadratic regression model in the case \lr (without interaction terms), while we consider a simple regression model for \mlr.

\subsubsection{Fully Connected Neural Networks}

In the case of fully connected networks, we consider a simple architecture, described in detail in \Cref{sec:fcnn}, with a traditional \mse loss function.
We introduce both $\ell_1$ and $\ell_2$ regularisations for the parameters of the network.
The latter is defined both independently and as weight decay~\cite{Loshchilov:2017:Decoupled} in the gradient descent optimiser.
The global loss function is, thus:
\begin{equation}
    \cL_{\text{glob}}\qty( \bfy, \hbfy;\, \Omega;\, \alpha, \beta)
    =
    \Ev{\qty( \hbfy - \bfy )^2}
    +
    \alpha\, \cL_1\qty( \Omega )
    +
    \beta\, \cL_2\qty( \Omega ),
\end{equation}
where $\cL_p\qty( \Omega ) = \sqrt[\uproot{2}p]{ \sum_{\omega \in \Omega} \abs{\omega}^p }$ is the $\ell_p$ norm of the set of parameters of the network $\Omega$.
In the expression of the loss, $\bfy$ and $\hbfy$ represent the ground truth values and their predictions.
The two hyperparameters $\alpha$ and $\beta$  control the entity of the penalties.
The learning objective is thus a convex optimisation problem, in principle freeing training from weight initialisation issues.
We use the \adamw optimiser~\cite{Loshchilov:2017:Decoupled}, which has been shown to grant better generalisation abilities with respect to other implementations (e.g.\ the traditional \adam algorithm~\cite{Kingma:2014:Adam}).
We leave the default hyperparameters $\beta_1$ and $\beta_2$ in its definition, and an initial learning rate defined by the optimisation.

\subsubsection{Multitask Architectures}\label{sec:arch_mt}

The \mt loss presents additional peculiarities, with respect to the previous case:
\begin{equation}
    \cL_{\text{MT}}\qty( \bfY, \hbfY;\, \Omega;\, \bfv, \alpha, \beta)
    =
    \bfv \cdot \bfL\qty( \bfY, \hbfY;\, \Omega )
    +
    \alpha\, \cL_1\qty( \Omega )
    +
    \beta\, \cL_2\qty( \Omega ).
\end{equation}
The vector $\bfL\qty( \bfY, \hbfY;\, \Omega ) \in \R^p$ contains the losses of each prediction output.
With this notation, $\bfY \in \R^{n \times h}$ and $\hbfY \in \R^{n \times h}$ are matrix representations of the ground truths and their predictions, respectively, where $h$ is the number of outputs of the model.

In this formulation, $\bfv$, $\beta$ and $\gamma$ are hyperparameters of the model.
As we deal with the multiple outputs of the network, $\bfv \in \R^h$ is the vector of coefficients of the linear combination of single-output loss functions.
In this article, we adopt the strategy to weight equally the secondary outputs, and to perform an adaptive adjustment of the scalar weight $\hatv$ of the principal output.
In particular, we start the learning process with a coefficient $0 < \hatv_0 < 1$ for the main output.
After each epoch $t$ (full iteration of the training dataset), we implement an exponential decrease of the learning coefficient $\hatv_{t+1} = \eta\, \hatv_t$.
Secondary outputs are all weighted by a factor $1 - \hatv_t$.
In other words, we consider the hyperparameter vector $\bfv_t = \qty( \hatv_t, \qty(1 - \hatv_t) \1_{h-1})$ for each epoch $t$, where $\1_d$ is a $d$-dimensional unit vector.
Both $\hatv_0$ and $0 < \eta < 1$ are subject to hyperparameter optimisation.
The result is an algorithm capable of first learning the most difficult output (usually, the concentration of the analyte), and then fine-tuning on the simpler predictions (the integral intensities of the emission).

\begin{figure}[t]
    \centering
    \includegraphics[width=0.75\linewidth]{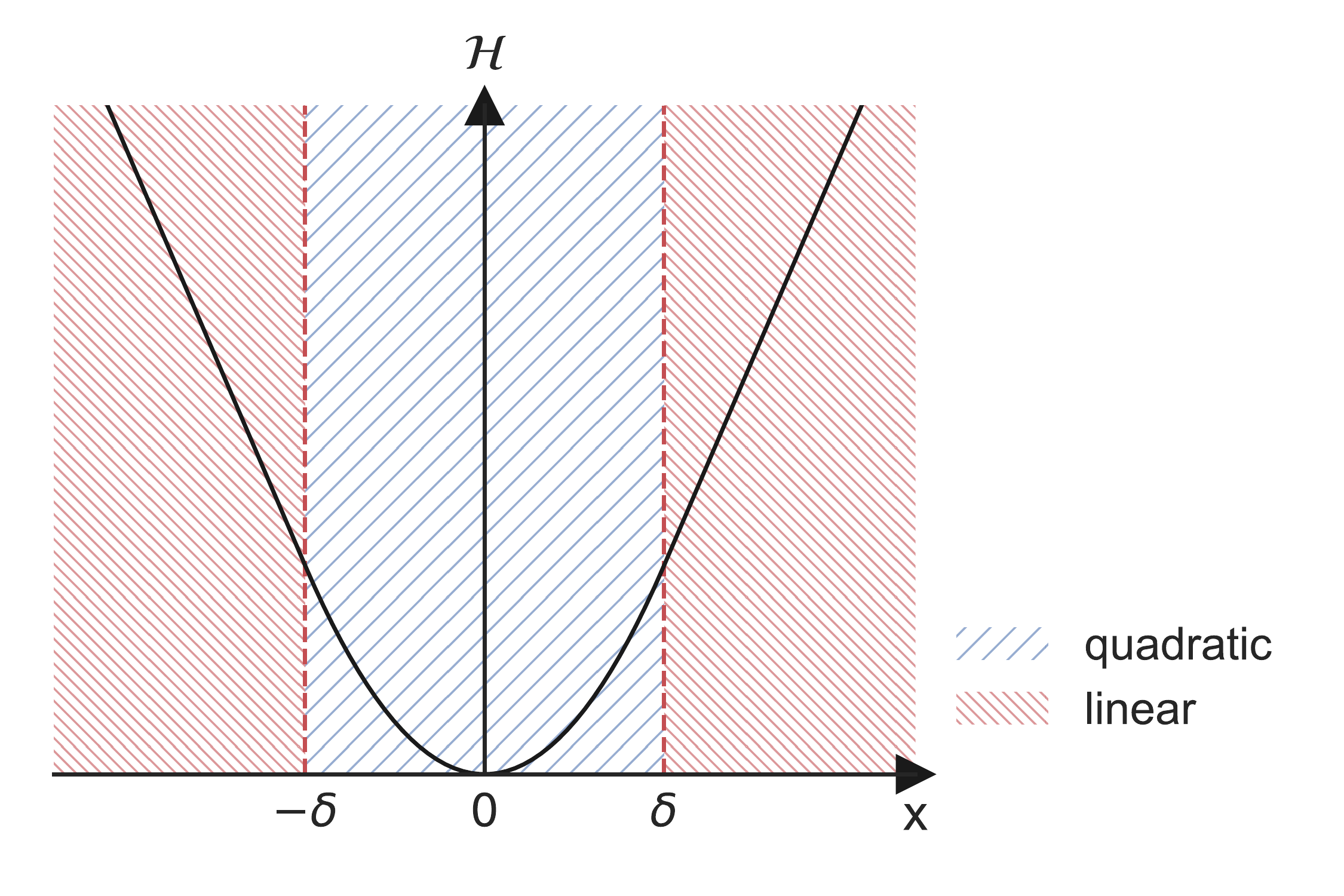}
    \caption{%
        \emph{Huber loss function.}
        The Huber loss is a continuous interpolation of a quadratic curve with a linear behaviour, due to the presence of the hyperparameter $\delta$.
    }
    \label{fig:huber_loss}
\end{figure}

We adopt the \mse loss function for the secondary outputs of the \mt architecture.
However, for the main output (the concentration of the analyte), we introduce a \emph{Huber} loss~\cite{Huber:1964:Robust} to ease the learning of the principal output.
Such loss function is defined as:
\begin{equation}
    \cH\qty( \bfy, \hbfy;\, \delta)
    =
    \begin{cases}
        \frac{1}{2} \Ev{\qty( \hbfy - \bfy )^2} &\qif \abs{\hbfy - \bfy} < \delta
        \\
        \delta\, \Ev{\abs{\hbfy - \bfy} - \frac{\delta}{2}} &\qif \abs{\hbfy - \bfy} \ge \delta
    \end{cases}.
\end{equation}
This formulation is a continuous and differentiable interpolation between a quadratic (\mse) and linear (\mae) loss, depending on the value of a hyperparameter $\delta$ (see \Cref{fig:huber_loss}).
The loss is less sensitive to outliers, given its linear behaviour for deviations greater than $\delta$.
It collapses to the \mse for $\delta = 0$.

As in the \fcnns case, we use the \adamw optimiser with default hyperparameters.
The initial learning rate is defined during the optimisation.
When dealing with different numbers of spectra in the simulated training sets, we adjust the mini-batch size of the stochastic gradient descent to fit learning needs and \gpu memory.
When the mini-batch size is modified by a given multiplicative factor, we also scale the learning rate of the same factor, to improve the chances of convergence.
We also implement an exponential learning rate scheduling: after each training epoch $t$, the learning rate $q_t$ is multiplied by a factor $0 < \varphi < 1$ as $q_{t+1} = \varphi\, q_t$.

\subsubsection{Hyperparameter Optimisation Strategy}

In the case of the \lr and \mlr, no optimisation is required, hence no validation dataset is used in those cases.
Training is directly performed on the dedicated set, before computing the final predictions on the test set.
For \pls[1], we optimise the number of components used to infer the results.
We build a 5-fold cross-validation strategy in order to seek the best hyperparameter.
We use a grid search approach (see \Cref{tab:pls_comps} in \Cref{app:details} for the number of components used in the \pls[1] models).
Differently, in the case of \nns, we use a separate holdout validation set for the development of the models.
We train several models in parallel with different hyperparameters, and we retain the architecture with highest evaluation score on the validation set.
For both experimental datasets, we use the \mse on the validation set of the predicted concentrations as scalar metric to evaluate the best model.

The values of the hyperparameters are sampled using a tree-structured \emph{Parzen} estimator~\cite{Bergstra:2011:Algorithms}.
It enables to use Gaussian mixtures to evaluate the posterior distribution of the hyperparameters, making the optimisation easier.
We run 200 cycles of optimisation of at most 200 epochs.
In all cases, an early stopping strategy is employed where no improvement on the validation loss occurred for at least 100 epochs.
Given the different nature of the data, we optimise the \fcnn and the \mt architecture independently for the alloy and cement matrices.
However, for metallic samples, we run the full optimisation only on the Al matrix, for simplicity.
For the \mt network, we run the optimisation cycle on synthetic data: we optimise the architecture for \num{5000} synthetic spectra per sample, and a noise parameter $\beta = 0.05$.
We fix the mini-batch size to 128.
Once the hyperparameter are chosen, we separately re-train the best model for different sizes of the simulated training sets and noise parameters.

\subsubsection{Training of the Neural Networks}

We train the \fcnn for \num{1000} epochs, with an early stopping after 500 epochs without improvements on the validation loss, and a mini-batch size of 4 spectra.
The \mt networks are trained for 500 epochs, with an early stopping after 250 epochs without improvements on the validation loss.
In this case, we choose to adjust the mini-batch size to \numlist{16; 32; 64; 128} spectra when working with \numlist{100; 1000; 5000; 10000} simulated spectra per sample.

\subsection{Neural Network Architectures}

In this analysis, we use two different kinds of networks: a classical \fcnn as baseline, and a \mt \cnn.
In this section, we detail the implementation of these architectures as used in the analysis of the \libs data.

\subsubsection{Fully Connected Neural Networks}\label{sec:fcnn}

The first architecture we consider is a simple multi-layered perceptron~\cite{Rosenblatt:1958:Perceptron} with a single output unit.
The $n$-th layer in the architecture computes:
\begin{equation}
    \bfx^{(n+1)}
    =
    \cB\qty( f\qty( \cD_r\qty( \bfx^{(n)} \cdot \bfW^{(n+1)} + \bfb^{(n+1)} ) ) ),
\end{equation}
where $\bfx^{(n)} \in \R^p$ is the input vector and $\bfx^{(n+1)} \in \R^q$ is the activated output.
In the expression, $\bfW^{(n+1)} \in \R^{p \times q}$ and $\bfb^{(n+1)} \in \R^q$ are the weight and bias of the $n$-th layer.
A dropout operation $\cD_r$ with rate $0 \le r < 1$ is used as regularisation strategy.
Finally, $f$ is the activation function.
The function $\cB$ is a batch normalisation (\bn) operation~\cite{Ioffe:2015:Batch}, used to accelerate the convergence of the network.
That is, for each mini-batch $\qty{\bfz_i \in \R^p,\, i = 1,\, 2,\, \dots,\, B}$ of data in the gradient descent, the inputs are rescaled:
\begin{equation}
    \hbfz_i = \frac{\bfz_i - \bfmu_B}{\bfsigma_B},
    \qquad
    i = 1,\, 2,\, \dots,\, B,
\end{equation}
where
\begin{equation}
    \bfmu_B = \frac{1}{B - 1} \sum\limits_{i = 1}^B \bfz_i,
    \qquad
    \bfsigma_B^2 = \frac{1}{B - 1} \sum\limits_{i = 1}^B \qty( \bfz_i - \bfmu_B )^2
\end{equation}
are the empirical mean and variance of the mini-batch.
\bn is defined as:
\begin{equation}
    \begin{tabular}{ccccc}
        $\mathrm{BN}_{\beta, \gamma}$ & $\colon$ & $\R^{B \times p}$ & $\to$     & $\R^{B \times p}$
        \\
                                      &          & $\bfZ$                   & $\mapsto$ & $\mathrm{BN}_{\beta, \gamma}\qty( \bfZ )$
    \end{tabular},
\end{equation}
where
\begin{equation}
    \mathrm{BN}_{\beta, \gamma}\qty( \bfZ ) = \gamma\, \hbfZ + \beta.
\end{equation}
In the last equation, $\gamma$ and $\beta$ are learnable parameters of the network.
Other than speeding up training, \bn can also restrict the growth the values of the activated outputs of the layers, thus making it easier to train complex models.

\begin{figure}[t]
    \centering
    \includegraphics[width=0.75\linewidth]{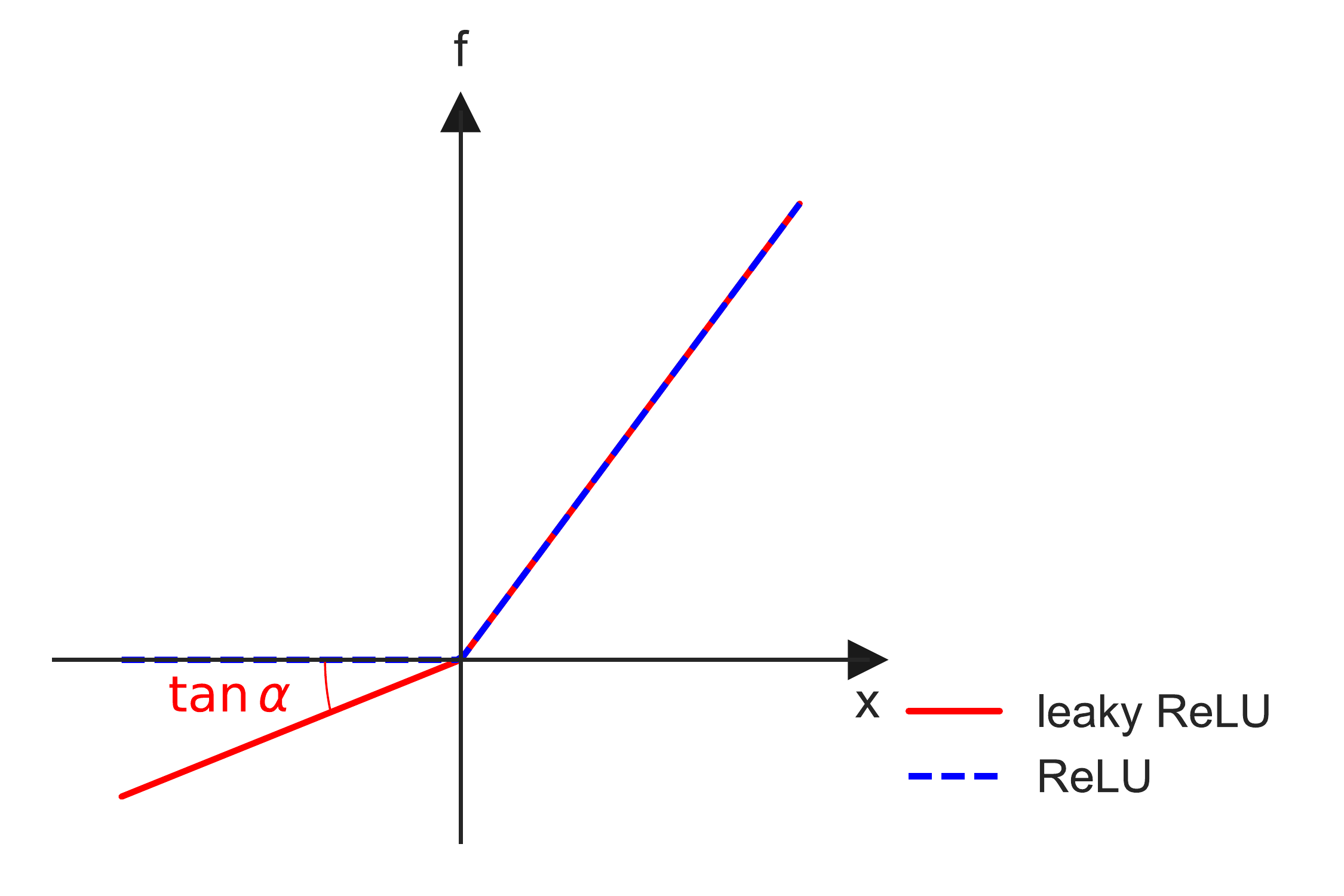}
    \caption{%
        \emph{Leaky \relu activation.}
        Differently from the usual \relu function, the hyperparameter $\alpha$ of the leaky \relu represents its negative slope.
    }
    \label{fig:leaky_relu}
\end{figure}

For the analysis of the alloy matrices, the network contains 4 hidden layers with \numlist{136; 56; 328; 392} units, respectively.
The dropout rate is \num{0.23}.
The activation function used in the architecture is a leaky \emph{REctified Linear Unit} (\relu) with a slope $\alpha = 0.03$.
Defined as $z \mapsto \max(z, \alpha z)$, it becomes a simple \relu for $\alpha = 0$, as in \Cref{fig:leaky_relu}.
The initial learning rate is \num{e-3}, and the weight decay is \num{3e-2}.
In the case of cement samples, the network has 2 hidden layers with \numlist{424; 152} units.
The dropout rate is fixed to \num{0.17}.
The activation function is, again, a leaky \relu with a slope of \num{5.5e-2}.
For both types of experimental dataset, the optimisation procedure leads to vanishing hyperparameters for $\ell_1$ and $\ell_2$ regularisations.

\subsubsection{Multitask Neural Networks}

The multitask architecture is fundamentally built on strided convolution operations.
Given two vectors $\bfx \in \R^p$ and $\bfv \in \R^k$, and an integer \emph{stride} $s$, the operation can be written in components:
\begin{equation}
    \qty( \bfx *_s \bfv )_i
    =
    \sum\limits_{n = 0}^{k - 1} x_{i\, s + n} v_n,
    \qquad
    i = 1,\, 2,\, \dots,\, q,
    \qquad
    q = \left\lfloor 1 + \frac{p - k}{s} \right\rfloor.
\end{equation}
The $n$-th convolutional layer, thus computes:
\begin{equation}
    \bfx^{(n+1)}
    =
    f\qty( \cD_r\qty( \cB\qty( \bfx^{(n)} *_s \bfv^{(n+1)} + \bfb^{(n+1)} ) ) ),
\end{equation}
where $\bfv^{(n+1)}$ and $\bfb^{(n+1)} \in \R^q$ are learnable parameters of the $n$-th layer.
We do not use pooling layers, since, in the case of spectral, they may result in the loss of relevant information among neighbouring wavelength channels (as also pointed out in~\cite{Zhang:2019:DeepSpectra}).

In the case of the alloy matrices, the shared representation network contains 3 convolutional layers, while the task-specific networks present 4 convolutional and 1 fully connected layer.
The strides are particularly wide, in order to reduce significantly the size of the outputs of the layers.
In this case, \bn was used throughout the architecture, to contain the values of the activations.
In the implementation, a $\ell_2$ regularisation loss was also added to constrain the growth of the free parameters.

For the cement samples, the shared representation is the result of 2 convolutional layers.
The task-specific networks are made of 3 hidden convolutional layers.
Only the heads of the network use \bn operations.
The final regressor networks are made of a single hidden layer.
All layers present a dropout operation, and the activations vary between pure \relu and leaky \relu.

For both architectures, different values of the initial learning rate are also used.
We also notice that the optimisation procedure eventually chooses $\delta = 0$ in the loss function, thus opting for a classical \mse loss, to assign a stronger penalty to the possibly remaining outliers.
The full list of hyperparameters is available in \Cref{tab:mt_hyper}.

\subsection{Software and Hardware Setups}

\begin{figure}[t]
    \centering
    \includegraphics[width=0.75\linewidth]{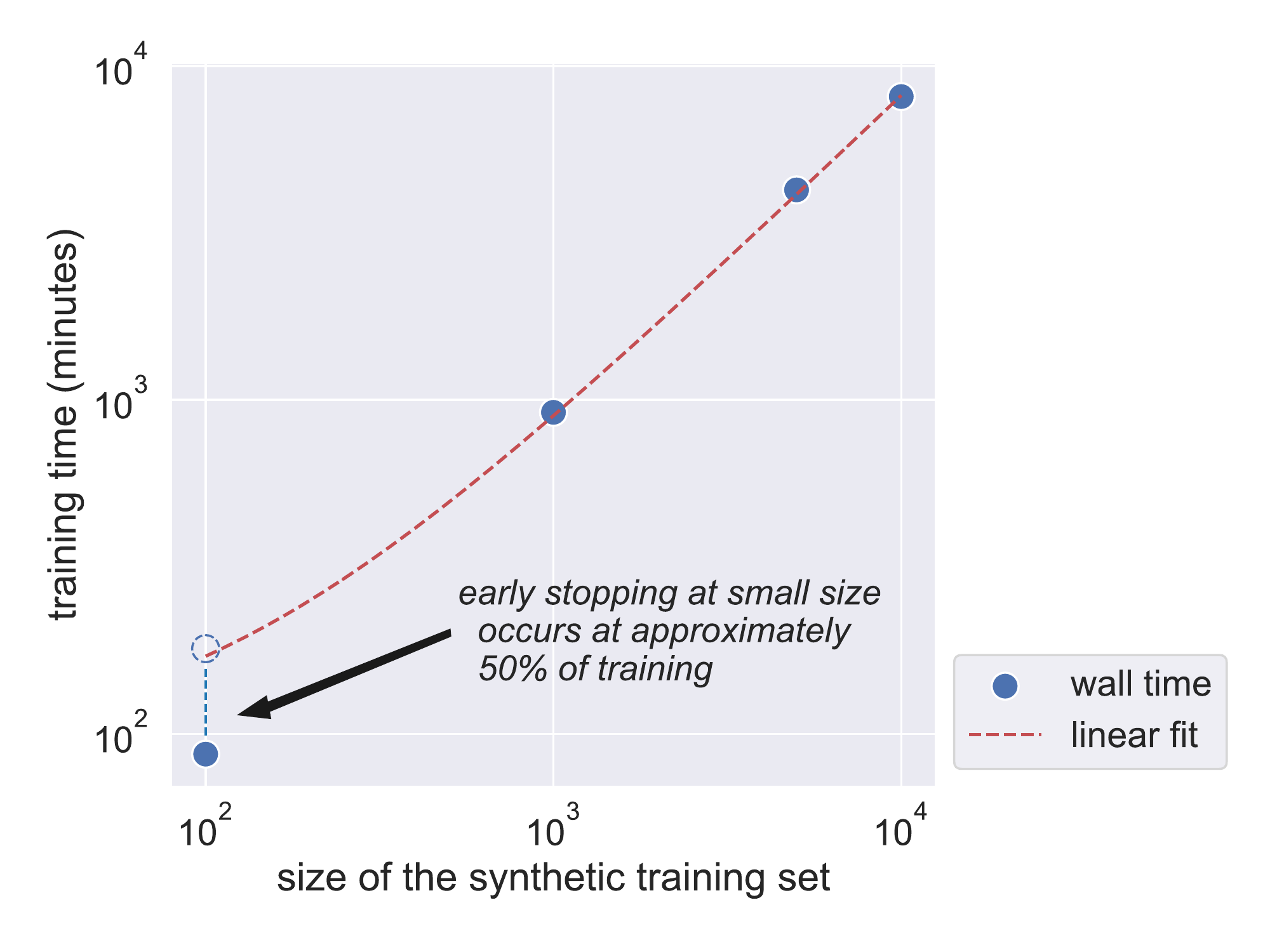}
    \caption{%
        \emph{Wall time of the training process.}
        At the smallest synthetic set size, the training is interrupted halfway through the complete training, due to the early stopping strategy.
        Accounting for this effect leads to a linear relation between size of the training set and training time.
        }
    \label{fig:walltime}
\end{figure}

For the analysis, we use the Python libraries \emph{Scikit-learn}~\cite{Pedregosa:2011:Scikit} and \emph{PyTorch}~\cite{Paszke:2019:PyTorch} (and its wrapper \emph{Pytorch Lightning}~\cite{Falcon:2019:Pytorch}) for the development of shallow and \dl algorithms, respectively.
\emph{NumPy}~\cite{Harris:2020:Numpy} is used for the synthesis of artificial data.
We choose \emph{Optuna}~\cite{Akiba:2019:Optuna} for the optimisation of the hyperparameters of the networks.
Computations of linear models and \fcnns are performed on a laptop with \SI{32}{\giga\byte} \ram and Intel\textsuperscript{\textcopyright}~Core\texttrademark~i7-10875H \cpu and a NVidia\textsuperscript{\textcopyright}~\emph{Quadro}~T2000 \gpu with \SI{4}{\giga\byte} \gddr{6} memory.
The optimisation and training of the \mt architecture are performed in parallel on NVidia\textsuperscript{\textcopyright} V100 cards with \SI{16}{\giga\byte} or \SI{32}{\giga\byte} \hbm[2] memory, and NVidia\textsuperscript{\textcopyright} P5000 cards with \SI{16}{\giga\byte} \gddr{5X} memory.
The setup is able to provide virtually immediate results for the shallow \ml algorithms.
The optimisation cycle of the \fcnns usually takes a few hours to complete (approximately 9 hours for each dataset we consider in the analysis, using up to 4 \gpus in parallel).
Training takes less than an hour (approximately 45 minutes) for each matrix.
On the other hand, the \mt architecture is longer to train, though not dramatically more involved.
The hyperparameter optimisation takes a few days, depending on the number of cycles and the size of the search space (\num{4.5} days in our case, 4 \gpu cards in parallel).
Training on a longer number of epochs takes hours to days to complete, as shown schematically in \Cref{fig:walltime} (the average training time between different experimental matrices has been considered).
Notice that the training wall time scales linearly with the size of the training set, as expected (early stopping must be taken into account).
Computations were carried out using a private infrastructure, namely the \emph{FactoryIA} supercomputer, located in the \^{I}le-de-France region in France, with a carbon efficiency of \SI[per-mode=symbol]{0.1}{\kilo\carb\per\kWh} (national average).
Considering only the most power consuming card and the \mt model (NVidia\textsuperscript{\textcopyright} V100 \SI{32}{\giga\byte} consumes up to \SI{300}{\watt}), we estimate the cost of optimisation to \SI{13}{\kilo\carb}, separately for the cement samples and the alloy matrices.
Training simultaneously on the different sizes of the synthetic training set impacted for approximately \SI{7}{\kilo\carb} for each experimental matrix~\cite{Lacoste:2019:Quantifying}, approximately equivalent to \SI{120}{\kilo\meter} driven by an average petrol-powered passenger vehicle~\cite{EPA:2010:Light-Duty}.
Experiments repeated during several months, amounting to several return journeys from Paris, France, to Turin, Italy ($2 \times \SI{800}{\kilo\meter} = \SI{1600}{\kilo\meter}$).
Some specific \libs applications (e.g.\ inference on a simple matrix) may require smaller \mt models.
In this case, less synthetic spectra are needed, thus making the analysis feasible even on less performing hardware.
On the other hand, once the architecture is trained, predictions are very fast to compute (from less than a second to a few seconds, depending on size of the batch of spectra).
Models can thus be easily deployed and used for online inference.

\section{Results and Discussion}\label{sec:results}

In what follows, we compare the results of the different analysis techniques on the experimental matrices.
We first introduce the predictions made by the shallow \ml algorithms and the \fcnns.
We then discuss in detail the \mt architecture, and the aspect of validation of its predictions.

\subsection{Baseline Analysis}

In the analysis, we consider different types of baselines, in order to provide a complete comparison of the proposed technique with the \sota.
Specifically, we show the results of \pls[1], which does not require the extraction of information from the experimental spectra.
Differently, models such as \lr, \mlr and \fcnns need \emph{a priori} knowledge of the spectra and a screening process to extract the integral intensities used for training.
Similarly to the first, our proposed \mt model does not require expertise in selecting input data, since the important variables are learnt during training.
It requires a degree knowledge of the emission lines of the analyte to choose the secondary outputs.
However, as the addition of several outputs to the network is, in principle, always beneficial (see \Cref{sec:mt_cnn}), there is no need for an accurate selection of these lines.
A simple choice is to consider the full set of persistent lines of the analyte in the spectral range available, since spectral interference on single lines (or bands) is taken care of by the \cnn architecture.

For the comparison with the \mt model, we group the results of \lr, \mlr and \fcnns into a single \emph{ensemble} meta-model, based on the \rmse of the predictions.
For these algorithms, we only display the best result among the three of them in the main text.
We refer to \Cref{app:results} for the separate results.
Details on the traditional analysis on experimental data are given in \Cref{sec:trad_an}, where we also provide the results of the baseline algorithms on synthetic data.
We also include the results of the \mt model on experimental data, without synthetic data augmentation in \Cref{sec:trad_mt}, as a separate term of comparison with other algorithms.

\subsubsection{Traditional Algorithms}\label{sec:trad_an}

As shown in \Cref{tab:trad_an} in \Cref{app:results}, \fcnns seem to be more versatile, and provide, in general, the best generalisation results, with respect to the linear models used as baselines.
Simpler algorithms, such as \lr or \mlr, struggle at very low concentrations of the analyte, displaying error fractions which can reach and surpass the unit.
As expected, the linear models also present a typical heteroscedastic behaviour, with uncertainties which tend to proportionally decrease at higher concentrations of the analyte.
Globally, the \pls[1] model presents larger uncertainties, with respect to previous models.
However, in some cases, the implicit dimensionality reduction of the algorithm, helps in partially solving the spectral interference in the data, and provides good results.
As already anticipated, these results are summarised in a meta-model, which therefore displays different behaviours on the entire range of variation of the concentration of the analyte.

\begin{figure}[t]
    \centering
    \includegraphics[width=0.75\linewidth]{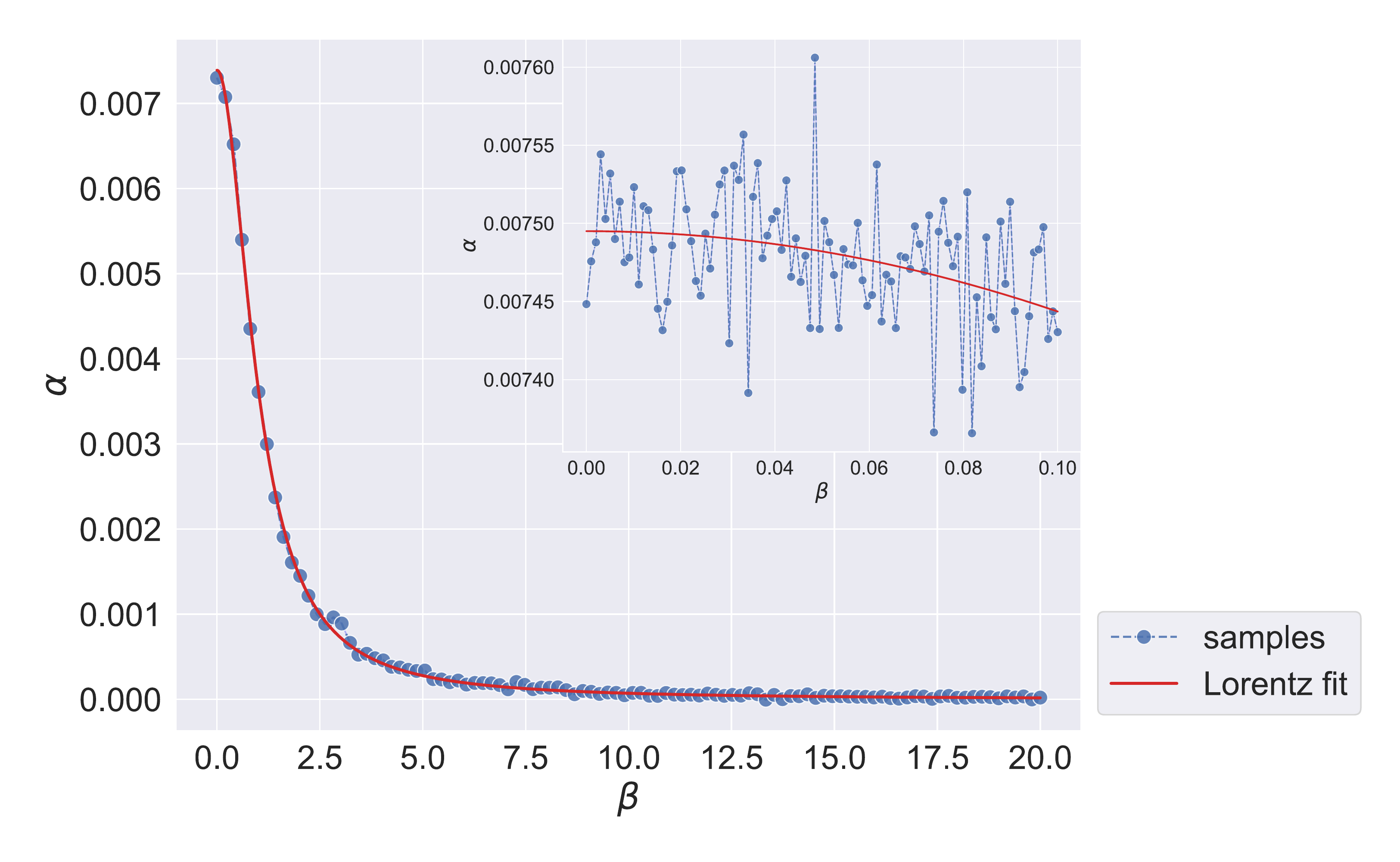}
    \caption{%
        \emph{Regression coefficient as a function of the noise parameter.}
        The large plot represents the functional behaviour of the parameter $\alpha$ as a function of the confounder $\beta$.
        The inset plot shows the same behaviour for the explored values of $\beta$: modifications are relevant starting from the fourth significant figure.
    }
    \label{fig:lorentz_fit}
\end{figure}

In \Cref{tab:trad_an_synth} in \Cref{app:results}, we show the results of the classical algorithms on synthetic data, for comparison with \mt learning.
In this case, we simply replace the experimental training set with a synthetically generated dataset, made of \num{1000} spectra per sample for the alloy matrices, and \num{100} spectra for each cement sample.
We notice that linear models are, in general, only slightly affected by the presence of synthetic data.
Uncertainties computed at different sizes of the training set are compatible with those found using only the experimental data.
As predictable, the \fcnns behave differently: more complicated matrices, with strong differences between samples, such as Al, Co and Zr, show an improvement.
However, matrices in the presence of strongly interfered emission lines, often show a decrease in performance, when comparing \emph{a posteriori} the predictions of the concentration of the analyte.
This is most probably due to the size of the model not being able to account for all local variations in the training distribution to ensure a good generalisation to unknown data.
In the case of cement samples, no significant differences are registered.

For the traditional algorithms, we also performed a scan in the $\beta$ parameter at different sizes of the training set.
For the values explored, the results of the linear models do not vary as functions of the intensity of the noise.
This can be seen by considering, without loss of generality, a simple regression model $y_i = \alpha x_i + \varepsilon$ with predictions $\haty_i$, where $i = 1,\, 2,\, \dots,\, N$ and $\varepsilon \sim \cN\qty( 0, 1 )$ and $x_i,\, y_i \in \R$.
We have
\begin{equation}
    \alpha \qty( \frac{1}{N} \sum\limits_{i = 1}^N x_i^2 - \barx^2 ) - \frac{1}{N} \sum\limits_{i = 1}^N x_i \haty_i - \barx \bary = 0,
\end{equation}
where
\begin{equation}
    \barx = \frac{1}{N} \sum\limits_{i = 1}^N x_i,
    \qquad
    \bary = \frac{1}{N} \sum\limits_{i = 1}^N \haty_i.
    \label{eq:lr}
\end{equation}
In our implementation of the synthetic data augmentation, the role of the predictor $x_i$ is represented by the intensity of the emission $x_i \defeq I_i = z_i\, p_i$, where $z_i \sim \cN\qty( 1,\, \beta )$ and $p_i \sim \cP( \barI )$ as in \eqref{eq:synthesis} ($\barI$ is expressed as a pure number in arbitrary units).
The dependent variable $y_i$ is the concentration $C_i$ of the analyte.
From \eqref{eq:lr}, the parameter $\alpha$ shows a dependence
\begin{equation}
    \alpha
    \propto
    \frac{1}{\qty(1 + \beta^2) \qty( 1 + \barI ) \barI - \barI^2}
    \propto
    \frac{1}{1 + \qty(1 + \barI) \beta^2},
\end{equation}
since
\begin{equation}
\begin{split}
    \frac{1}{N} \sum\limits_{i = 1}^N x_i^2
    & =
    \Ev{z_i^2}\, \Ev{p_i^2}
    \\
    & =
    \qty( \Ev{z_i}^2 + \Var{z_i} )\, \qty( \Ev{p_i} + \Ev{p_i \qty( p_i - 1 )} )
    \\
    & =
    \qty( 1 + \beta^2 )\, \qty( 1 + \barI )\, \barI.
\end{split}
\end{equation}
Hence, the intensity of the noise $\beta$ directly influences the learning of the parameter $\alpha$, which follows a Lorentzian profile.
The procedure can be generalised to \mlr or more complex models, with the same principle.
In \Cref{fig:lorentz_fit}, we show a simulation on synthetic data, generated as in the analysis: the fit shows the Lorentz-like behaviour of the type
\begin{equation}
    \alpha
    =
    \frac{a}{b + \beta^2}
    =
    \frac{a}{b} - \frac{a}{b^2} \beta^2 + \cO\qty( \beta^4 ),
\end{equation}
where $a$ and $b$ are fixed by the least squares.
For the values explored in the analysis $\beta \ll 1$, hence the behaviour of $\alpha$ is dominated by the constant term, as confirmed also by the analysis.
In fact, the decrease in $\alpha$ is substantially less than \SI{1}{\percent} for the explored values $0 \le \beta \le \num{e-1}$.
However, the derivation is no longer valid for models using mini-batch gradient descent, notably \nns, since, in these cases, only a subset of samples is considered at each iteration of the gradient descent.
Therefore, the contribution of the $\beta$ parameter is different and directly impacts the predictions.

\subsubsection{Multitask Learning Without Synthetic Data Augmentation}\label{sec:trad_mt}

As a separate baseline, we also trained the \mt model on experimental data.
The columns \texttt{Exp} in \Cref{tab:size} reports the results for all samples considered in the analysis.
For the alloy matrices, the \mt model is not able to provide accurate predictions, as its parameters do not adapt correctly to the data.
The synthesis of data is thus fundamental to increase the generalisation ability of the model, and to provide accurate predictions.
Differently from alloys, in the case of the cement samples, the \mt model is actually able to provide an acceptable performance even with experimental data.
Results are, however, not at the same level as using synthetic data, especially at higher concentration of the analyte.
As shown in the next section, in the case of cement matrices, the best performance is actually reached with just a few synthetic samples.

\subsection{Performance Review of the Multitask Approach}

We focus on various aspects such as the dependency on the size of the synthetic training set, the level of the random noise $\beta$ in \eqref{eq:synthesis}, and the validation of the predictions.
We compare the results of the baseline techniques using the best result of the ensemble model and \pls[1].

\subsubsection{Training Set Size}

\begin{figure}[t]
    \centering
    \includegraphics[width=0.49\linewidth]{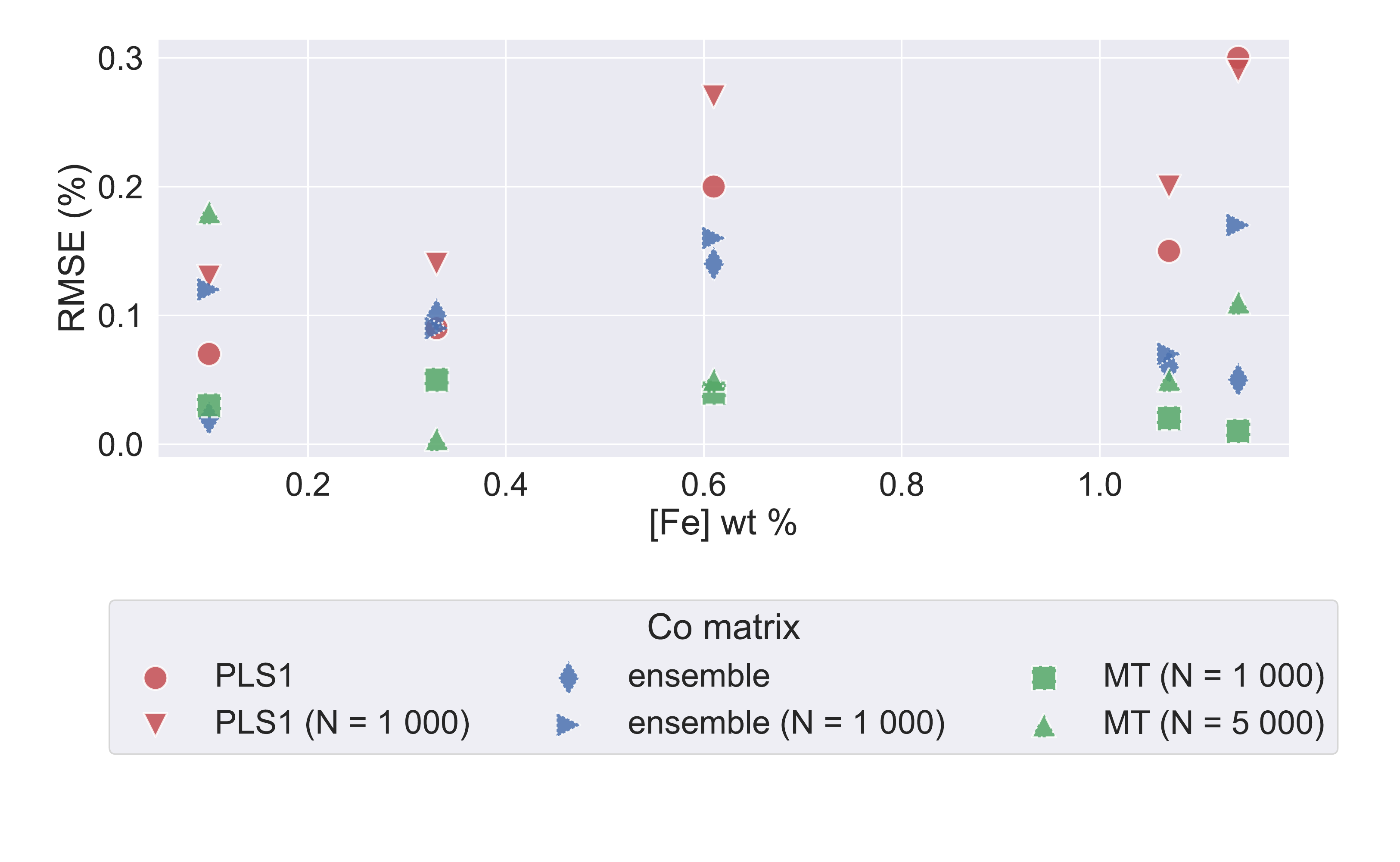}
    \hfill
    \includegraphics[width=0.49\linewidth]{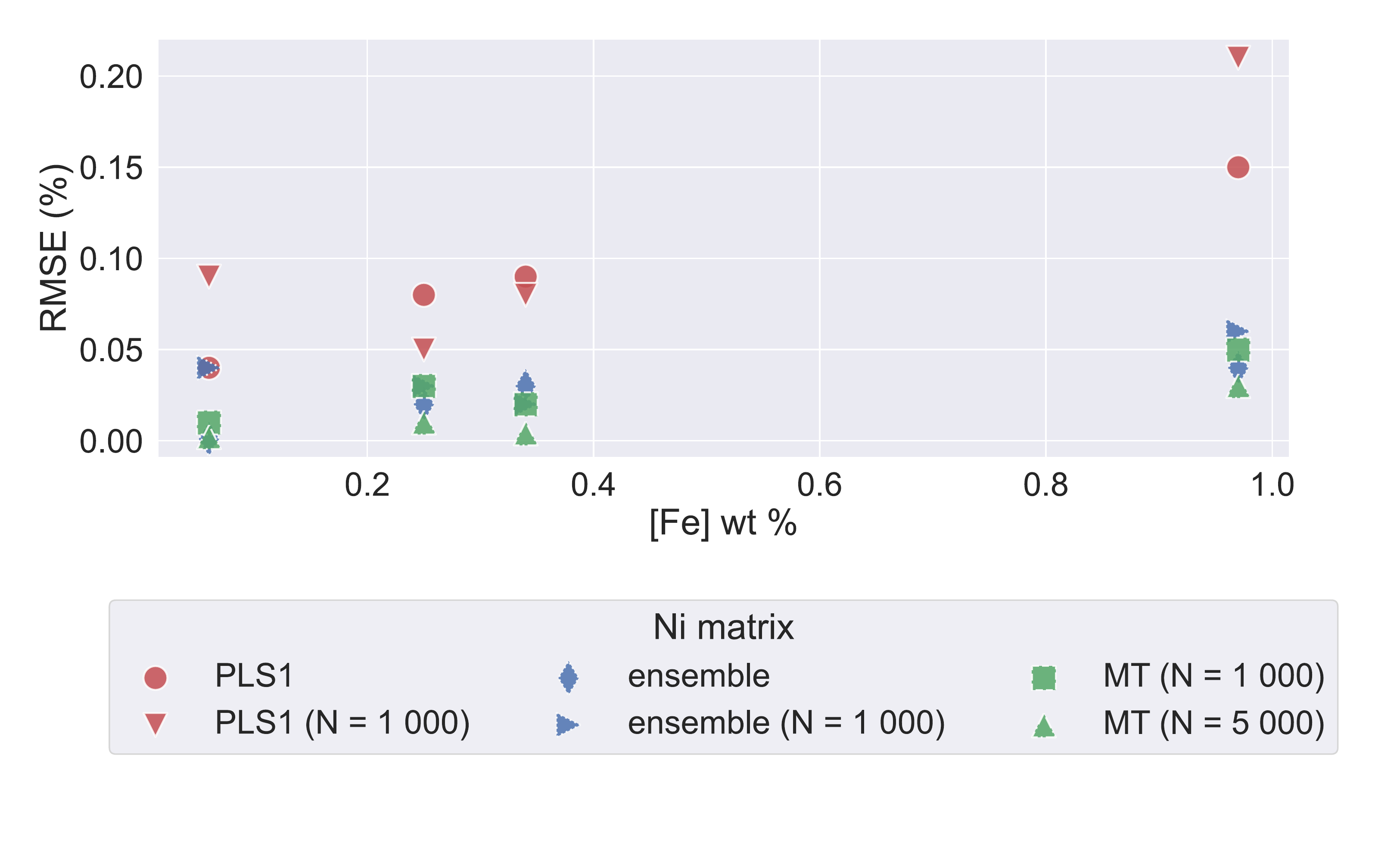}
    \\
    \includegraphics[width=0.49\linewidth]{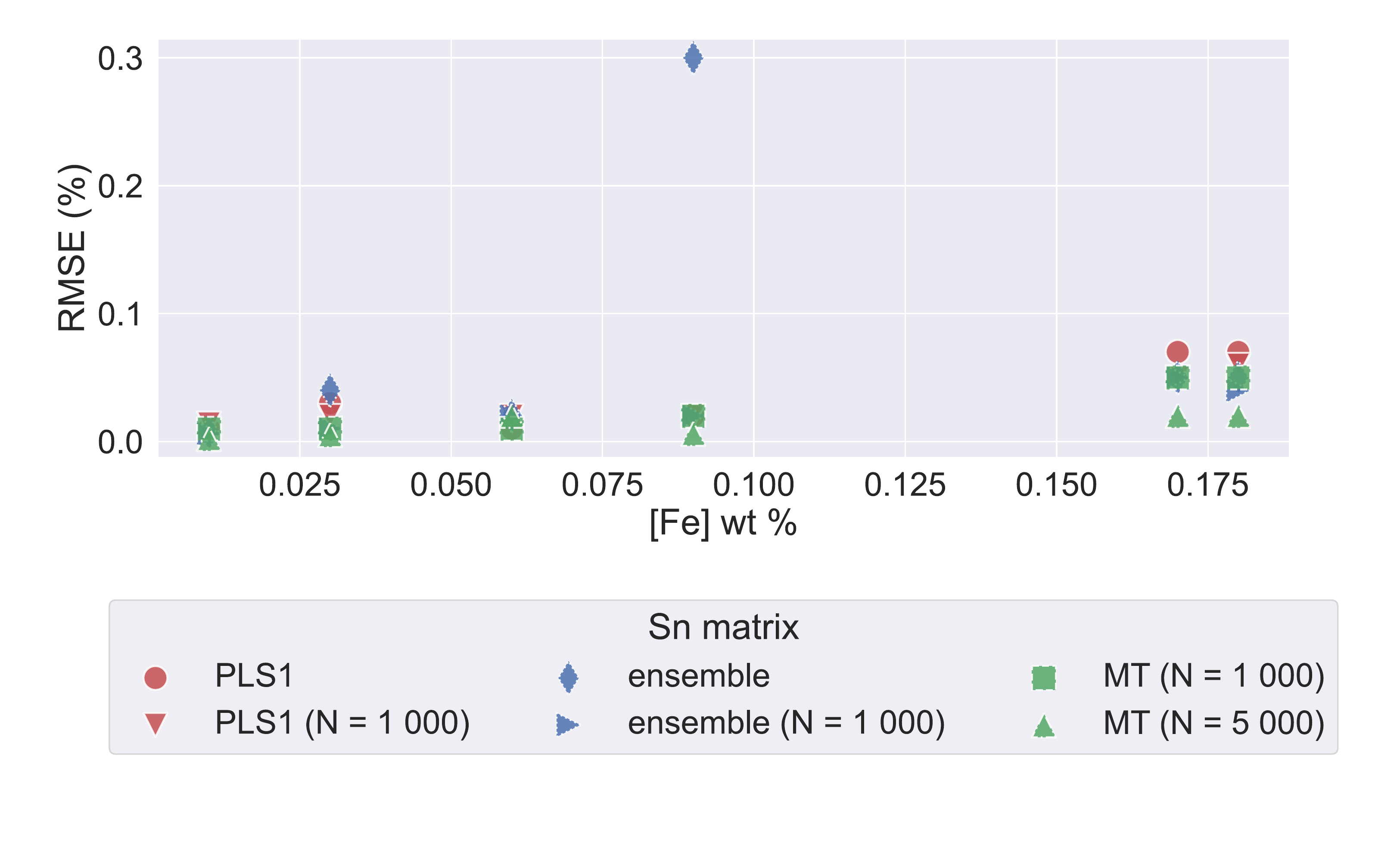}
    \hfill
    \includegraphics[width=0.49\linewidth]{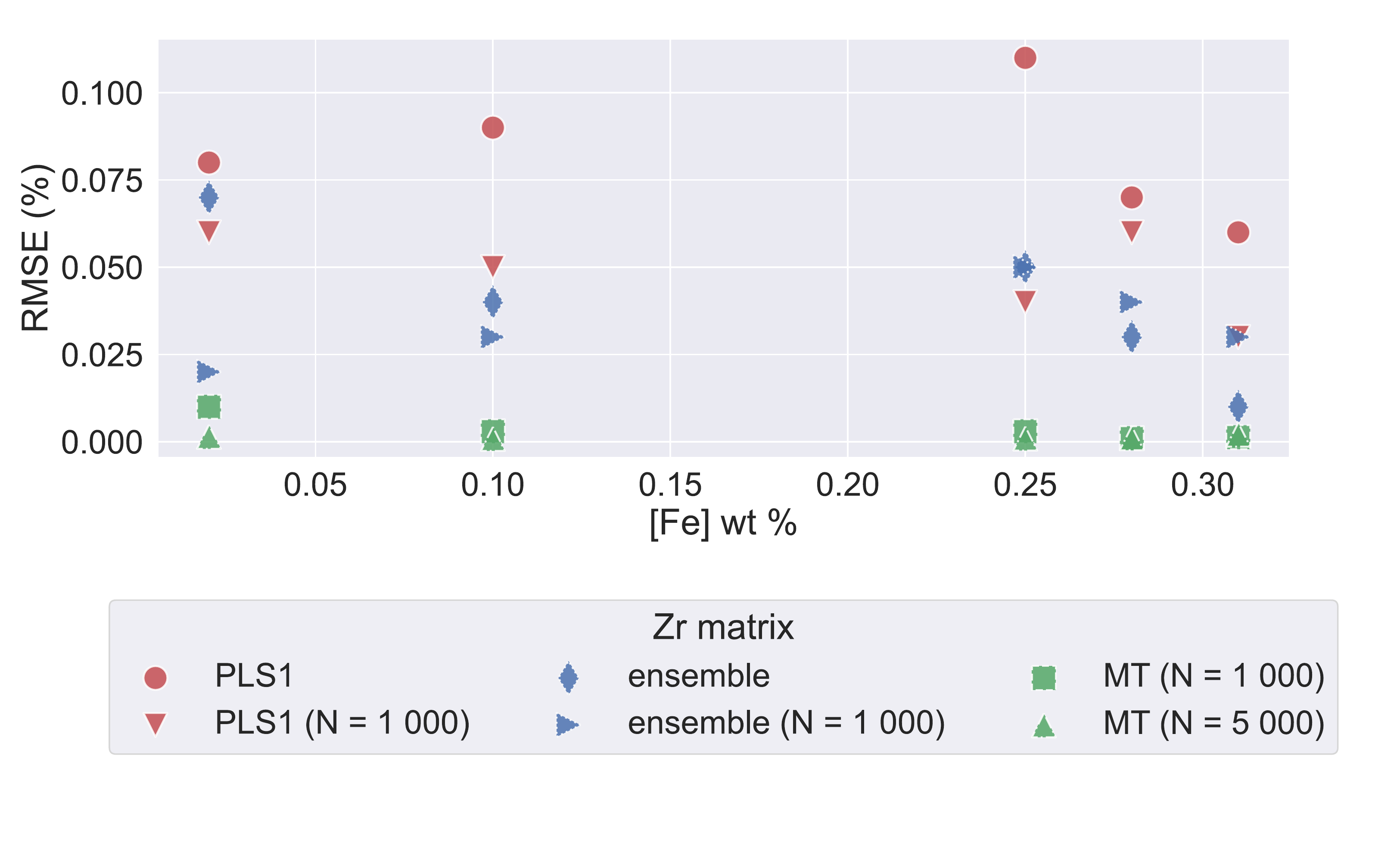}
    \caption{%
        \emph{Prediction uncertainties of the \mt model on alloy matrices as a function of training set size.}
        Uncertainties are expressed as \rmse on the concentration of the analyte.
        For brevity, results of the \mt architecture are reported for \numlist{1000;5000} synthetic spectra per sample, as they represent the best terms of comparison in the dataset.
    }
    \label{fig:mt_size}
\end{figure}

\begin{figure}[t]
    \centering
    \includegraphics[width=0.75\linewidth]{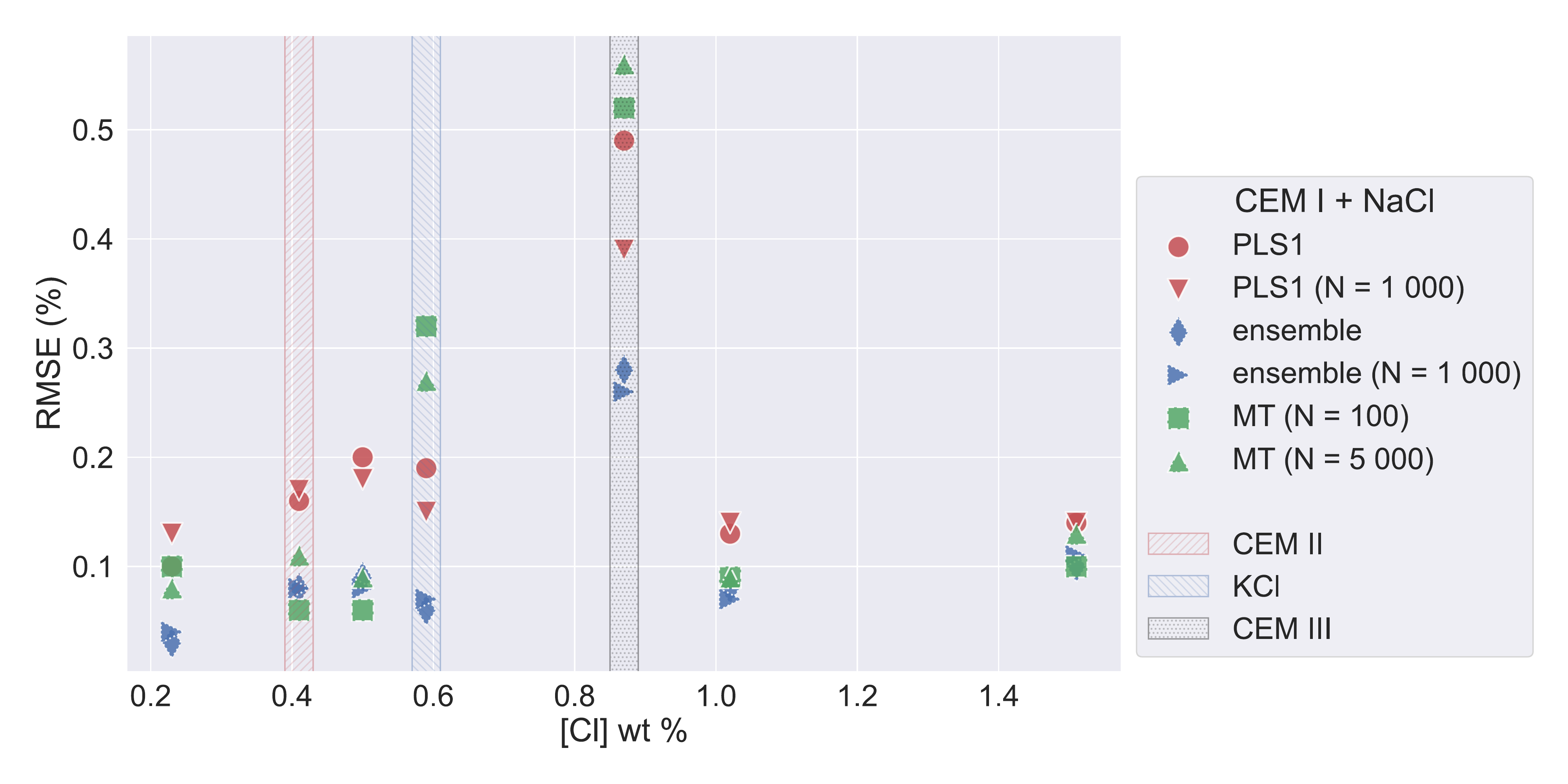}
    \caption{%
        \emph{Prediction uncertainties of the \mt model on cement samples as a function of training set size.}
        Uncertainties are expressed as \rmse on the concentration of the analyte.
        For brevity, results of the \mt architecture are reported for \numlist{100;5000} synthetic spectra per sample, as they represent the best terms of comparison in the dataset.
        Anomalous samples are highlighted in the plot, with the description of the different elements in their fabrication.
    }
    \label{fig:cl_size}
\end{figure}

We first consider the dependence on the size of the synthetic training set: we train different models generating \numlist{100;1000;5000;10000} synthetic spectra for each sample.
For these considerations, we fix the noise parameter $\beta$ to the best fitting choices for the different samples.
We choose $\beta = 0.10$ for the alloy matrices, and $\beta = 0.03$ for the cement samples, in order to avoid introducing an additional dependency (see \Cref{fig:qqplots,,fig:co_coverage,fig:ni_coverage,fig:sn_coverage,fig:zr_coverage} in \Cref{app:figures}, and their relative details in \Cref{sec:simulation}).
For brevity, in \Cref{tab:size} in \Cref{app:results}, we report the full inference results, which we graphically summarise in \Cref{fig:mt_size} for the alloy matrices, and in \Cref{fig:cl_size} for the cement samples.

The possibility to increase arbitrarily the size of the training set is key to increase the overall performance of the model.
We see, for instance, that, in terms of robustness, the \mt architecture is capable of delivering a homoscedastic performance for all concentrations of the analyte (apart from anomalies, which are discussed in the following).
Moreover, the prediction uncertainties are usually comparable with or better than the ensemble model, which already presents good results.
The \mt architecture is thus capable of selecting the information contained in the data to base its own predictions.
The use of synthetic samples enables to capture the fluctuations at low concentrations, where the \pls[1] model struggles to give accurate results.
In this case, the analysis usually benefits from a manual simplification of the data by an expert in the field, as shown by the meta-model.
However, the \mt model seems to provide at least comparable performance, by selecting automatically spectral ranges of relevance for the predictions.
As a general remark, the optimal number of synthetic spectra varies for each matrix, depending on the degree of spectral interference, noise, and sparsity (e.g.\ \num{5000} spectra per sample in the case of the Zr matrix, or \num{1000} spectra per sample for the Cu matrix).
The ideal size of the synthetic training set is, however, quite challenging to know \emph{a priori}.
Rules of thumb in known \dl competitions, such as \emph{ImageNet}~\cite{Russakovsky:ILSVRC:2015}, seem to suggest a number $\cO(\num{e3})$ of samples per class, in order to reach good classification or regression results.
Data augmentation is usually the solution to the lack of training data, provided some conditions on the augmentation procedure are respected~\cite{Balestriero:2022:Data}.
Research on the topic shows different functional dependencies of standard classification and regression metrics, which usually grow as a function of the size of the training set~\cite{Figueroa:2012:Predicting, Cho:2015:Much, Zhu:2016:We, Shahinfar:2020:How}.
Imbalanced datasets~\cite{Juba:2019:Precision} may, however, spoil such behaviour.
Heuristically, we noticed that the creation of a large number of synthetic spectra impacts on training time for less than \SI{1}{\percent} of the total training time, while the latter grows linearly (see \Cref{fig:walltime}).
Such behaviour makes it usually possible to experiment with a few options, in order to determine, using the validation set or the experimental training samples, the best trade-off between the performance of the model and the computational power available.

Finally, notice that, depending on the matrix of the samples and the size of the \mt model (number of hidden layers, size of the kernels, etc.), there may be regimes in which the addition of too many synthetic samples may hurt the performance.
Recent studies on the double descent phenomenon show an additional trade-off between the size of the model and the amount of data used for training.
Nevertheless, the mechanism behind this behaviour remains an open research question~\cite{Advani:2020:High, Belkin:2018:Reconciling, Geiger:2019:Jamming, Nakkiran:2021:Deep}.
In the ``classical'' \dl regime, consisting in the optimisation of the bias-variance trade-off, we look for one model, which fits the training set and generalises well to other samples.
Larger \dl architectures enable more models to fit the same training set, and use the additional parameters to improve the generalisation.
This appears as a second descent in the test loss, as a function of the number of free parameters in the model.
However, in the phase transition between the two regimes, more synthetic samples may not necessarily lead to better results, since the architectures may not have enough parameters to model correctly the variance of the training set.
A simple solution is experimenting with the number of synthetic samples, for instance, looking at the inference results on the experimental training set, or on a synthetic validation set.
A second solution, depending on the computational power available, is to study the performance as a function of the size of the \dl models (e.g.\ the number of free parameters).
This should allow targeting explicitly the double descent of the loss function, rather than the optimisation of bias and variance.

\subsubsection{Random Noise}

\begin{figure}[t]
    \centering
    \includegraphics[width=0.49\linewidth]{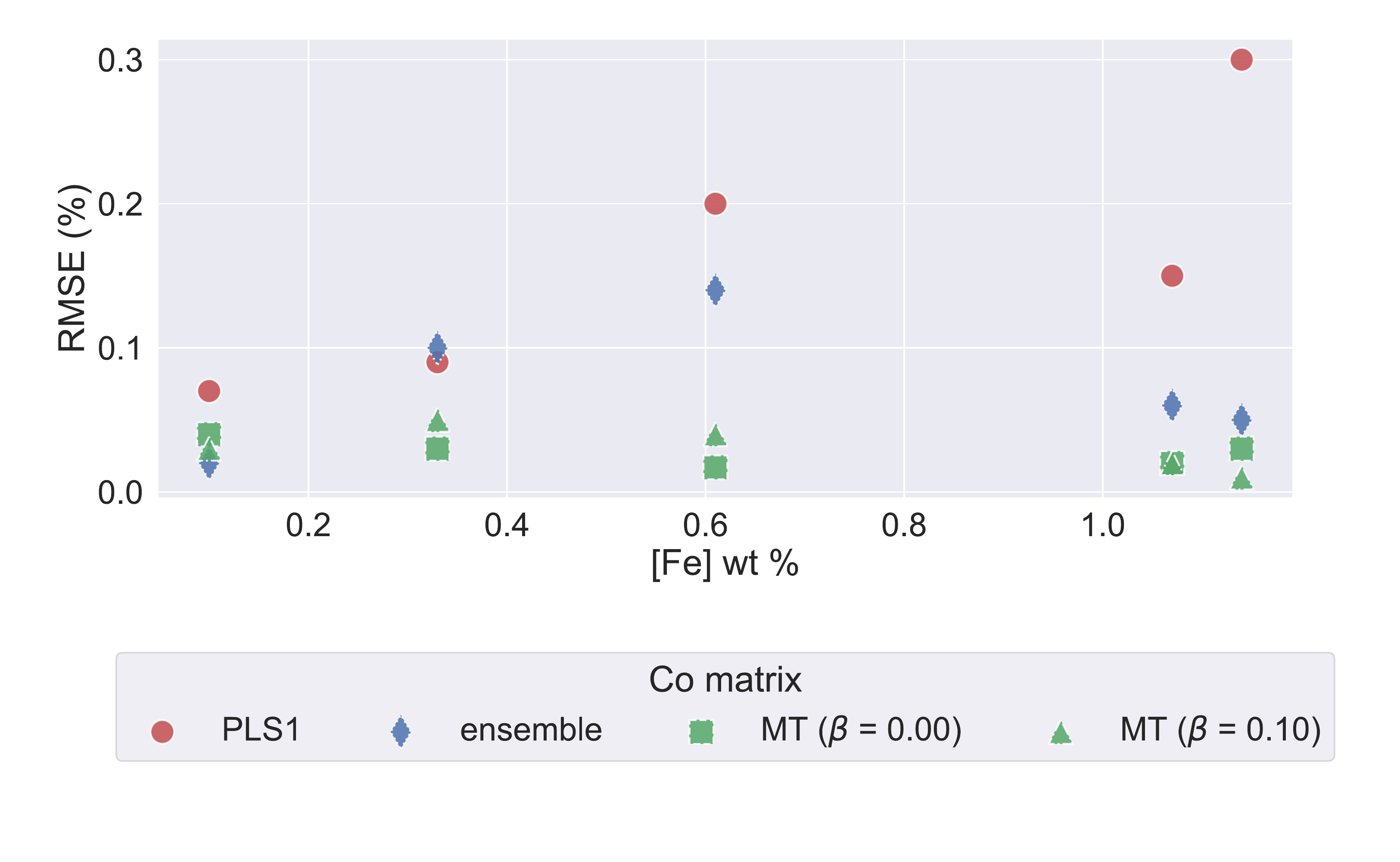}
    \hfill
    \includegraphics[width=0.49\linewidth]{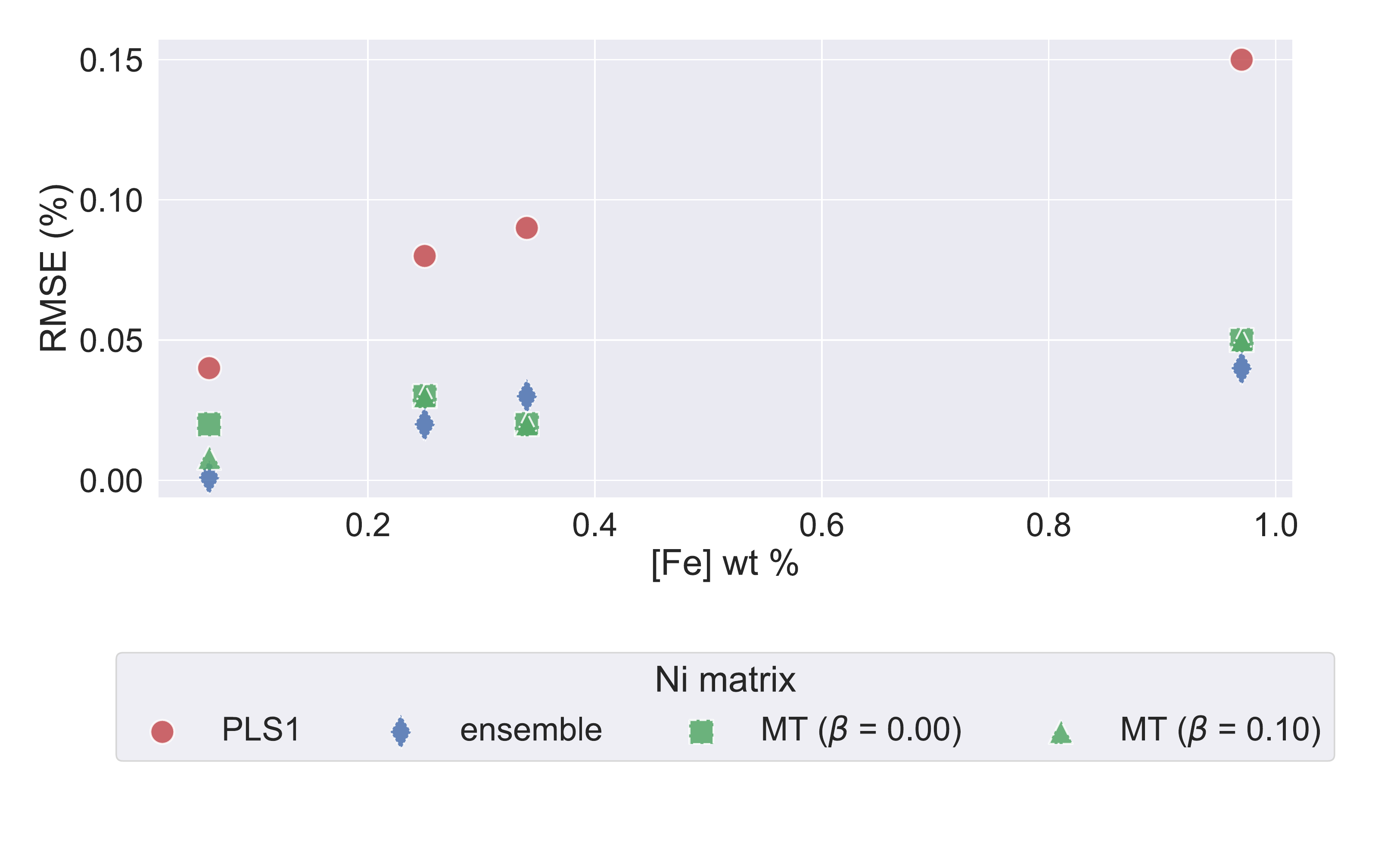}
    \\
    \includegraphics[width=0.49\linewidth]{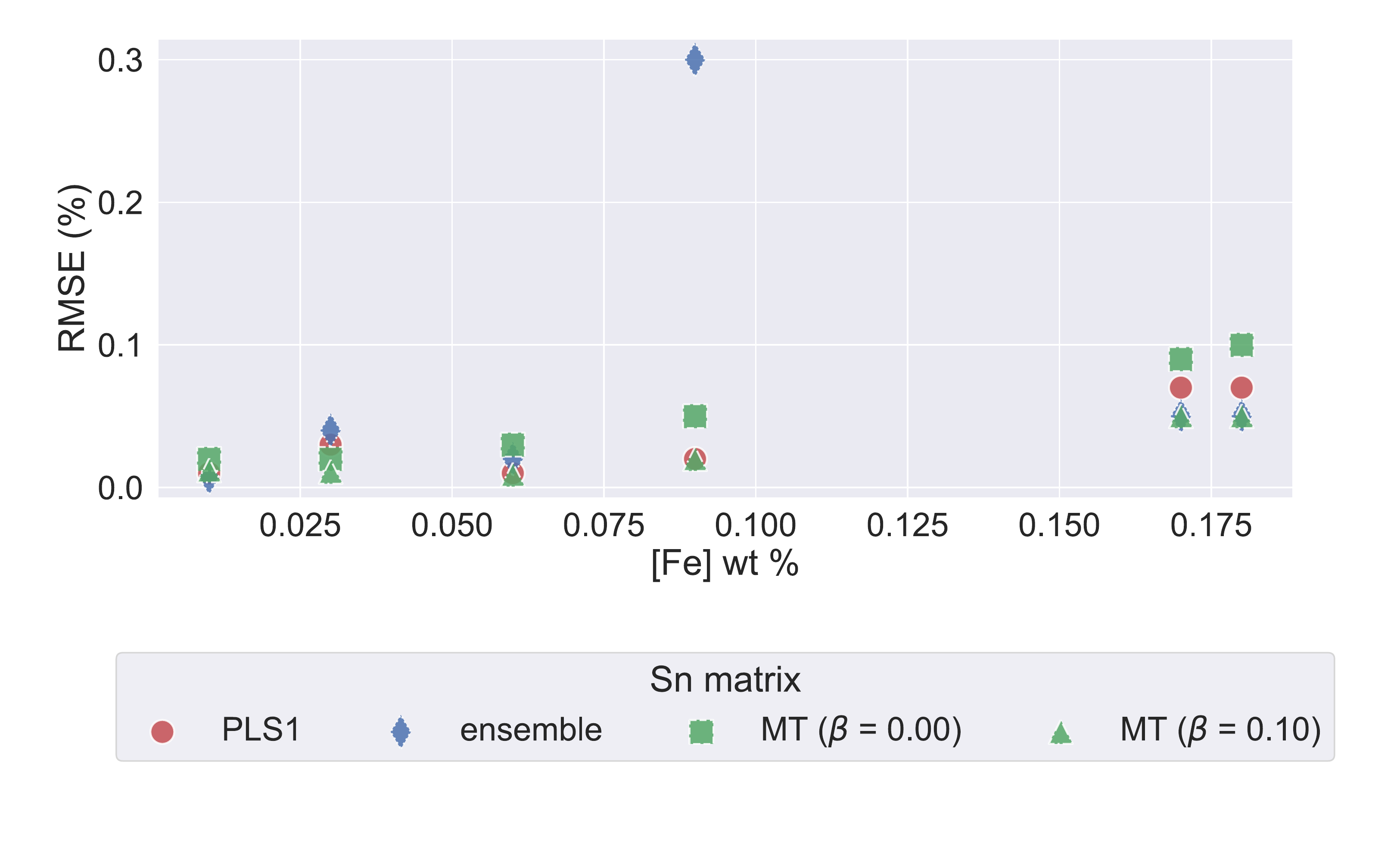}
    \hfill
    \includegraphics[width=0.49\linewidth]{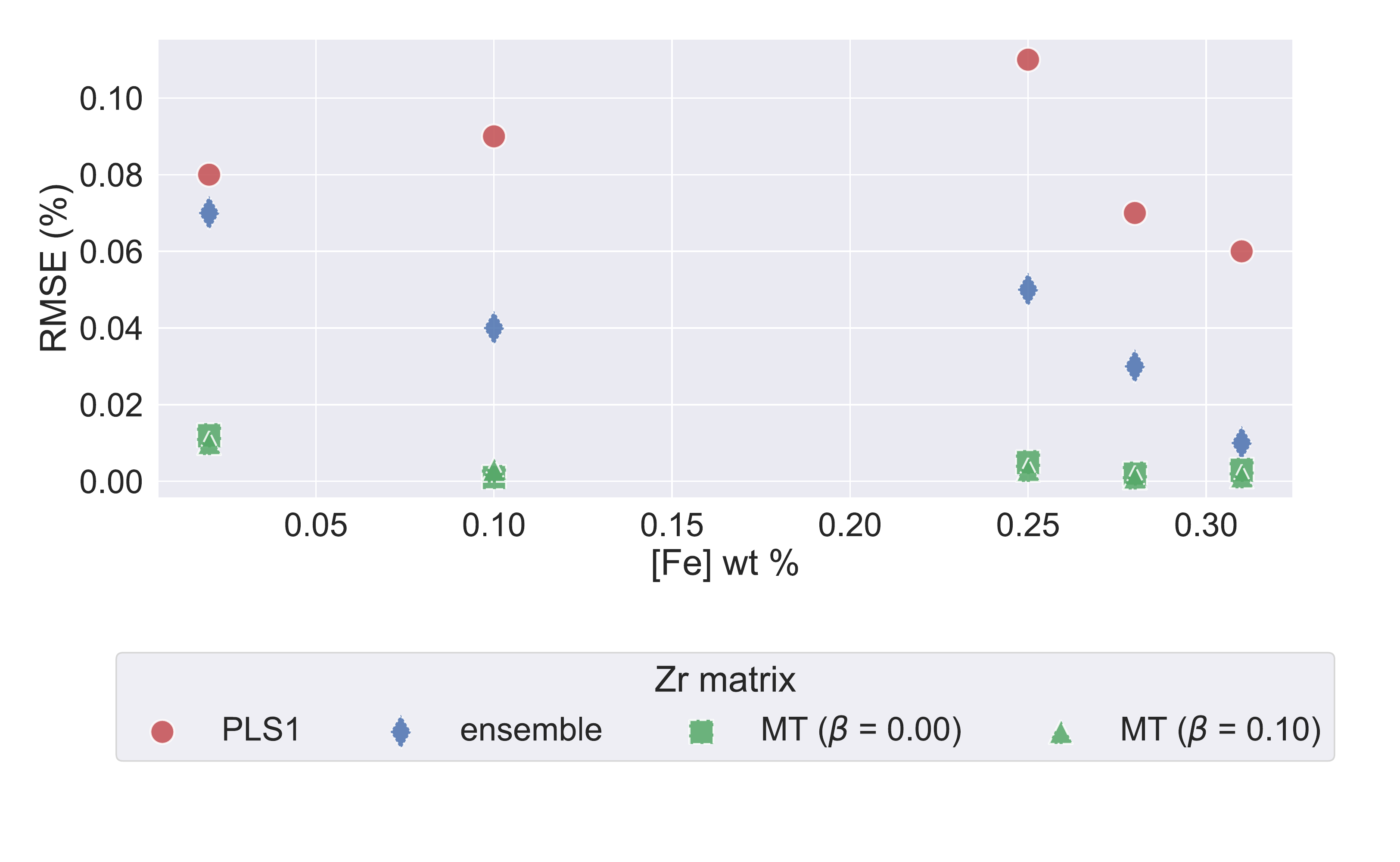}
    \caption{%
        \emph{Prediction uncertainties of the \mt model on alloy matrices as a function of the noise parameter.}
        Uncertainties are expressed as \rmse on the concentration of the analyte.
        Results of the \mt architecture are reported for $\beta = 0.00$ and $\beta = 0.10$, for illustration purposes.
    }
    \label{fig:mt_noise}
\end{figure}

\begin{figure}[t]
    \centering
    \includegraphics[width=0.75\linewidth]{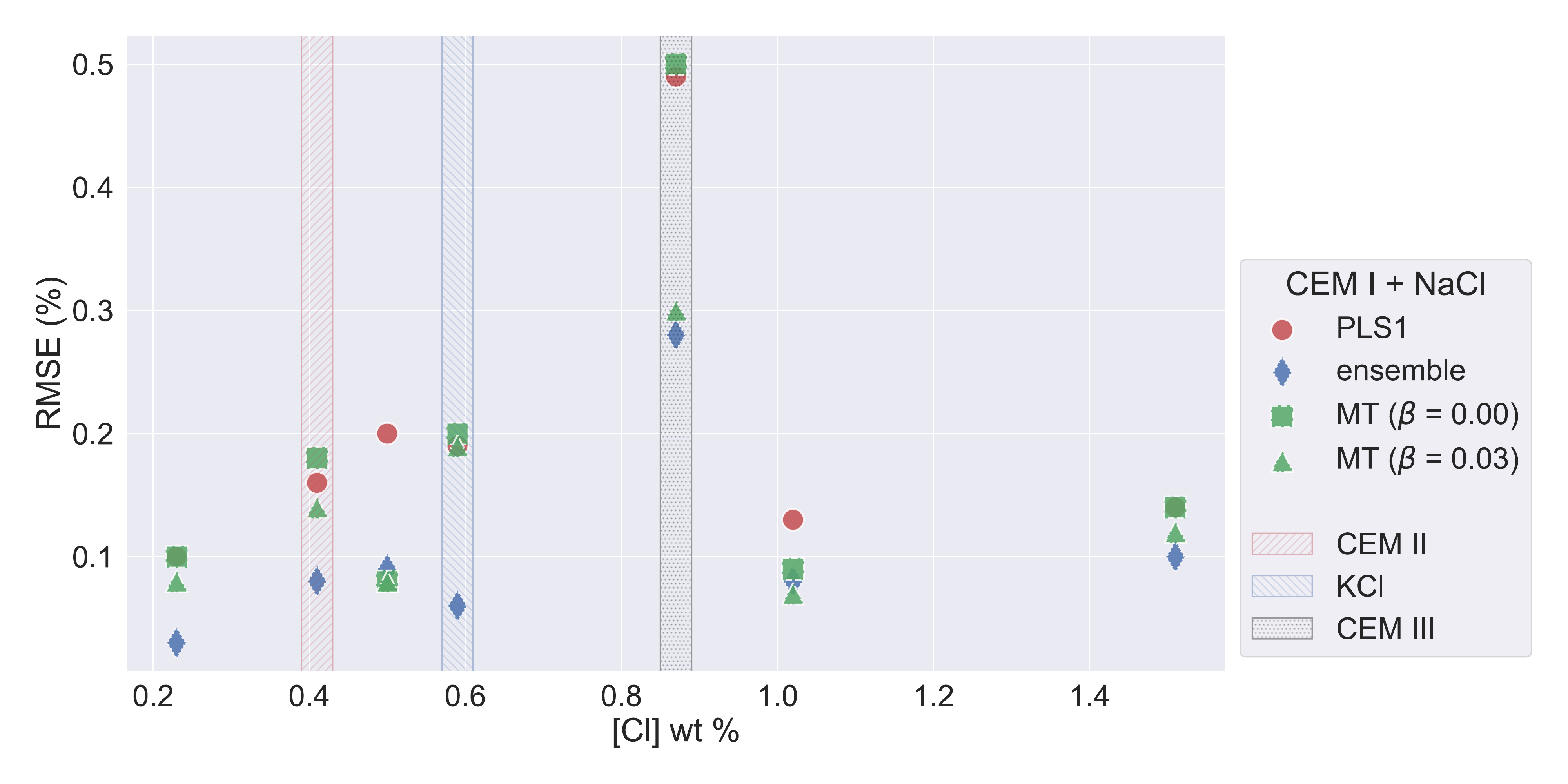}
    \caption{%
        \emph{Prediction uncertainties of the \mt model on cement samples as a function of the noise parameter.}
        Uncertainties are expressed as \rmse on the concentration of the analyte.
        Results of the \mt architecture are reported for $\beta = 0.00$ and $\beta = 0.03$, for illustration purposes.
    }
    \label{fig:cl_noise}
\end{figure}

We then focus on the dependence of the random noise level on the correct representation of the experimental data using synthetic samples.
As previously shown, the noise parameter $\beta$ influences the distribution of the intensities of the emission lines or molecular bands.
Thus, the parameter can be chosen deterministically to suit different needs, for instance by optimising the \rs score of the coverage diagrams, as shown in \Cref{sec:simulation}, either globally, or on a sample basis.
In this section, we do not specifically look for an optimum of the confounder parameter, since we explicitly report results for different values of the noise $\beta$.
However, we highlight the best choice of such parameter in the range of its explored values.
In order to avoid introducing additional variability, we fix the number of synthetic spectra to \num{1000} per sample, which seems to be a sensible choice in terms of size of the training set, memory requirements, and training time.
The detailed results of the analysis are shown in \Cref{tab:mt_noise} in \Cref{app:results}.
We present a graphical summary of the salient results in \Cref{fig:mt_noise}, for the alloy matrices, and \Cref{fig:cl_noise}, for the cement samples.
The results suggest that the addition of random noise influences the performance and the robustness of the model.

In particular for the cement samples, the optimal results are achieved using $\beta = 0.03$, which also represents the best fitting choice of the coverage diagrams.
As a consequence, the correct modelling of the spectra reduces the average deviations of the samples, even for the out-of-distribution cases.
It also improves the robustness of the model at different scales, seen as an overall decrease in the error metrics and a homoscedastic behaviour of the model (with respect, for instance, to the simple \lr).

In the case of the alloy matrices, the \mt model presents a similar behaviour.
The addition of noise to create realistic synthetic samples improves, in general, the performance and the robustness, even though the effect is less intense than previously observed for the cement samples.
One of the reasons is experimentally observable in the coverage plots (see \Cref{fig:co_coverage,fig:ni_coverage,fig:sn_coverage,fig:zr_coverage} in \Cref{app:figures} for a reference).
Spectral interference, more pronounced in these samples, strongly affects different kinds of matrices, and introduces variability at different values of the analyte concentration.
Larger, though acceptable, deviations are displayed by difficult cases, such as the Sn matrices, for which the choice $\beta = 0.10$ is not optimal for all samples.
On the other hand, matrices, whose samples are well modelled by such a value of the noise parameter (such as Ni and Zr), present both a good and robust out-of-the-box performance.
In some alloy matrices, we also notice a negligible impact of differences in the choice of $\beta$.
This is, in general, due to the \dl architecture which is able to leverage the presence of a strong signal, with the help of a large number of heads in the architecture.
As previously shown, the MT architecture outputs the intensity of several Fe lines, each competing to improve the main prediction and the other secondary outputs (see \Cref{sec:mt_cnn}).
This behaviour enables to mitigate the effects of a sub-optimal modelling of the spectra, as well as providing good generalisation capabilities.

\subsubsection{Validation of the Results}\label{sec:validation}

For the analysis of the trustworthiness of the model, we use the predictions of the \mt model on the independent test set of the alloy matrices and the cement samples.
In this analysis, ground truth values of the concentration of the analyte are available for a direct comparison with the predictions of the model.
However, in a field application reference values are not available.
The analysis of the secondary outputs of the \mt architecture is a tool to assess the confidence of the predictions and detection of the anomalous samples or modifications in the experimental conditions.
In this case, for the comparison, we fix the number of synthetic spectra per sample to \num{5000} for both cement and alloy matrices, and we use the $\beta$ parameter best fitting the local coverage plots.

\begin{table}[t]
\centering
\begin{tabular}{@{}c|cc@{}}
\toprule
               & \multicolumn{2}{c}{\textbf{C.I. MAPE (\%)}}               \\ \cmidrule(l){2-3} 
\textbf{[Cl] wt \%}     & \textbf{593.46~nm}    & \textbf{617.74~nm} \\ \midrule
0.06           & 3~\textpm~2 & 0.9~\textpm~0.4   \\
0.19           & 2.1~\textpm~0.9 & 0.7~\textpm~0.3   \\
0.32           & 1.8~\textpm~0.6 & 0.6~\textpm~0.3   \\
0.46           & 1.3~\textpm~0.5 & 1.0~\textpm~0.6   \\
0.59           & 1.3~\textpm~0.4 & 1.0~\textpm~0.4   \\
0.72           & 1.1~\textpm~0.5 & 0.7~\textpm~0.3   \\
0.85           & 0.7~\textpm~0.3 & 0.4~\textpm~0.1 \\
0.98           & 0.9~\textpm~0.3 & 0.6~\textpm~0.4 \\
1.15           & 1.0~\textpm~0.3 & 1.0~\textpm~0.5 \\
1.43           & 1.9~\textpm~1.0 & 0.9~\textpm~0.3 \\
1.71           & 1.3~\textpm~0.6 & 1.4~\textpm~0.8 \\
1.95           & 1.6~\textpm~0.4 & 1.2~\textpm~0.3 \\ \bottomrule
\end{tabular}
\caption{%
    \emph{Confidence intervals of the cement samples.}
    Confidence intervals (C.I.) measured as in \eqref{eq:conf} on the experimental set using $\nu = 25$ degrees of freedom and a confidence level $1 - \alpha = 0.975$ ($\hatt_{1 - \alpha}^{\,\nu} = 2.06$).
}
\label{tab:ci_cement}
\end{table}

We first focus on the cement matrices ($\beta = 0.03$).
We choose two CaCl molecular bands, used also in \mlr and \fcnn to predict the concentration of Cl in the samples, for the analysis of the trustworthiness of the predictions.
This represents an easy choice, as the two bands are the two most intense in the spectral range considered.
Following \Cref{sec:trust}, we compute the confidence intervals on the experimental training data (unseen by the model, which is trained on synthetic spectra) as in \eqref{eq:conf}.
For the analysis, we consider $\nu = 25$ degrees of freedom at a confidence level of $1 - \alpha = 0.975$, in order to take well into account possible outliers or defects on the surface of the matrices.
Nonetheless, the predictions of the \mt model can be studied as a function of $\alpha$, at different levels of confidence.
We show the corresponding values in \Cref{tab:ci_cement}.
We consider these values as reference in the analysis of the trustworthiness of the model, since they represent known standards, whose labels are available.
In \Cref{tab:cl_valid}, we present the analysis of the inference on the test cement samples.
\emph{A posteriori}, we notice that the predictions of the \mt architecture are all compatible with the respective ground truths, even though some samples present larger uncertainties, which may indicate faulty values.
However, in the absence of reference values, predictions alone are not sufficient to measure the trustworthiness of the model.

The predictions of the secondary values show that the band at \SI{593.46}{\nano\meter} is able to display a pattern which identifies some anomalies in the prediction of the integral intensities.
As previously shown (see \Cref{fig:validation_cases} for a reference), this pattern of \mape and $t$-value is typical of anomalous samples, for which the model does not provide precise predictions.
In hindsight, the analysis of the secondary outputs identifies the three out-of-distribution samples present in the dataset (different matrix and salt).
To measure the trustworthiness of the predictions, we then use a standard \emph{Student}’s $t$ two-tailed test (confidence $1 - \alpha = 0.95$ and 25 degrees of freedom) to assess the predictions of the molecular emission bands.
We notice that, although the samples register as anomalies, the variance of the predictions is such to include the true values of the secondary outputs in the error intervals with good confidence (case 4 in \Cref{fig:validation_cases}).
Given the inter-dependencies of the \mt model previously discussed, the confidence on the predictions of the secondary output influences directly the confidence on the main prediction, the concentration of the analyte.
The precise quantification of the confidence level is, nonetheless, not trivial because of the large number of parameters involved in the computation.
In the case at hand, we can interpret the result by noting that the \mt model is still capable of providing trustworthy predictions on the concentration of the analyte.
However, its precision is highly affected in the presence of anomalous samples: the predictions of the main output contain the true value, within the uncertainty, with high probability.
In other words, in this scenario, the predicted values of the concentration of the analyte can be considered compatible with the reference values, provided by the supplier.
Further investigation on three anomalous samples remains necessary.

\begin{table}[t]
\centering
\begin{tabular}{@{}cccccc@{}}
\toprule
\textbf{Ground truth}   & \textbf{Prediction}                    & \multicolumn{2}{c}{\textbf{MAPE (\%)}}  & \multicolumn{2}{c}{\textbf{t-value}}    \\
\textbf{[Cl] wt \%}     & \textbf{[Cl] wt \%}                    & \textbf{593.46~nm} & \textbf{617.74~nm} & \textbf{593.46~nm} & \textbf{617.74~nm} \\ \midrule
0.23                    & 0.25~\textpm~0.08 & 1.4                & 0.7                & 0.69               & 0.34                                    \\
0.41 (CEM II)           & 0.41~\textpm~0.11 & \textbf{4.1}       & 0.8                & 0.98               & 0.32                                    \\
0.50                    & 0.54~\textpm~0.09 & 1.7                & 0.7                & 0.48               & 0.26                                    \\
0.59 (KCl)              & 0.33~\textpm~0.27 & \textbf{4.2}       & 0.8                & 1.29               & 0.31                                    \\
0.87 (CEM III)          & 1.40~\textpm~0.56 & \textbf{4.7}       & \textbf{1.4}       & 0.77               & 0.41                                    \\
1.02                    & 0.97~\textpm~0.09 & 1.3                & 0.4                & 0.38               & 0.16                                    \\
1.51                    & 1.54~\textpm~0.13 & 1.5                & 0.8                & 0.43               & 0.30                                    \\ \bottomrule
\end{tabular}
\caption{%
    \emph{Trustworthiness of the predictions on cement samples.}
    Analysis of the predictions and the secondary outputs of the \mt model for the cement samples (uncertainties reported represent the \rmse).
    Possible anomalies are highlighted in bold characters.
}
\label{tab:cl_valid}
\end{table}

\begin{figure}[t]
    \centering
    \includegraphics[width=0.49\linewidth]{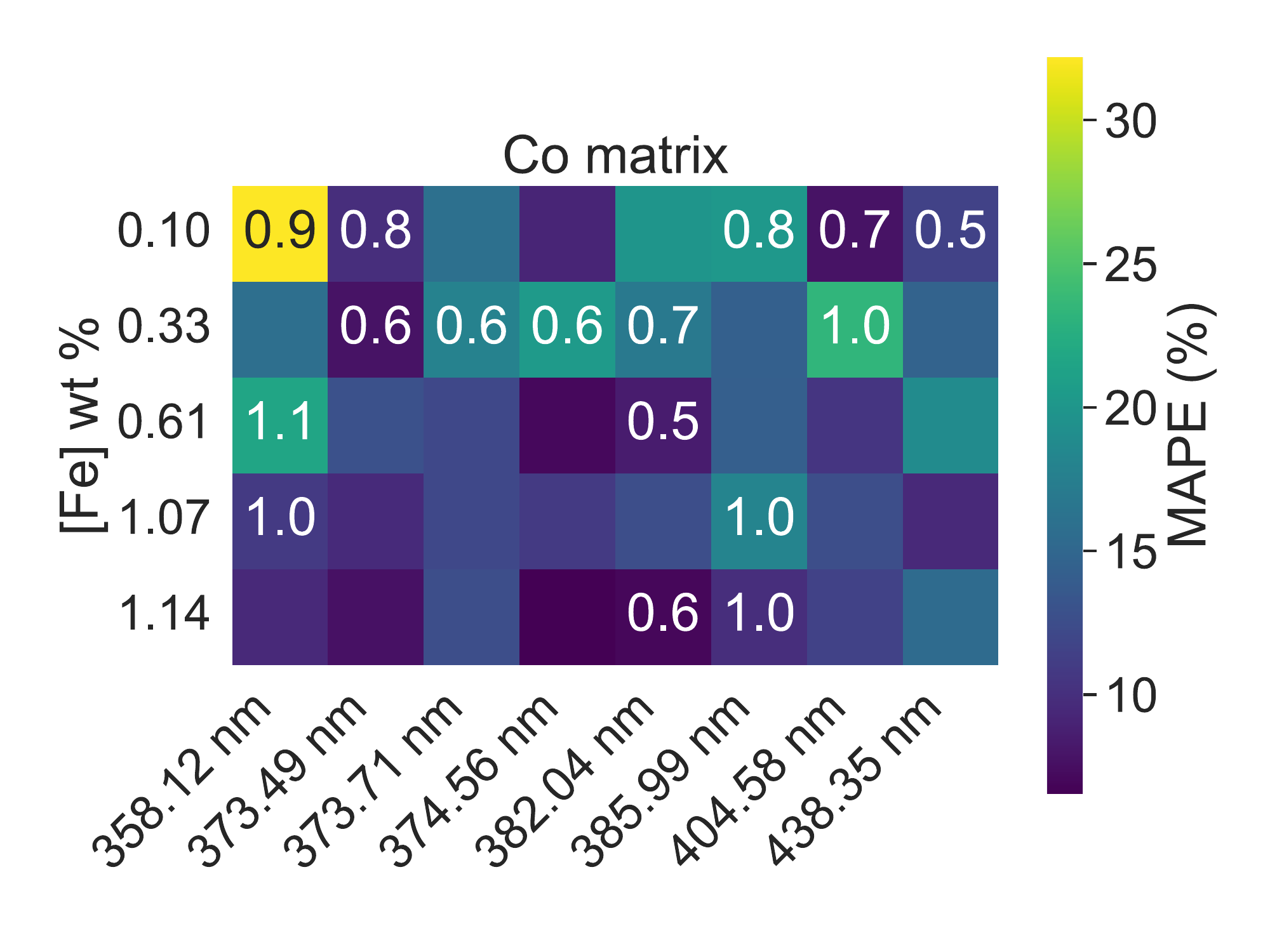}
    \hfill
    \includegraphics[width=0.49\linewidth]{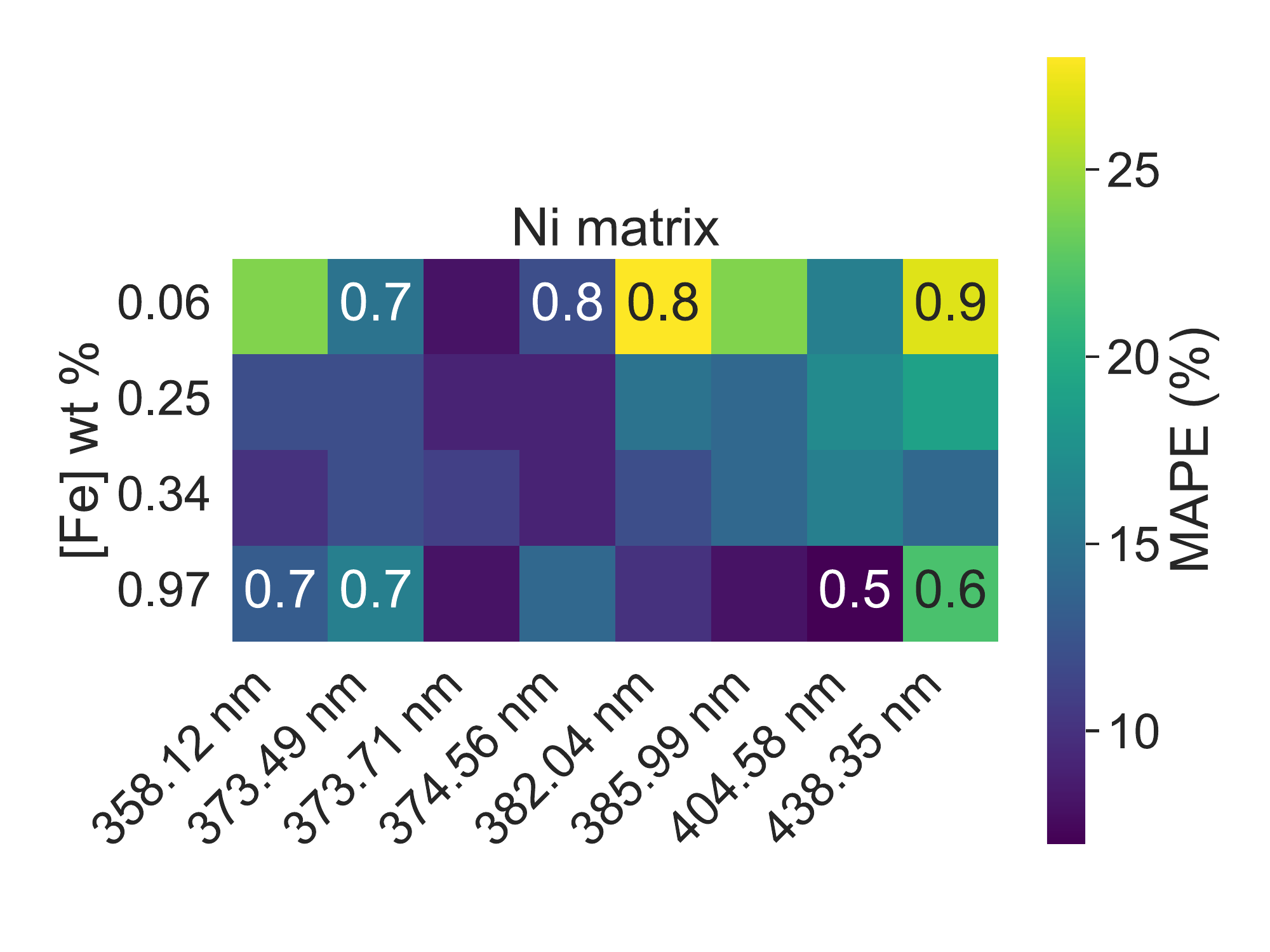}
    \\
    \includegraphics[width=0.49\linewidth]{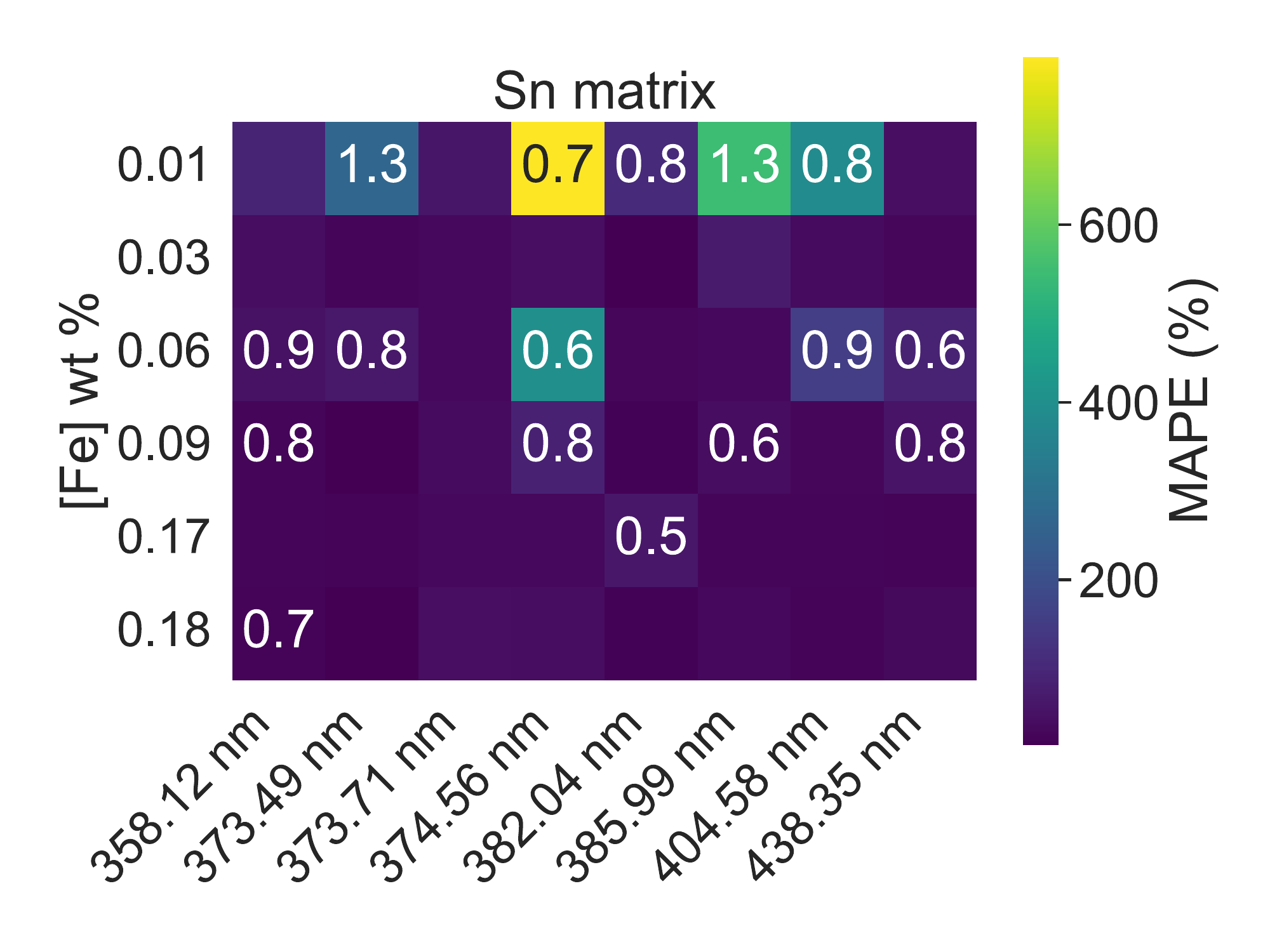}
    \hfill
    \includegraphics[width=0.49\linewidth]{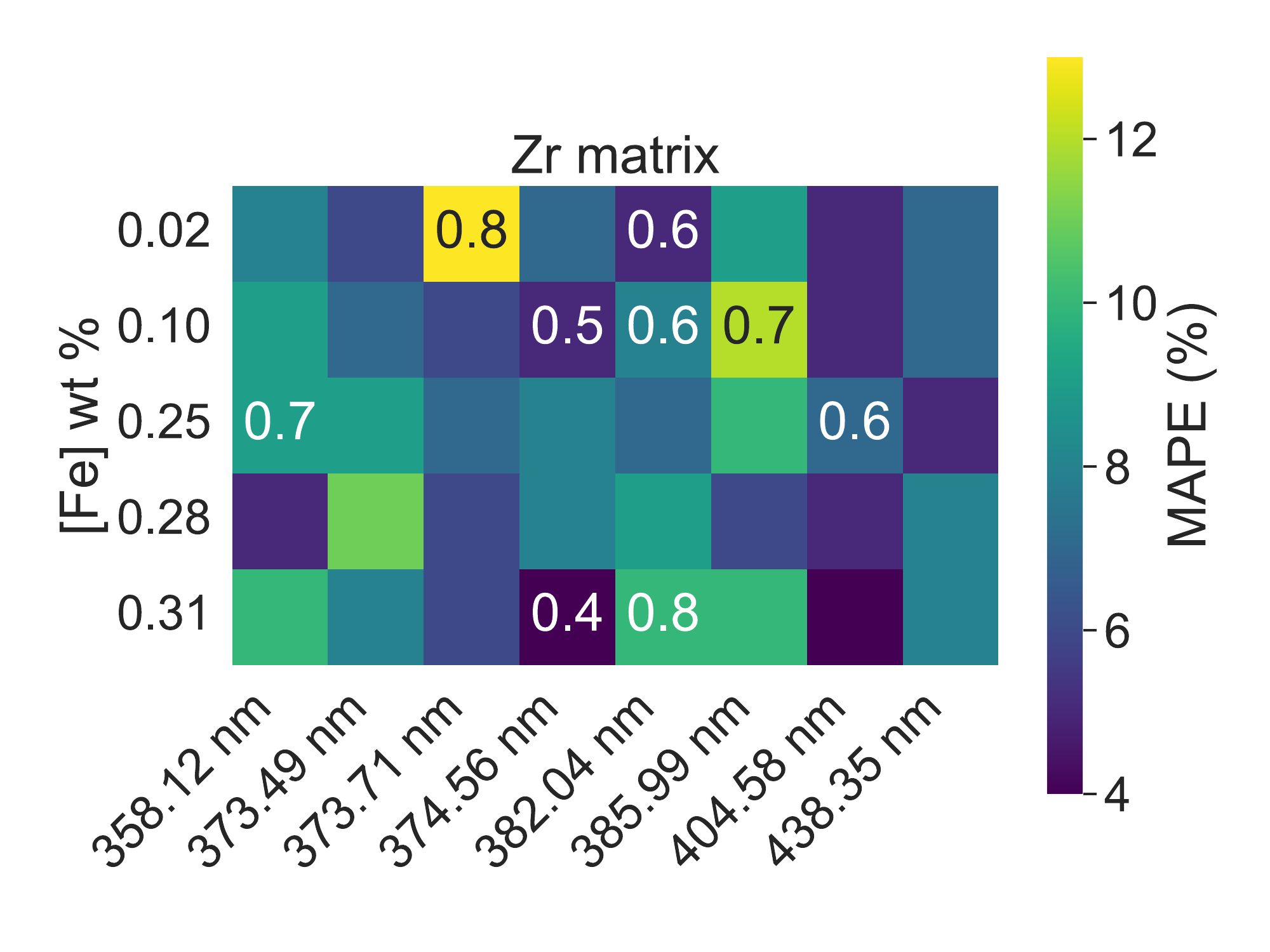}
    \caption{%
        \emph{Summary of the confidence analysis on the alloy matrices.}
        The colour map represents the \mape.
        The $t$ variable is indicated for values outside the confidence intervals measured in the experimental set, for the corresponding concentrations of the analyte, using $\nu = 25$ degrees of freedom and confidence $1 - \alpha = 0.95$ ($\hatt_{1 - \frac{\alpha}{2}}^{\,\nu} = 2.06$).
    }
    \label{fig:alloy_t_mape}
\end{figure}

In \Cref{fig:alloy_t_mape}, we graphically summarise the same analysis for the alloy matrices (see \Cref{tab:mt_valid} in \Cref{app:results} for the full results).
They reasonably present quite different behaviours with respect to the cement samples.
We notice that the Sn matrix displays very high values of the \mape in correspondence of lower Fe concentrations, even though the variance of the predictions is such to ensure the compatibility of the predictions.
This indicates, again, a \mt model whose interpolation is possible even at small concentration of the analyte, even though precision may strongly be impacted.
As previously observed, iron lines are strongly affected by spectral interference, and represents a difficult case for the architecture.
As in the Zn matrix, the interference with Cu and Ni on the entire spectral range leads to a complex computation for the model.
The first element, specifically, is present in the matrix in strongly greater concentrations with respect to the analyte (18 to 160 times the concentration of Fe in the Sn matrix, 60 to 170 in the Zn matrix).
Since both Sn and Zn matrices were not normalised on the contributions of Cu (contrary, for instance, to the Cu matrix itself), the presence of the element restricts the generalisation ability of the model.
We also notice that, for the Sn matrix, the uncertainties of the \mt architecture do not differ from the performance of the ensemble model.
A similar discussion holds also for the Zn matrix (see \Cref{tab:mt_valid} in \Cref{app:results}), for which the spectral interference produces large uncertainties in the prediction of the secondary outputs.
However, the large variance of the predictions of the secondary outputs ensures compatibility with the true values: the model is still able to extrapolate the predictions, even in the presence of a large number of outliers.
In turn, the predictions of the concentration of the analyte can be considered trustworthy (\emph{a posteriori}, they are indeed compatible with the respective ground truths, and present quite small uncertainties, too).
For instance, to improve the performance on these matrices, it would be interesting to increase the training time of the \ai architecture, or fine-tune the models using a the experimental spectra.
As an alternative, more experimental spectra may provide additional elements to model the synthetic data.
Other matrices present only local spikes in the \mape metric, in correspondence of a few Fe emission lines, and mostly for low concentrations of the analyte.
The corresponding $t$-values are well within the limits of standard statistical tests, as shown in \Cref{fig:alloy_t_mape}.
In general, the \mt models, trained on the different matrices, can be deemed trustworthy with high probability.
Moreover, no specific samples stand out as anomalies.
No sample presents incompatible $t$-values or frequent (e.g.\ on all secondary outputs) and large \mape (values are at most at the limit of the confidence intervals computed on the experimental data).
This provides a way to assess the robustness of the \mt architecture on the entire range of variability of the Fe concentration.
Notice that the analysis of the secondary outputs is quantitative: for instance, a $p$-value of the prediction is computable (though, generally, not necessary).
This enables to assess the performance of the model at different concentration levels.
Moreover, different choices of the confidence $1 - \alpha$ allow comparing the model as functions of the confidence level.
Finally, this analysis is always possible, with any sample, since the information contained in the secondary outputs of the model is directly comparable with the experimental data.
This is different from the concentration of the analyte itself, which is known only for standard samples.
This is achievable, in general, with a multi-output model, capable of providing different related quantities as predictions.
A \mt architecture is particularly suited, though more complex, since it first processes the input spectra into a common latent representation (in the hard parameter sharing implementation).
It then proceeds to achieve the best possible performance on each task separately, in turn improving the overall performance and reducing the risk of overfit of the training data (better generalisation).

\subsubsection{Anomaly Detection Mechanism: Calibration Transfer}

\begin{figure}[t]
    \centering
    \includegraphics[height=0.38\linewidth]{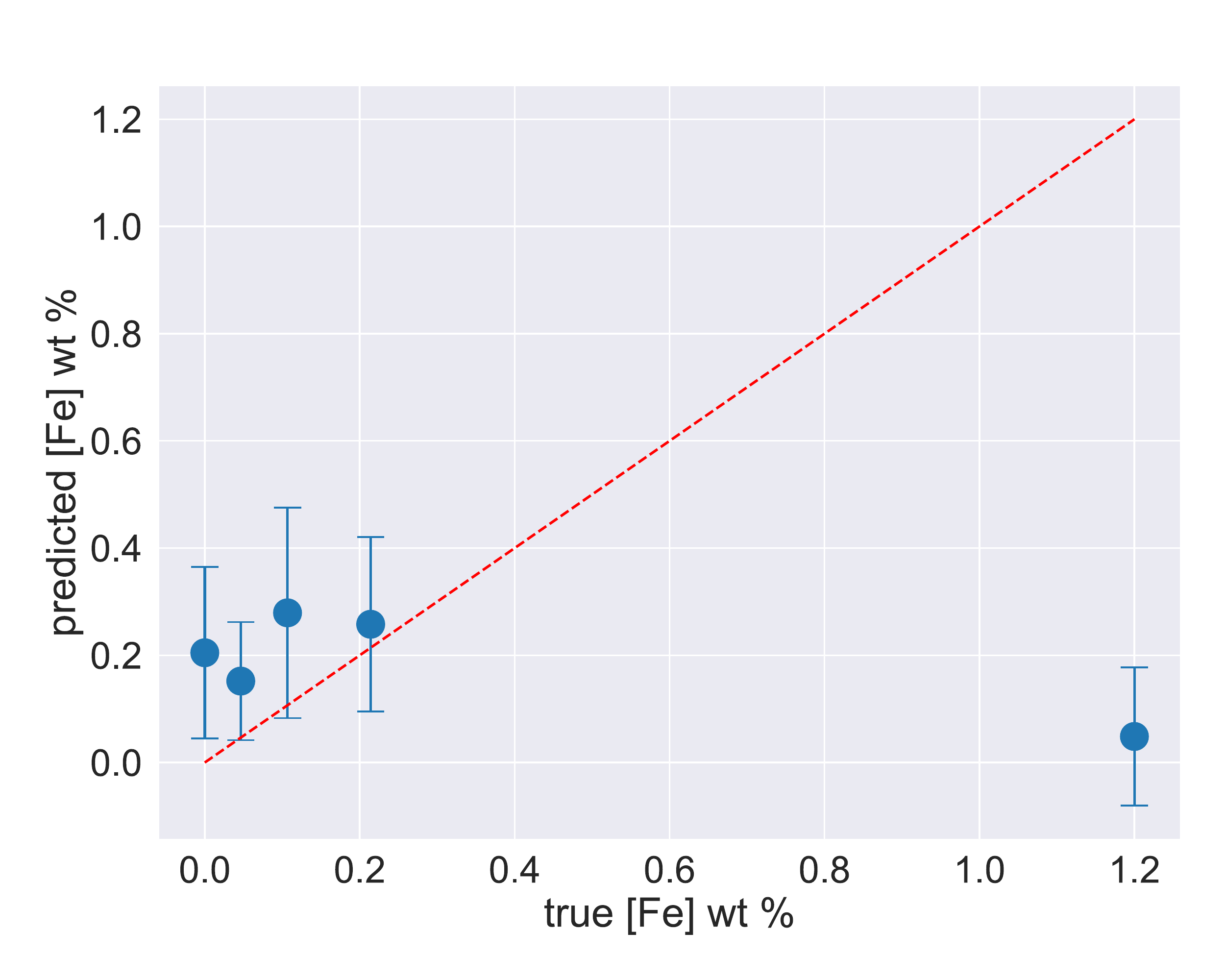}
    \hfill
    \includegraphics[height=0.38\linewidth]{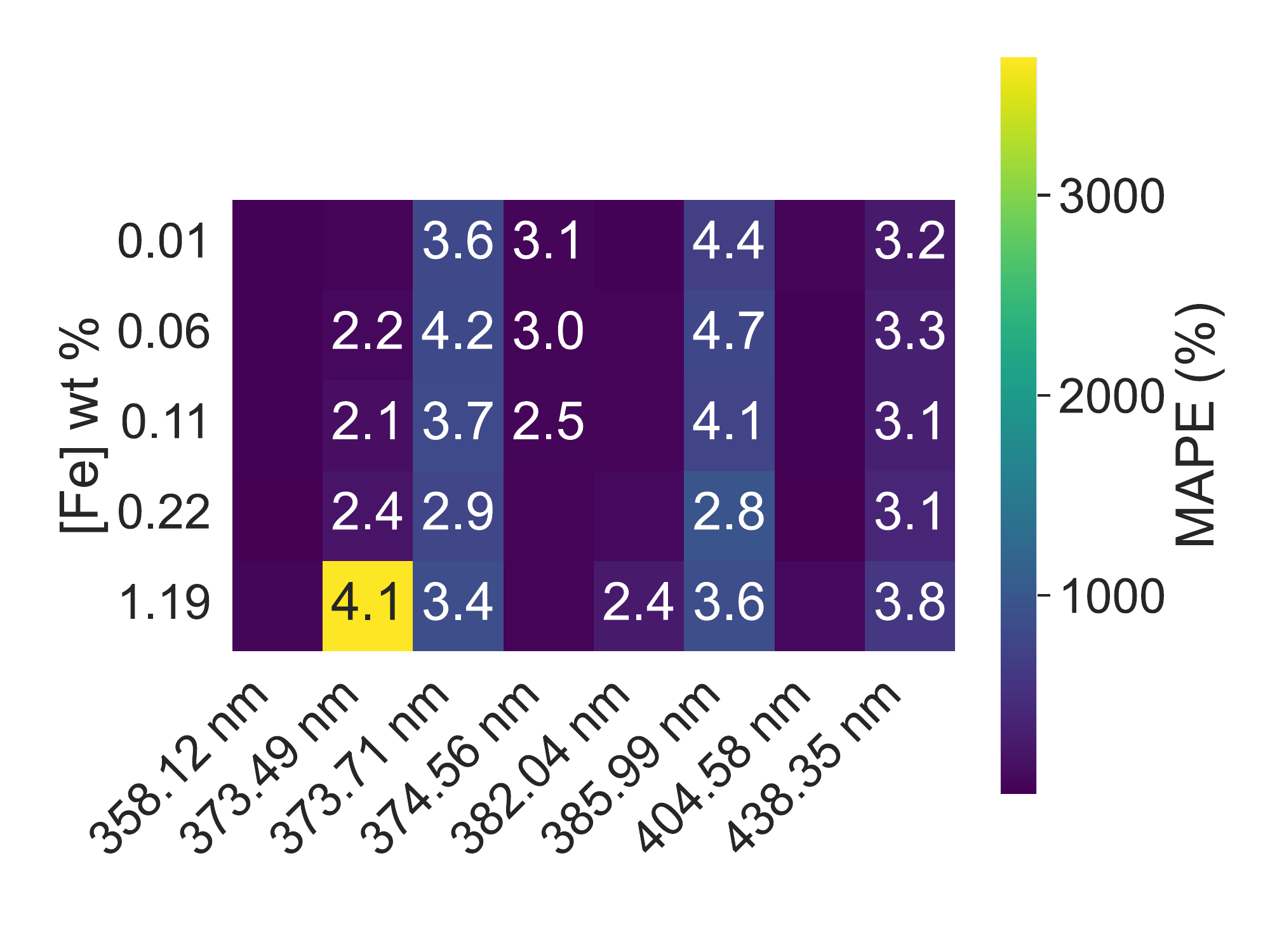}
    \caption{%
        \emph{Calibration transfer of the \mt model.}
        The plot on the left compares the ground truth values of the concentration of the analyte in the Ti matrix with the predictions of the \mt model trained on the Co matrix.
        The summary on the right displays both the values of the error fraction (\mape) of the secondary outputs and the incompatible values of the $t$ variable.
    }
    \label{fig:transfer}
\end{figure}

In order to test the whole procedure, we consider the \mt model trained on a matrix, and compute the predictions on different matrices, to illustrate the anomaly detection mechanism.
We test the trustworthiness of the predictions in the presence of a change in the distribution of the samples.
Specifically, we consider the \mt model trained on the Co matrix, and we perform the inference on the Ti and Zn matrices.
In \Cref{fig:transfer}, we show the true and predicted concentrations of the analyte, generally known only \emph{a posteriori}.
The model does not generalise among different matrices, hence it is not usable for calibration transfer.
In the same image, we also show the statistical analysis on the secondary outputs.
The Ti matrix presents predictions characterised by large values of the error fractions and by incompatible values of the $t$ variable, whose threshold is $\hatt_{1 - \frac{\alpha}{2}}^{\,\nu} = 2.06$ for $\nu = 25$ degrees of freedom and a confidence $1 - \alpha = 0.95$.
For instance, this shows that the predictions of the concentration levels \SIlist{0.11;0.22}{wt \percent}, though compatible with their respective ground truths, should not be considered trustworthy.
The same analysis on the Zn matrix shows \mape larger than \num{e4} on the predictions of the emission lines (see \Cref{fig:transfer_2} in \Cref{app:figures}).
According to \Cref{fig:validation_cases}, any value of the $t$ variable would then be considered anomalous.
The \mt model is thus capable of detecting anomalies or modifications in the experimental conditions, which is key to assessing correctly the ability of the model to provide trustworthy predictions,.
Especially when \emph{a posteriori} analyses are not feasible.

\subsubsection{Towards the Resolution of Spectral Interference}

\begin{figure}[t]
    \centering
    \includegraphics[width=0.49\linewidth]{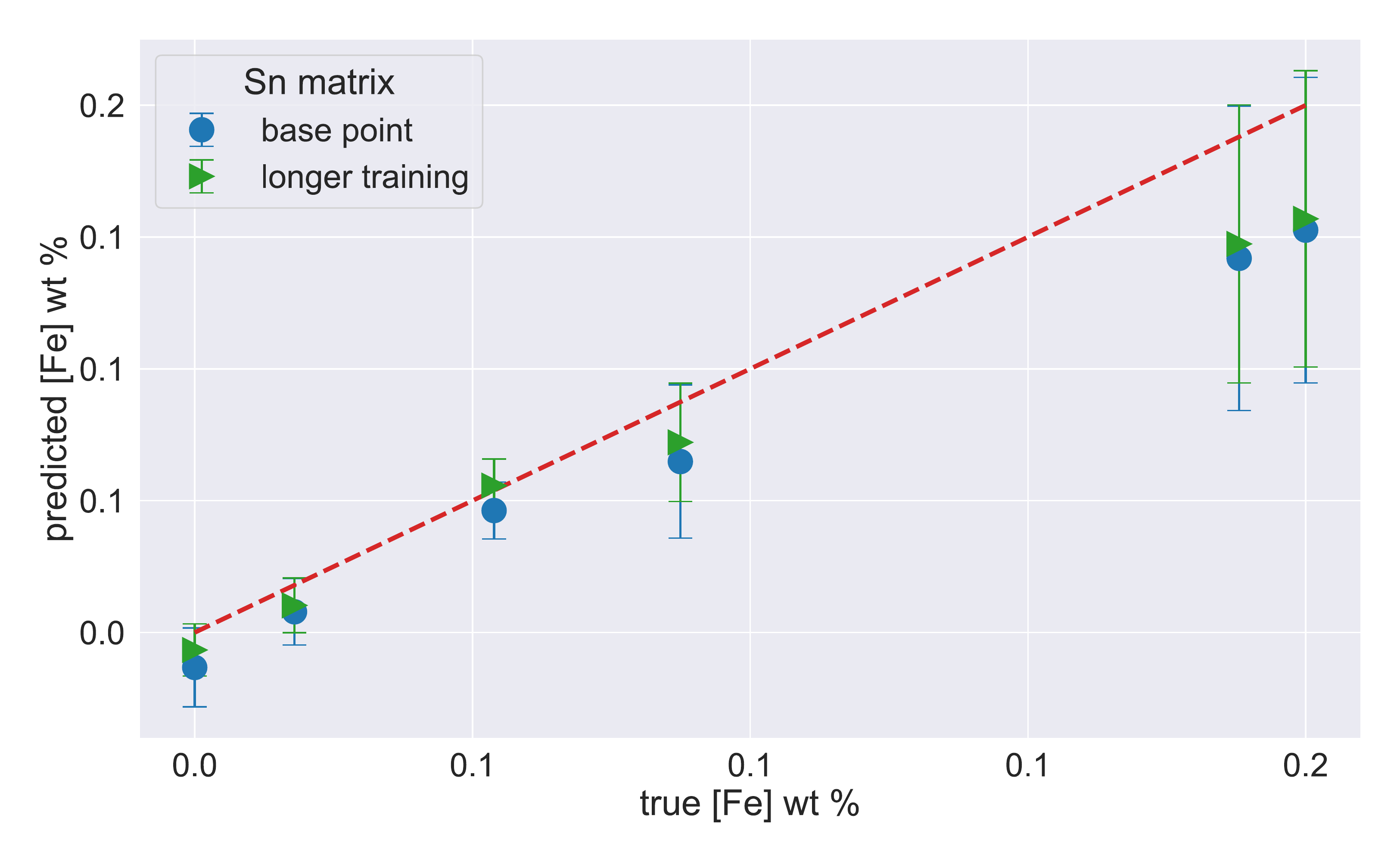}
    \hfill
    \includegraphics[width=0.49\linewidth]{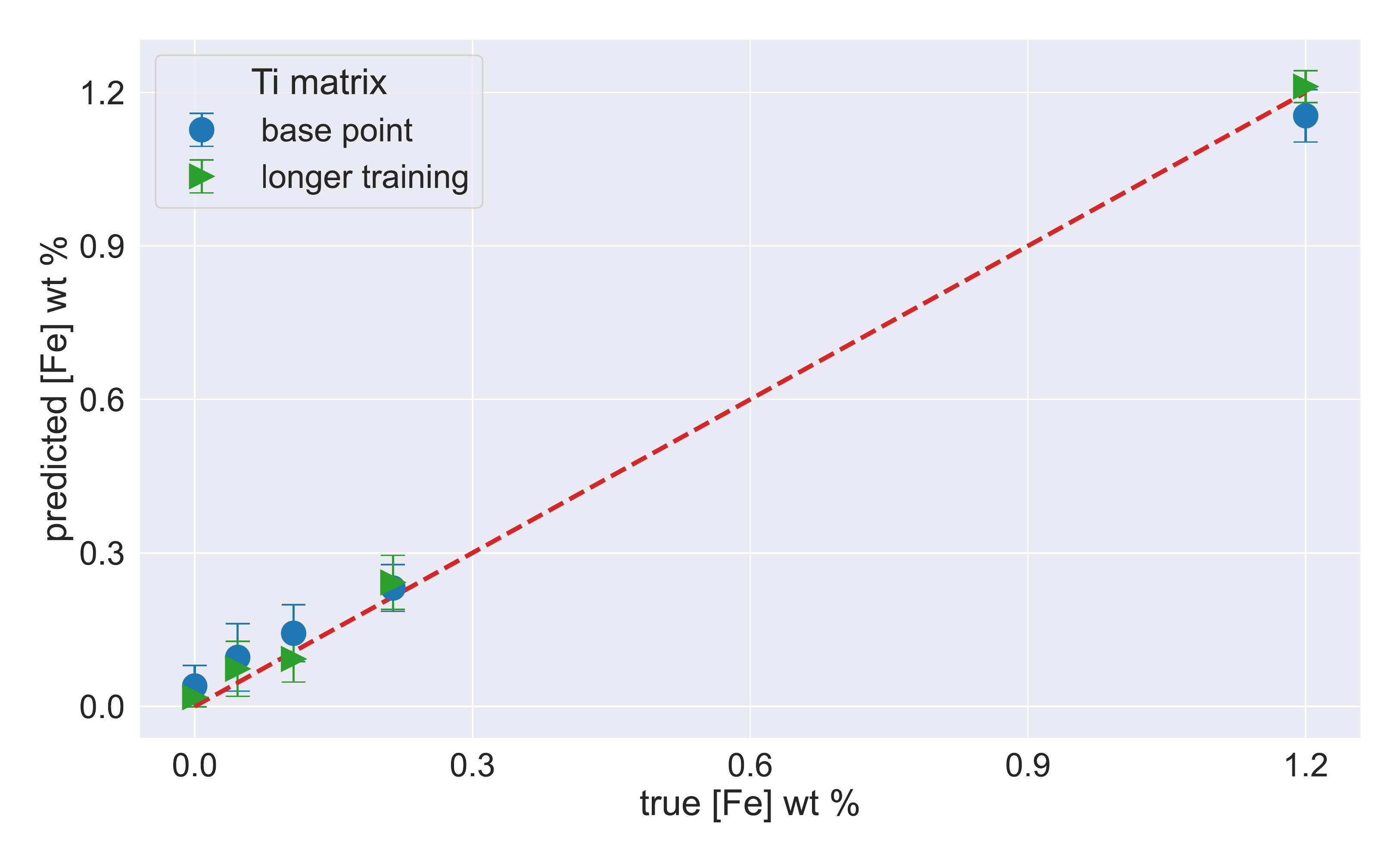}
    \\
    \includegraphics[width=0.49\linewidth]{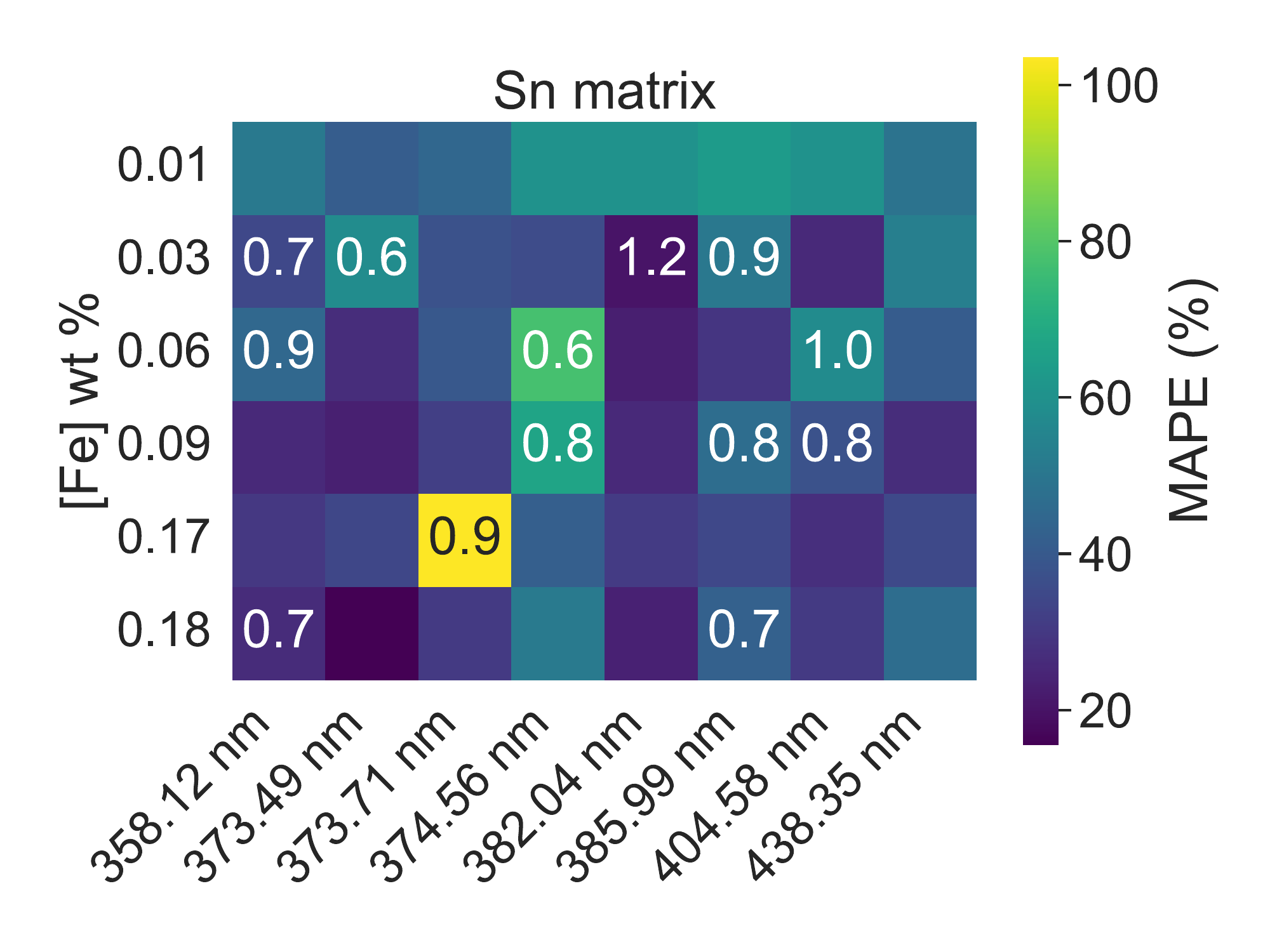}
    \hfill
    \includegraphics[width=0.49\linewidth]{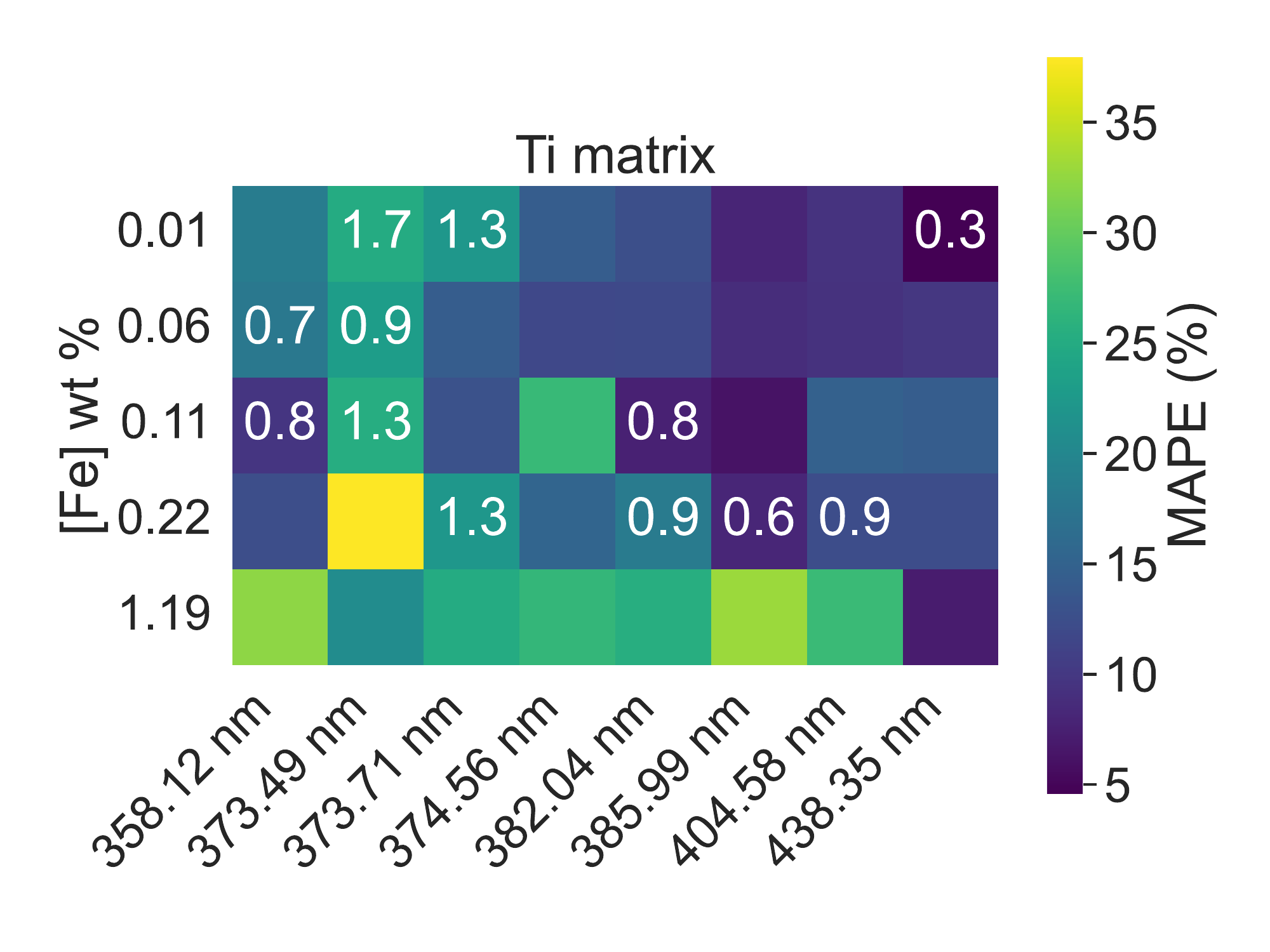}
    \caption{%
        \emph{Impact of longer training on matrices with sparser spectra.}
        The top portion of the figure represents the comparison of the predicted values of the concentration of Fe in the matrices with the corresponding ground truths.
        At the bottom, we show the results of the confidence analysis of the secondary outputs: $t$ values are annotated for values of \mape outside the confidence intervals computed as in \eqref{eq:conf}.
    }
    \label{fig:longer_training}
\end{figure}

Finally, we quickly consider the case of matrices, such as Sn and Ti, in the presence of spectral interference of Fe with elements such as Cu (see \Cref{sec:validation}) for the Sn matrix or the matrix itself in the case of Ti.
Despite the sparser spectra (see \Cref{fig:alloys_avg}), due to these phenomena, the \mt models show higher uncertainty in the prediction of the line intensities, even though the variance of the predictions ensures compatible results.
Given the training technique adopted in the analysis, the fine-tuning of the model to the intensity of the emission lines happens later in training, due to the growth of the associated parameter in the loss (see \Cref{sec:training}).
We consider the Sn and Ti matrix with \num{1000} synthetic spectra per sample, as it represents a good trade-off between performance and wall time needed to train the \mt model.
We perform a longer \num{2000} epochs training.
We also introduce a \num{e-3} $\ell_1$ regularisation factor to enforce sparsity in the model, to focus training only on necessary parameters.
In \Cref{fig:longer_training}, we show the analysis of the main output of the \mt model.
As far as the prediction of the concentration of the analyte is concerned, longer training and sparsity seem to be only slightly beneficial to the trueness of the model.
This is the expected behaviour, since the learning of the main output happens sooner in training.
The major improvement is rather on the reduction of the error fraction of the secondary outputs, which contain more precise predictions.
The \ai model has improved its understanding of the dependencies between the variables (the wavelength channels) contained in the data (spectra).
This has a two-fold effect on the overall performance of the model.
On the one hand, it contributes to slightly improve the predictions of the principal output.
On the other, it grants better generalisation, by improving the adaptation of all outputs to the specific type of data.
In turn, the trustworthiness and confidence of the model improves.
With respect to \Cref{fig:alloy_t_mape}, the Sn matrix in \Cref{fig:longer_training} displays lower \mape for all secondary outputs, and for all concentrations of Fe.
It also displays lower values of the $t$-variable, in correspondence of \mape near the limits of the confidence intervals.
Though not shown previously as an image (results are available in \Cref{tab:mt_valid} in \Cref{app:results}), the Ti matrix displays the same behaviour.
We thus notice that the validation technique needs a good balance in learning of all tasks, in order to provide trustworthy predictions and the corresponding means to check their predictions.

\section{Conclusions}

In this work, we use \dl techniques to address the quantitative analysis of \libs data, the prediction of the concentration of an analyte, using a multivariate calibration procedure.
We focus on complementary aspects: the creation of a synthetic set of spectra as a data augmentation technique to increase the number of samples available for training, the construction of a robust \mt learning model based on deep \cnns, and the analysis of the confidence of the predictions.
We use the entire experimental emission spectra as inputs, without the need for a preselection of variables or dimensionality reduction.
These \dl architectures are, however, more complicated and longer to train than usual univariate or multivariate analysis.
In some cases, they may require large computational power, though we quickly show that specific applications may be satisfied by simpler setups.
Nevertheless, we leverage the robustness and performance of \cnns with the possibility to provide a tool to assess the trustworthiness of the predictions of the model, even for unknown data.
Moreover, once the models are trained, the \mt architecture represents a fast way to perform inference.

To this end, we introduce a \mt learning architecture, based on hard parameter sharing.
The model is capable of predicting the concentration of the analyte and the integral intensities of relevant emission lines (or molecular bands), at the same time.
Given the size and complexity of the \dl model, we introduce a data augmentation technique, based on the simulation of \libs spectra, to create an arbitrary number of input spectra, statistically representative of the experimental data.
In the analysis, we study the role of different elements entering both the theoretical modelling of the spectra and the training of the \nns.
We show that synthetic spectra can indeed be used to train \dl models with good generalisation ability.
The \mt architectures display robustness (homoscedastic behaviour) across the range of variation of the analytes.
It also presents an overall good performance, usually better or, at least, comparable to the \sota, though without the need for any preselection or reduction of the input data.
Finally, the presence of the secondary outputs, allows us to introduce a statistical analysis, based on the mutual dependencies of the parameters of the \ai architecture, which enables the assessment of the trustworthiness of the model.
Comparisons of the predicted values with the intensities found in the experimental spectra, always available, can be used to study the predictions of the concentration of the analyte, at a given level of confidence of the model.
In turn, this grants the ability to assess the extrapolation abilities of the \dl model.
The technique can then be used to identify anomalous samples, or to detect a change in the experimental conditions, with respect to the training samples.

Given the versatility of \mt learning and its complementarity to other techniques already explored in the literature, new investigations may be possible in the future. 
In particular, the model used to generate synthetic spectra may be further developed to account for different experimental aspects.
For instance, it could be used to model different noise sources as different confounder parameters in the likelihood function in \eqref{eq:likelihood}.
Moreover, advanced generative models, such as adversarial networks~\cite{Goodfellow:2014:Advances} or the more recently successful diffusion models~\cite{SohlDicstein:2015:Deep, Ramesh:2021:Zero, Saharia:2022:Photorealistic}, may be employed to model the distribution of the \libs spectra and generate realistic synthetic datasets~\cite{Alaa:2022:Faithful}, with the possibility to control several experimental and physical parameters.
\mt also offers the possibility to provide multiple predictions of concentrations at once.
In principle, different analytes could be studied together (as already explored in the literature) and a measure of global confidence could be attached to the model, using the secondary outputs as a basis for the statistical analysis.
Moreover, interpretable-\ai methods, either based on the study of the hidden structures in \nns~\cite{Zhang:2020:Understanding, Zhao:2021:Interpretable} or quantitative statistical tools~\cite{Lipovetsky:2001:Analysis, Strumbelj:2014:Explaining, Ribeiro:2016:Why}, could be coupled to our technique to provide a more complete analysis.
Finally, as training sets of arbitrary size can be synthesised, the impact of larger architectures could be explored at length.
This could provide the possibility to use more complex and larger models, based on transformers or attention mechanisms~\cite{Vaswani:2017:Attention}, which may also give some insights on interpretability of the decision mechanisms.

\section*{Author Contributions}

\textbf{Riccardo Finotello}: conceptualization, data curation, formal analysis, investigation, methodology, software, validation, visualization, writing -- original draft, writing -- review \& editing;
\textbf{Daniel L'Hermite}: investigation, methodology, resources, validation, writing -- review \& editing;
\textbf{Celine Quéré}: investigation, methodology, validation, writing -- review \& editing;
\textbf{Benjamin Rouge}: investigation, methodology, validation;
\textbf{Mohamed Tamaazousti}: funding acquisition, project administration, resources, supervision, writing -- review \& editing;
\textbf{Jean-Baptiste Sirven}: data curation, funding acquisition, project administration, resources, supervision, writing -- review \& editing.

\section*{Conflicts of Interests}

There are no conflicts to declare.

\section*{Acknowledgements}

We acknowledge the financial support of the \emph{Cross-Disciplinary Programme on Instrumentation and Detection} of CEA, the French \emph{Alternative Energies and Atomic Energy Commission}.
We thank G.\ Gallou for proposing this collaboration.
This publication was made possible by the use of the \emph{FactoryIA} supercomputer, financially supported by the \emph{Ile-De-France Regional Council}.

\printbibliography[heading=bibintoc]

\clearpage

\appendix\break\pagenumbering{roman}\renewcommand{\thepage}{\roman{page}}

\section{Compositions of the Matrices}\label{app:matrices}

\begin{table}[h]
\centering
\begin{tabular}{@{}cccc@{}}
\toprule
                                   & \textbf{Type}   & \textbf{Salt} & \textbf{{[}Cl{]} wt\%} \\ \midrule
\multirow{12}{*}{\rotatebox{90}{training samples}} & CEM I           & NaCl          & 0.06                   \\
                                   & CEM I           & NaCl          & 0.19                   \\
                                   & CEM I           & NaCl          & 0.32                   \\
                                   & CEM I           & NaCl          & 0.46                   \\
                                   & CEM I           & NaCl          & 0.59                   \\
                                   & CEM I           & NaCl          & 0.72                   \\
                                   & CEM I           & NaCl          & 0.85                   \\
\bottomrule
\end{tabular}
\caption{%
    \emph{Elemental composition of the cement matrices.}
    The type of cement and the salt used in fabrication are specified on a sample basis.
}
\label{tab:cement_samples}
\end{table}

\newgeometry{margin=1.5cm}

\begin{landscape}
\begin{table}[h]
\centering
\resizebox{\columnwidth}{!}{%
\begin{tabular}{@{}cc|ccccc|ccccc|ccccc|cccc@{}}
\toprule
                                   & \textbf{Matrix} & \multicolumn{5}{c|}{\textbf{Aluminium}}                                               & \multicolumn{5}{c|}{\textbf{Cobalt}}                                                          & \multicolumn{5}{c|}{\textbf{Copper}}                                            & \multicolumn{4}{c}{\textbf{Nickel}}                              \\ \midrule
\textbf{}                          & \textbf{Sample} & \textbf{51XG00H2} & \textbf{35831} & \textbf{63698} & \textbf{52963} & \textbf{68404} & \textbf{IMZ 188} & \textbf{BS171C} & \textbf{ECRM3781D} & \textbf{BSMP35N} & \textbf{IMZ 186} & \textbf{CT1}  & \textbf{CT2} & \textbf{CT3}   & \textbf{CT4}   & \textbf{CT5}   & \textbf{B7011} & \textbf{B7012} & \textbf{B7004} & \textbf{F294} \\ \midrule
\multirow{44}{*}{\rotatebox{90}{Composition wt \%}} & Ag              &                   &                &                &                &                &                  &                 &                    &                  &                  & 0.057         & 0.042        & 0.026          & 0.016          & 0.0062         &                &                &                &               \\
                                   & Al              & 98.35             & 97.49          & 95.08          & 98.73          & 98.00          & (0.005)          & (0.04)          &                    & 0.047            & 0.28             &               &              &                &                &                & 1.2            & 1.25           & 1.64           & 0.05          \\
\bottomrule
\end{tabular}%
}
\caption{%
    \emph{Elemental composition of the alloy matrices (part 1).}
    The concentration of the analyte is in bold characters.
}
\label{tab:alloys_part1}
\end{table}
\end{landscape}

\begin{landscape}
\begin{table}[h]
\centering
\resizebox{\columnwidth}{!}{%
\begin{tabular}{@{}cc|cccccc|ccccc|ccccc|ccccc@{}}
\toprule
                                   & \textbf{Matrix} & \multicolumn{6}{c|}{\textbf{Tin}}                                                                               & \multicolumn{5}{c|}{\textbf{Titanium}}                                                 & \multicolumn{5}{c|}{\textbf{Zinc}}                                                           & \multicolumn{5}{c}{\textbf{Zirconium}}                                                      \\ \midrule
\textbf{}                          & \textbf{Sample} & \textbf{IMNL891} & \textbf{IMNL892} & \textbf{IMNL893} & \textbf{IMNL894} & \textbf{IMNL895} & \textbf{IMNL896} & \textbf{BST26} & \textbf{BST80}  & \textbf{BST81}  & \textbf{BST22} & \textbf{BSTSU-1} & \textbf{43XZ5B} & \textbf{43XZ9A} & \textbf{43XGALF1A} & \textbf{43XSC2A} & \textbf{43XSC3A} & \textbf{ET1}     & \textbf{ET2}    & \textbf{ET3}     & \textbf{ET4}     & \textbf{ET5}     \\ \midrule
\multirow{44}{*}{\rotatebox{90}{Composition wt \%}} & Ag              &                  &                  &                  &                  &                  &                  &                &                 &                 &                &                  & 0.0254          &                 &                    &                  &                  &                  &                 &                  &                  &                  \\
                                   & Al              &                  &                  &                  &                  &                  &                  & 0.002          & 0.0746          & 0.0664          & 0.004          & 6.11             & 3.164           & 3.17            & 4.68               & 3.41             & 3.14             & 0.00133          & 0.00158         & 0.01329          & 0.04384          & 0.09514          \\
\bottomrule
\end{tabular}%
}
\caption{%
    \emph{Elemental composition of the alloy matrices (part 2).}
    The concentration of the analyte is in bold characters.
}
\label{tab:alloys_part2}
\end{table}
\end{landscape}

\restoregeometry

\clearpage

\section{Implementation Details}\label{app:details}

\begin{table}[h]
\centering
\begin{tabular}{@{}ccccc@{}}
\toprule
\textbf{Wavelength   (nm)} & \textbf{Integration interval (nm)} & \textbf{Relative intensity} & \textbf{Spectrum} \\ \midrule
358.12                     & 358.11 -- 358.14                   & 600                         & Fe I              \\
373.49                     & 373.47 -- 373.51                   & 700                         & Fe I              \\
373.71                     & 373.69 -- 373.73                   & 600                         & Fe I              \\
374.56                     & 374.54 -- 374.58                   & 600                         & Fe I              \\
382.04                     & 382.02 -- 382.06                   & 500                         & Fe I              \\
385.99                     & 385.97 -- 386.01                   & 500                         & Fe I              \\
404.58                     & 404.56 -- 404.60                   & 300                         & Fe I              \\
438.35                     & 438.33 -- 438.37                   & 200                         & Fe I              \\ \bottomrule
\end{tabular}
\caption{%
    \emph{Wavelengths corresponding to persistent emission lines of Fe}.
    Integration intervals used as secondary outputs of the multitask architecture are also presented.
}
\label{tab:fe_lines}
\end{table}

\begin{table}[h]
\centering
\begin{tabular}{@{}ccc@{}}
\toprule
\textbf{Matrix} & \textbf{Wavelength (nm)} & \textbf{Integration interval (nm)} \\ \midrule
Al              & 396.15                   & 396.13 -- 396.18                   \\
Co              & 345.35                   & 345.33 -- 345.67                   \\
Cu              & 324.75                   & 324.74 -- 324.77                   \\
Ni              & 352.45                   & 352.43 -- 352.46                   \\
Sn              & 326.23                   & 326.21 -- 326.25                   \\
Ti              & 399.86                   & 399.84 -- 399.87                   \\
Zn              & 334.50                   & 334.49 -- 334.52                   \\
Zr              & 360.12                   & 360.10 -- 360.14                   \\ \bottomrule
\end{tabular}
\caption{%
    \emph{Normalisation wavelengths of the alloy matrices.}
    Integration intervals are specified for each matrix.
}
\label{tab:normal}
\end{table}

\begin{table}[h]
\centering
\begin{tabular}{@{}ccc@{}}
\toprule
                & \multicolumn{2}{c}{\textbf{PLS1 Components}}                    \\ \midrule
\textbf{Matrix} & \textbf{Experimental data} & \textbf{Synthetic data (N = 1000)} \\ \midrule
Al              & 2                          & 2                                  \\
Co              & 6                          & 8                                  \\
Cu              & 1                          & 8                                  \\
Ni              & 1                          & 1                                  \\
Sn              & 1                          & 1                                  \\
Ti              & 7                          & 4                                  \\
Zn              & 3                          & 3                                  \\
Zr              & 4                          & 3                                  \\ \midrule
Cement          & 2                          & 2                                  \\ \bottomrule
\end{tabular}
\caption{%
    \emph{Number of components used in the \pls[1] model.}
    5-fold cross validation was used to determine the best value of the hyperparameter.
}
\label{tab:pls_comps}
\end{table}

\begin{table}[h]
\centering
\begin{tabular}{@{}cccc@{}}
\toprule
                                   &                                                    & \multicolumn{2}{c}{\textbf{Value}}                          \\
                                   & \textbf{Hyperparameter}                            & \textbf{Alloys}               & \textbf{Cement}             \\ \midrule
                                   & $\varphi$                                          & 0.995                         & 0.995                       \\
                                   & $\ell_1$                                           & 0                             & 0                           \\
                                   & $\ell_2$                                           & \num{3e-5}   & 0                           \\
                                   & learning rate                                      & \num{e-4}    & \num{e-3}  \\
                                   & $\halpha_0$                                        & 0.91                          & 0.93                        \\
                                   & $\eta$                                             & 0.99                          & 0.99                        \\
\bottomrule
\end{tabular}
\caption{%
    \emph{List of hyperparameters used for the \mt model.}
    Symbols are introduced in \Cref{sec:training}.
}
\label{tab:mt_hyper}
\end{table}

\clearpage

\section{Results}\label{app:results}

\begin{table}[h]
\centering
\resizebox{\columnwidth}{!}{%
\begin{tabular}{@{}ccc|cccc|cccc|cccc|cccc@{}}
\toprule
                     &                         &                        & \multicolumn{4}{c|}{\textbf{Predicted [X] wt \%}}      & \multicolumn{4}{c|}{\textbf{RMSE wt \%}}                        & \multicolumn{4}{c|}{\textbf{MAE wt \%}}                          & \multicolumn{4}{c}{\textbf{MAPE (\%)}}                     \\ \cmidrule(l){4-19}
\textbf{X}           & \textbf{Matrix}         & \textbf{[X] wt \%} & \textbf{LR} & \textbf{MLR} & \textbf{PLS1} & \textbf{FCNN} & \textbf{LR}    & \textbf{MLR}  & \textbf{PLS1} & \textbf{FCNN}  & \textbf{LR}    & \textbf{MLR}  & \textbf{PLS1}  & \textbf{FCNN}  & \textbf{LR} & \textbf{MLR} & \textbf{PLS1} & \textbf{FCNN} \\ \midrule
\multirow{40}{*}{\rotatebox{90}{Fe}} & \multirow{5}{*}{Al}     & 0.11                   & 0.13        & 0.15         & 0.17          & 0.13          & 0.05           & 0.05          & 0.07          & \textbf{0.04}  & 0.04           & 0.04          & 0.06           & \textbf{0.02}  & 36          & 39           & 57            & \textbf{22}   \\
                     &                         & 0.36                   & 0.39        & 0.34         & 0.37          & 0.35          & 0.10           & \textbf{0.06} & 0.07          & 0.08           & 0.10           & \textbf{0.05} & 0.06           & 0.07           & 28          & \textbf{15}  & 18            & 20            \\
                     &                         & 0.41                   & 0.45        & 0.43         & 0.45          & 0.39          & 0.09           & 0.06          & 0.08          & \textbf{0.04}  & 0.08           & 0.05          & 0.06           & \textbf{0.04}  & 18          & 13           & 16            & \textbf{10}   \\
\bottomrule
\end{tabular}%
}
\caption{
    \emph{Results of the analysis for traditional algorithms with experimental data.}
    Results of inference are relative to the independent test set.
    The columns show the Root Mean Squared Error (\rmse), the Mean Absolute Error (\mae), and the Mean Absolute Percentage Error (\mape).
    The algorithms refer to Linear Regression (\lr), Multivariate Linear Regression (\mlr), Partial Least Squares (\pls[1]), and Fully Connected Neural Networks (\fcnn).
    Bold characters represent the best estimates.
}
\label{tab:trad_an}
\end{table}

\begin{table}[h]
\centering
\resizebox{\columnwidth}{!}{%
\begin{tabular}{@{}ccc|cccc|cccc|cccc|cccc@{}}
\toprule
                     &                         &                        & \multicolumn{4}{c|}{\textbf{Predicted [X] wt \%}}      & \multicolumn{4}{c|}{\textbf{RMSE wt \%}}                   & \multicolumn{4}{c|}{\textbf{MAE wt \%}}                    & \multicolumn{4}{c}{\textbf{MAPE (\%)}}                     \\ \cmidrule(l){4-19}
\textbf{X}           & \textbf{Matrix}         & \textbf{[X] wt \%} & \textbf{LR} & \textbf{MLR} & \textbf{PLS1} & \textbf{FCNN} & \textbf{LR} & \textbf{MLR} & \textbf{PLS1} & \textbf{FCNN} & \textbf{LR} & \textbf{MLR} & \textbf{PLS1} & \textbf{FCNN} & \textbf{LR} & \textbf{MLR} & \textbf{PLS1} & \textbf{FCNN} \\ \midrule
\multirow{40}{*}{\rotatebox{90}{Fe}} & \multirow{5}{*}{Al}     & 0.11                   & 0.10        & 0.13         & 0.12          & 0.11          & 0.07        & 0.05         & 0.04          & 0.03          & 0.06        & 0.04         & 0.04          & 0.02          & 52          & 38           & 33            & 20            \\
                     &                         & 0.36                   & 0.34        & 0.34         & 0.32          & 0.35          & 0.09        & 0.07         & 0.07          & 0.07          & 0.09        & 0.05         & 0.05          & 0.05          & 24          & 15           & 15            & 15            \\
                     &                         & 0.41                   & 0.45        & 0.43         & 0.42          & 0.44          & 0.07        & 0.05         & 0.05          & 0.04          & 0.05        & 0.04         & 0.04          & 0.03          & 12          & 9            & 9             & 7             \\
                     &                         & 0.56                   & 0.64        & 0.61         & 0.60          & 0.62          & 0.10        & 0.10         & 0.09          & 0.08          & 0.09        & 0.09         & 0.08          & 0.06          & 16          & 16           & 14            & 11            \\
                     &                         & 0.81                   & 0.62        & 0.67         & 0.66          & 0.65          & 0.21        & 0.16         & 0.17          & 0.18          & 0.19        & 0.14         & 0.15          & 0.16          & 24          & 17           & 19            & 20            \\ \cmidrule(l){2-19} 
                     & \multirow{5}{*}{Co}     & 0.10                   & 0.15        & 0.16         & 0.02          & 0.18          & 0.21        & 0.12         & 0.13          & 0.12          & 0.19        & 0.09         & 0.11          & 0.08          & 191         & 89           & 110           & 83            \\
                     &                         & 0.33                   & 0.36        & 0.36         & 0.22          & 0.40          & 0.13        & 0.11         & 0.14          & 0.09          & 0.11        & 0.10         & 0.11          & 0.08          & 33          & 31           & 32            & 23            \\
                     &                         & 0.61                   & 0.52        & 0.50         & 0.36          & 0.55          & 0.17        & 0.16         & 0.27          & 0.18          & 0.16        & 0.15         & 0.25          & 0.16          & 27          & 25           & 41            & 27            \\
                     &                         & 1.07                   & 1.06        & 1.09         & 0.96          & 1.10          & 0.15        & 0.16         & 0.20          & 0.07          & 0.13        & 0.13         & 0.17          & 0.07          & 12          & 12           & 16            & 6             \\
                     &                         & 1.14                   & 0.97        & 1.00         & 0.87          & 1.07          & 0.22        & 0.17         & 0.29          & 0.09          & 0.17        & 0.14         & 0.27          & 0.07          & 15          & 12           & 24            & 6             \\ \cmidrule(l){2-19} 
                     & \multirow{5}{*}{Cu}     & 0.02                   & 0.04        & 0.03         & 0.02          & 0.02          & 0.03        & 0.014        & 0.01          & 0.01          & 0.02        & 0.01         & 0.008         & 0.009         & 130         & 68           & 52            & 56            \\
                     &                         & 0.05                   & 0.05        & 0.04         & 0.04          & 0.05          & 0.02        & 0.02         & 0.02          & 0.03          & 0.02        & 0.014        & 0.02          & 0.02          & 36          & 31           & 38            & 49            \\
                     &                         & 0.08                   & 0.09        & 0.10         & 0.10          & 0.11          & 0.03        & 0.06         & 0.06          & 0.04          & 0.03        & 0.04         & 0.04          & 0.03          & 31          & 45           & 46            & 31            \\
                     &                         & 0.10                   & 0.08        & 0.09         & 0.09          & 0.10          & 0.02        & 0.03         & 0.03          & 0.02          & 0.02        & 0.02         & 0.03          & 0.01          & 20          & 22           & 26            & 12            \\
                     &                         & 0.17                   & 0.13        & 0.15         & 0.15          & 0.14          & 0.05        & 0.05         & 0.05          & 0.05          & 0.04        & 0.04         & 0.04          & 0.04          & 26          & 24           & 25            & 24            \\ \cmidrule(l){2-19} 
                     & \multirow{4}{*}{Ni}     & 0.06                   & 0.10        & 0.11         & 0*            & 0.07          & 0.04        & 0.05         & 0.09          & 0.009         & 0.04        & 0.05         & 0.08          & 0.007         & 66          & 80           & 142           & 11            \\
                     &                         & 0.25                   & 0.26        & 0.26         & 0.21          & 0.28          & 0.05        & 0.03         & 0.05          & 0.04          & 0.04        & 0.03         & 0.05          & 0.03          & 16          & 10           & 18            & 12            \\
                     &                         & 0.34                   & 0.38        & 0.35         & 0.27          & 0.33          & 0.07        & 0.04         & 0.08          & 0.02          & 0.06        & 0.03         & 0.07          & 0.01          & 19          & 9            & 20            & 4             \\
                     &                         & 0.97                   & 0.89        & 0.89         & 0.79          & 0.93          & 0.18        & 0.12         & 0.21          & 0.06          & 0.15        & 0.10         & 0.18          & 0.04          & 15          & 10           & 19            & 5             \\ \cmidrule(l){2-19} 
                     & \multirow{6}{*}{Sn}     & 0.01                   & 0.01        & 0.02         & 0.03          & 0.02          & 0.02        & 0.01         & 0.014         & 0.006         & 0.014       & 0.008        & 0.01          & 0.004         & 105         & 63           & 92            & 34            \\
                     &                         & 0.03                   & 0.04        & 0.05         & 0.05          & 0.04          & 0.03        & 0.02         & 0.02          & 0.01          & 0.02        & 0.02         & 0.02          & 0.01          & 84          & 70           & 80            & 29            \\
                     &                         & 0.06                   & 0.06        & 0.06         & 0.06          & 0.04          & 0.02        & 0.02         & 0.02          & 0.02          & 0.02        & 0.02         & 0.02          & 0.01          & 29          & 29           & 30            & 25            \\
                     &                         & 0.09                   & 0.10        & 0.10         & 0.09          & 0.12          & 0.03        & 0.02         & 0.02          & 0.05          & 0.03        & 0.02         & 0.013         & 0.04          & 30          & 23           & 15            & 49            \\
                     &                         & 0.17                   & 0.14        & 0.14         & 0.14          & 0.14          & 0.05        & 0.06         & 0.05          & 0.05          & 0.04        & 0.05         & 0.05          & 0.04          & 21          & 31           & 27            & 21            \\
                     &                         & 0.18                   & 0.13        & 0.13         & 0.13          & 0.16          & 0.05        & 0.06         & 0.06          & 0.04          & 0.05        & 0.05         & 0.05          & 0.02          & 25          & 30           & 30            & 14            \\ \cmidrule(l){2-19} 
                     & \multirow{5}{*}{Ti}     & 0.01                   & 0.04        & 0*           & 0*            & 0.02          & 0.10        & 0.04         & 0.05          & 0.25          & 0.07        & 0.03         & 0.04          & 0.015         & 787         & 371          & 467           & 158           \\
                     &                         & 0.06                   & 0.13        & 0.09         & 0.08          & 0.08          & 0.16        & 0.07         & 0.07          & 0.08          & 0.13        & 0.07         & 0.07          & 0.05          & 237         & 127          & 124           & 82            \\
                     &                         & 0.11                   & 0.15        & 0.15         & 0.13          & 0.11          & 0.08        & 0.07         & 0.06          & 0.11          & 0.05        & 0.06         & 0.05          & 0.03          & 46          & 49           & 43            & 24            \\
                     &                         & 0.22                   & 0.27        & 0.30         & 0.29          & 0.22          & 0.09        & 0.11         & 0.10          & 0.14          & 0.08        & 0.08         & 0.07          & 0.03          & 37          & 38           & 33            & 14            \\
                     &                         & 1.19                   & 1.19        & 1.12         & 1.11          & 1.10          & 0.25        & 0.14         & 0.14          & 0.22          & 0.20        & 0.12         & 0.1           & 0.11          & 17          & 10           & 11            & 9.1           \\ \cmidrule(l){2-19} 
                     & \multirow{5}{*}{Zn}     & 0.02                   & 0.05        & 0.06         & 0.05          & 0.06          & 0.04        & 0.04         & 0.03          & 0.04          & 0.04        & 0.04         & 0.03          & 0.04          & 196         & 218          & 153           & 223           \\
                     &                         & 0.05                   & 0.06        & 0.06         & 0.05          & 0.06          & 0.014       & 0.02         & 0.01          & 0.01          & 0.01        & 0.012        & 0.007         & 0.01          & 28          & 27           & 16            & 26            \\
                     &                         & 0.06                   & 0.03        & 0.09         & 0.09          & 0.06          & 0.05        & 0.07         & 0.08          & 0.003         & 0.03        & 0.04         & 0.05          & 0.002         & 47          & 59           & 86            & 4             \\
                     &                         & 0.07                   & 0.07        & 0.07         & 0.07          & 0.06          & 0.01        & 0.02         & 0.04          & 0.01          & 0.01        & 0.02         & 0.03          & 0.014         & 15          & 24           & 44            & 20            \\
                     &                         & 0.10                   & 0.07        & 0.07         & 0.07          & 0.06          & 0.03        & 0.05         & 0.04          & 0.04          & 0.03        & 0.04         & 0.04          & 0.04          & 28          & 42           & 40            & 41            \\ \cmidrule(l){2-19} 
                     & \multirow{5}{*}{Zr}     & 0.02                   & 0.05        & 0.04         & 0.07          & 0.01          & 0.07        & 0.03         & 0.06          & 0.02          & 0.05        & 0.03         & 0.05          & 0.02          & 240         & 117          & 226           & 80            \\
                     &                         & 0.10                   & 0.16        & 0.11         & 0.14          & 0.08          & 0.09        & 0.03         & 0.05          & 0.05          & 0.07        & 0.02         & 0.05          & 0.04          & 69          & 22           & 49            & 43            \\
                     &                         & 0.25                   & 0.22        & 0.23         & 0.26          & 0.25          & 0.06        & 0.05         & 0.04          & 0.06          & 0.04        & 0.04         & 0.03          & 0.04          & 17          & 17           & 14            & 16            \\
                     &                         & 0.28                   & 0.25        & 0.23         & 0.26          & 0.25          & 0.03        & 0.07         & 0.06          & 0.04          & 0.03        & 0.06         & 0.05          & 0.03          & 10          & 20           & 17            & 12            \\
                     &                         & 0.31                   & 0.27        & 0.33         & 0.36          & 0.30          & 0.05        & 0.11         & 0.12          & 0.03          & 0.04        & 0.10         & 0.10          & 0.02          & 14          & 32           & 32            & 6             \\ \midrule
\multirow{7}{*}{\rotatebox{90}{Cl}}  & \multirow{7}{*}{\rotatebox{90}{Cement}} & 0.23                   & 0.25        & 0.24         & 0.35          & 0.21          & 0.04        & 0.04         & 0.13          & 0.04          & 0.03        & 0.03         & 0.12          & 0.03          & 14          & 13           & 54            & 13            \\
                     &                         & 0.41                   & 0.45        & 0.45         & 0.56          & 0.41          & 0.08        & 0.09         & 0.17          & 0.08          & 0.07        & 0.07         & 0.16          & 0.06          & 16          & 17           & 38            & 16            \\
                     &                         & 0.50                   & 0.54        & 0.55         & 0.66          & 0.51          & 0.09        & 0.10         & 0.18          & 0.08          & 0.07        & 0.08         & 0.16          & 0.06          & 14          & 15           & 33            & 12            \\
                     &                         & 0.59                   & 0.60        & 0.61         & 0.72          & 0.58          & 0.07        & 0.07         & 0.15          & 0.07          & 0.06        & 0.06         & 0.13          & 0.05          & 9           & 10           & 23            & 9             \\
                     &                         & 0.87                   & 1.12        & 1.13         & 1.24          & 1.09          & 0.28        & 0.28         & 0.39          & 0.26          & 0.25        & 0.26         & 0.37          & 0.22          & 29          & 30           & 43            & 26            \\
                     &                         & 1.02                   & 1.01        & 1.03         & 1.14          & 0.98          & 0.07        & 0.07         & 0.14          & 0.08          & 0.06        & 0.06         & 0.13          & 0.06          & 6           & 6            & 13            & 6             \\
                     &                         & 1.51                   & 1.48        & 1.48         & 1.60          & 1.46          & 0.12        & 0.11         & 0.14          & 0.13          & 0.10        & 0.09         & 0.11          & 0.11          & 6           & 6            & 7             & 7             \\ \bottomrule
\end{tabular}%
}
\caption{
    \emph{Results of the analysis for traditional algorithms with synthetic data.}
    The size of the training set is fixed to \num{1000} spectra per sample for alloys, and \num{100} spectra per sample for cement matrices.
    Results of inference are relative to the independent test set.
    The columns show the Root Mean Squared Error (\rmse), the Mean Absolute Error (\mae), and the Mean Absolute Percentage Error (\mape).
    The algorithms refer to Linear Regression (\lr), Multivariate Linear Regression (\mlr), Partial Least Squares (\pls[1]), and Fully Connected Neural Networks (\fcnn).
    Bold characters represent the best estimates (*negative prediction, set to zero).
}
\label{tab:trad_an_synth}
\end{table}

\newgeometry{margin=2cm}

\begin{landscape}
\begin{table}[h]
\centering
\resizebox{\columnwidth}{!}{%
\begin{tabular}{@{}ccc|cccccc|cccccc|cccccc|cccccc@{}}
\toprule
                     &                         &                        & \multicolumn{6}{c|}{\textbf{Predicted [X] wt   \%}}                                      & \multicolumn{6}{c|}{\textbf{RMSE wt \%}}                                                             & \multicolumn{6}{c|}{\textbf{MAE wt \%}}                                                              & \multicolumn{6}{c}{\textbf{MAPE (\%)}}                                                       \\ \cmidrule(l){4-27} 
                     &                         &                        &               & \multicolumn{5}{c|}{\textbf{MT (training size)}}                             &                & \multicolumn{5}{c|}{\textbf{MT (training size)}}                                    &                 & \multicolumn{5}{c|}{\textbf{MT (training size)}}                                   &               & \multicolumn{5}{c}{\textbf{MT (training size)}}                              \\ \cmidrule(l){4-27} 
\textbf{X}           & \textbf{Matrix}         & \textbf{[X] wt \%} & \textbf{Best} & \textbf{Exp} & \textbf{100} & \textbf{1000} & \textbf{5000} & \textbf{10000} & \textbf{Best}  & \textbf{Exp}  & \textbf{100}  & \textbf{1000}   & \textbf{5000}   & \textbf{10000}  & \textbf{Best}   & \textbf{Exp}  & \textbf{100}  & \textbf{1000}  & \textbf{5000}   & \textbf{10000}  & \textbf{Best} & \textbf{Exp} & \textbf{100} & \textbf{1000} & \textbf{5000} & \textbf{10000} \\
\bottomrule
\end{tabular}%
}
\caption{%
    \emph{Complete analysis of the results of the \mt model as function of the size of the training set.}
    Results of inference concern the independent test set.
    The \mt model has been tested also on experimental data, independently of other models as separate baseline.
    Noise level is fixed at $\beta = 0.10$ for the alloy matrices, and $\beta = 0.03$ for the cement samples.
    Best results are selected from \Cref{tab:trad_an} according to the smallest \rmse per sample.
    The \mape has been used to discriminate possible equal values.
    Legend: \textsuperscript{1}\lr, \textsuperscript{2}\mlr, \textsuperscript{3}\pls[1], \textsuperscript{4}\fcnn.
}
\label{tab:size}
\end{table}
\end{landscape}

\restoregeometry

\newgeometry{margin=3.75cm}

\begin{landscape}
\begin{table}[h]
\centering
\resizebox{\columnwidth}{!}{%
\begin{tabular}{@{}ccc|cccccc|cccccc|cccccc|cccccc@{}}
\toprule
                     &                         &               & \multicolumn{6}{c|}{Predicted [X] wt   \%}   & \multicolumn{6}{c|}{RMSE wt \%}                   & \multicolumn{6}{c|}{MAE wt \%}                     & \multicolumn{6}{c}{MAPE (\%)}                \\ \cmidrule(l){4-27} 
                     &                         &               &         & \multicolumn{5}{c|}{MT (noise factor)} &       & \multicolumn{5}{c|}{MT (noise factor)}    &        & \multicolumn{5}{c|}{MT (noise factor)}    &      & \multicolumn{5}{c}{MT (noise factor)} \\ \cmidrule(l){4-27} 
X                    & Matrix                  & [X] wt \% & Best    & 0.00   & 0.01   & 0.03 & 0.05  & 0.10  & Best  & 0.00   & 0.01  & 0.03   & 0.05   & 0.10   & Best   & 0.00   & 0.01  & 0.03   & 0.05   & 0.10   & Best & 0.00  & 0.01  & 0.03  & 0.05  & 0.10  \\ \midrule
\multirow{40}{*}{\rotatebox{90}{Fe}} & \multirow{5}{*}{Al}     & 0.11          & 0.13\textsuperscript{4}  & 0.13   & 0.07   & 0.07 & 0.09  & 0.11  & 0.04  & 0.02   & 0.05  & 0.06   & 0.03   & 0.02   & 0.02   & 0.02   & 0.04  & 0.04   & 0.03   & 0.013  & 22   & 17    & 40    & 41    & 25    & 12    \\
                     &                         & 0.36          & 0.34\textsuperscript{2}  & 0.41   & 0.41   & 0.41 & 0.40  & 0.32  & 0.06  & 0.13   & 0.13  & 0.13   & 0.13   & 0.09   & 0.05   & 0.12   & 0.12  & 0.12   & 0.11   & 0.09   & 15   & 33    & 34    & 33    & 32    & 24    \\
                     &                         & 0.41          & 0.39\textsuperscript{4}  & 0.45   & 0.51   & 0.51 & 0.52  & 0.42  & 0.04  & 0.05   & 0.11  & 0.13   & 0.13   & 0.02   & 0.04   & 0.04   & 0.10  & 0.12   & 0.12   & 0.02   & 10   & 10    & 24    & 30    & 28    & 4     \\
                     &                         & 0.56          & 0.62\textsuperscript{4}  & 0.67   & 0.60   & 0.60 & 0.60  & 0.53  & 0.10  & 0.12   & 0.07  & 0.06   & 0.06   & 0.04   & 0.09   & 0.11   & 0.05  & 0.05   & 0.05   & 0.03   & 15   & 20    & 10    & 8     & 8     & 6     \\
                     &                         & 0.81          & 0.67\textsuperscript{2}  & 0.75   & 0.80   & 0.80 & 0.77  & 0.77  & 0.2   & 0.07   & 0.03  & 0.04   & 0.05   & 0.05   & 0.15   & 0.06   & 0.02  & 0.03   & 0.05   & 0.04   & 18   & 8     & 3     & 4     & 6     & 5     \\ \cmidrule(l){2-27} 
                     & \multirow{5}{*}{Co}     & 0.10          & 0.11\textsuperscript{4}  & 0.09   & 0.08   & 0.10 & 0.10  & 0.08  & 0.02  & 0.04   & 0.04  & 0.03   & 0.04   & 0.03   & 0.02   & 0.03   & 0.04  & 0.02   & 0.02   & 0.02   & 18   & 33    & 38    & 24    & 23    & 23    \\
                     &                         & 0.33          & 0.34\textsuperscript{4}  & 0.31   & 0.30   & 0.30 & 0.31  & 0.31  & 0.11  & 0.03   & 0.03  & 0.03   & 0.03   & 0.05   & 0.08   & 0.02   & 0.03  & 0.03   & 0.03   & 0.04   & 24   & 7     & 8     & 8     & 8     & 12    \\
                     &                         & 0.61          & 0.56\textsuperscript{4}  & 0.61   & 0.65   & 0.65 & 0.64  & 0.64  & 0.14  & 0.017  & 0.09  & 0.09   & 0.05   & 0.04   & 0.12   & 0.012  & 0.05  & 0.05   & 0.03   & 0.03   & 20   & 2     & 8     & 8     & 5     & 5     \\
                     &                         & 1.07          & 1.07\textsuperscript{1}  & 1.06   & 1.03   & 1.03 & 1.06  & 1.07  & 0.06  & 0.02   & 0.04  & 0.04   & 0.02   & 0.02   & 0.05   & 0.02   & 0.04  & 0.04   & 0.02   & 0.02   & 5    & 2     & 4     & 4     & 2     & 2     \\
                     &                         & 1.14          & 1.10\textsuperscript{4}  & 1.11   & 1.11   & 1.10 & 1.11  & 1.14  & 0.05  & 0.03   & 0.04  & 0.05   & 0.04   & 0.01   & 0.04   & 0.03   & 0.03  & 0.04   & 0.03   & 0.006  & 4    & 2     & 3     & 4     & 3     & 0.5   \\ \cmidrule(l){2-27} 
                     & \multirow{5}{*}{Cu}     & 0.02          & 0.03\textsuperscript{4}  & 0.04   & 0.03   & 0.03 & 0.04  & 0.02  & 0.02  & 0.02   & 0.015 & 0.02   & 0.02   & 0.012  & 0.014  & 0.02   & 0.013 & 0.019  & 0.02   & 0.010  & 85   & 141   & 83    & 116   & 135   & 63    \\
                     &                         & 0.05          & 0.05\textsuperscript{2}  & 0.06   & 0.04   & 0.05 & 0.07  & 0.05  & 0.013 & 0.02   & 0.010 & 0.012  & 0.024  & 0.009  & 0.010  & 0.02   & 0.009 & 0.009  & 0.02   & 0.007  & 22   & 42    & 21    & 21    & 50    & 16    \\
                     &                         & 0.08          & 0.10\textsuperscript{3}  & 0.10   & 0.08   & 0.11 & 0.10  & 0.11  & 0.03  & 0.02   & 0.011 & 0.03   & 0.03   & 0.03   & 0.02   & 0.02   & 0.010 & 0.03   & 0.02   & 0.03   & 27   & 25    & 12    & 31    & 28    & 31    \\
                     &                         & 0.10          & 0.11\textsuperscript{1}  & 0.14   & 0.11   & 0.14 & 0.16  & 0.13  & 0.04  & 0.05   & 0.02  & 0.04   & 0.06   & 0.03   & 0.03   & 0.04   & 0.014 & 0.04   & 0.06   & 0.03   & 27   & 42    & 14    & 42    & 63    & 32    \\
                     &                         & 0.17          & 0.17\textsuperscript{1}  & 0.16   & 0.15   & 0.18 & 0.18  & 0.17  & 0.05  & 0.015  & 0.022 & 0.009  & 0.019  & 0.011  & 0.04   & 0.012  & 0.017 & 0.007  & 0.016  & 0.010  & 25   & 7     & 10    & 4     & 9     & 6     \\ \cmidrule(l){2-27} 
                     & \multirow{4}{*}{Ni}     & 0.06          & 0.06\textsuperscript{4}  & 0.04   & 0.04   & 0.04 & 0.02  & 0.06  & 0.001 & 0.02   & 0.02  & 0.02   & 0.04   & 0.008  & 0.0009 & 0.02   & 0.019 & 0.018  & 0.04   & 0.007  & 1    & 30    & 31    & 30    & 59    & 12    \\
                     &                         & 0.25          & 0.26\textsuperscript{4}  & 0.26   & 0.25   & 0.25 & 0.26  & 0.24  & 0.02  & 0.03   & 0.02  & 0.02   & 0.03   & 0.03   & 0.017  & 0.03   & 0.02  & 0.019  & 0.02   & 0.02   & 7    & 12    & 8     & 8     & 9     & 8     \\
                     &                         & 0.34          & 0.32\textsuperscript{4}  & 0.33   & 0.30   & 0.30 & 0.31  & 0.33  & 0.03  & 0.02   & 0.04  & 0.05   & 0.04   & 0.02   & 0.02   & 0.02   & 0.04  & 0.04   & 0.03   & 0.02   & 7    & 6     & 12    & 13    & 10    & 5     \\
                     &                         & 0.97          & 0.95\textsuperscript{4}  & 0.98   & 0.91   & 0.93 & 0.93  & 0.94  & 0.04  & 0.05   & 0.10  & 0.07   & 0.08   & 0.05   & 0.02   & 0.05   & 0.07  & 0.05   & 0.06   & 0.04   & 2    & 5     & 7     & 6     & 6     & 4     \\ \cmidrule(l){2-27} 
                     & \multirow{6}{*}{Sn}     & 0.01          & 0.02\textsuperscript{3}  & 0*     & 0*     & 0*   & 0.002 & 0.002 & 0.008 & 0.02   & 0.06  & 0.02   & 0.012  & 0.012  & 0.006  & 0.02   & 0.05  & 0.018  & 0.011  & 0.011  & 45   & 139   & 403   & 140   & 81    & 85    \\
                     &                         & 0.03          & 0.05\textsuperscript{3}  & 0.01   & 0.01   & 0.01 & 0.01  & 0.02  & 0.03  & 0.02   & 0.02  & 0.02   & 0.015  & 0.011  & 0.02   & 0.014  & 0.02  & 0.014  & 0.014  & 0.009  & 74   & 51    & 80    & 49    & 51    & 33    \\
                     &                         & 0.06          & 0.06\textsuperscript{3}  & 0.03   & 0.02   & 0.04 & 0.05  & 0.05  & 0.009 & 0.03   & 0.04  & 0.019  & 0.012  & 0.009  & 0.008  & 0.03   & 0.04  & 0.018  & 0.009  & 0.007  & 13   & 45    & 66    & 32    & 16    & 12    \\
                     &                         & 0.09          & 0.08\textsuperscript{3}  & 0.03   & 0.02   & 0.04 & 0.06  & 0.07  & 0.02  & 0.05   & 0.07  & 0.04   & 0.03   & 0.02   & 0.017  & 0.05   & 0.06  & 0.04   & 0.02   & 0.019  & 19   & 62    & 74    & 51    & 28    & 22    \\
                     &                         & 0.17          & 0.14\textsuperscript{1}  & 0.08   & 0.07   & 0.09 & 0.07  & 0.13  & 0.05  & 0.09   & 0.10  & 0.09   & 0.10   & 0.05   & 0.03   & 0.09   & 0.10  & 0.08   & 0.10   & 0.04   & 19   & 53    & 57    & 48    & 56    & 24    \\
                     &                         & 0.18          & 0.14\textsuperscript{4}  & 0.09   & 0.07   & 0.11 & 0.07  & 0.14  & 0.05  & 0.10   & 0.11  & 0.09   & 0.12   & 0.05   & 0.04   & 0.09   & 0.11  & 0.07   & 0.11   & 0.04   & 22   & 50    & 59    & 39    & 63    & 22    \\ \cmidrule(l){2-27} 
                     & \multirow{5}{*}{Ti}     & 0.01          & 0.02\textsuperscript{4}  & 0.04   & 0.02   & 0.03 & 0.03  & 0.05  & 0.03  & 0.03   & 0.02  & 0.03   & 0.03   & 0.04   & 0.014  & 0.03   & 0.016 & 0.02   & 0.02   & 0.04   & 154  & 322   & 172   & 254   & 213   & 421   \\
                     &                         & 0.06          & 0.07\textsuperscript{4}  & 0.11   & 0.10   & 0.10 & 0.08  & 0.10  & 0.03  & 0.12   & 0.11  & 0.10   & 0.06   & 0.06   & 0.03   & 0.07   & 0.07  & 0.07   & 0.05   & 0.05   & 56   & 119   & 120   & 123   & 86    & 92    \\
                     &                         & 0.11          & 0.09\textsuperscript{4}  & 0.13   & 0.13   & 0.15 & 0.12  & 0.15  & 0.04  & 0.12   & 0.14  & 0.13   & 0.07   & 0.05   & 0.03   & 0.08   & 0.09  & 0.07   & 0.05   & 0.05   & 24   & 68    & 75    & 65    & 47    & 47    \\
                     &                         & 0.22          & 0.19\textsuperscript{4}  & 0.31   & 0.28   & 0.30 & 0.27  & 0.24  & 0.06  & 0.3    & 0.24  & 0.21   & 0.10   & 0.04   & 0.05   & 0.3    & 0.2   & 0.19   & 0.09   & 0.04   & 22   & 115   & 98    & 86    & 43    & 20    \\
                     &                         & 1.19          & 1.19\textsuperscript{4}  & 1.16   & 1.14   & 1.15 & 1.15  & 1.15  & 0.007 & 0.08   & 0.07  & 0.07   & 0.06   & 0.05   & 0.005  & 0.05   & 0.05  & 0.04   & 0.04   & 0.04   & 0.4  & 4     & 4     & 4     & 3     & 4     \\ \cmidrule(l){2-27} 
                     & \multirow{5}{*}{Zn}     & 0.02          & 0.05\textsuperscript{3}  & 0.03   & 0.03   & 0.02 & 0.04  & 0.02  & 0.03  & 0.013  & 0.015 & 0.008  & 0.019  & 0.012  & 0.03   & 0.010  & 0.015 & 0.007  & 0.018  & 0.009  & 156  & 54    & 82    & 39    & 100   & 50    \\
                     &                         & 0.05          & 0.05\textsuperscript{1}  & 0.04   & 0.04   & 0.03 & 0.04  & 0.04  & 0.005 & 0.010  & 0.010 & 0.014  & 0.009  & 0.012  & 0.004  & 0.007  & 0.010 & 0.013  & 0.008  & 0.008  & 9    & 15    & 21    & 29    & 18    & 18    \\
                     &                         & 0.06          & 0.06\textsuperscript{4}  & 0.06   & 0.05   & 0.06 & 0.06  & 0.06  & 0.004 & 0.003  & 0.013 & 0.030  & 0.011  & 0.008  & 0.004  & 0.002  & 0.011 & 0.02   & 0.011  & 0.007  & 7    & 3     & 18    & 41    & 17    & 11    \\
                     &                         & 0.07          & 0.06\textsuperscript{1}  & 0.07   & 0.07   & 0.07 & 0.08  & 0.07  & 0.014 & 0.010  & 0.008 & 0.011  & 0.008  & 0.012  & 0.012  & 0.006  & 0.006 & 0.008  & 0.006  & 0.008  & 17   & 8     & 9     & 11    & 8     & 11    \\
                     &                         & 0.10          & 0.06\textsuperscript{2}  & 0.09   & 0.08   & 0.10 & 0.09  & 0.09  & 0.04  & 0.006  & 0.02  & 0.02   & 0.012  & 0.008  & 0.04   & 0.006  & 0.02  & 0.02   & 0.010  & 0.007  & 40   & 6     & 20    & 20    & 10    & 7     \\ \cmidrule(l){2-27} 
                     & \multirow{5}{*}{Zr}     & 0.02          & 0.08\textsuperscript{4}  & 0.03   & 0.02   & 0.03 & 0.03  & 0.03  & 0.09  & 0.012  & 0.002 & 0.008  & 0.009  & 0.010  & 0.06   & 0.009  & 0.002 & 0.007  & 0.006  & 0.006  & 250  & 39    & 9     & 29    & 24    & 28    \\
                     &                         & 0.10          & 0.13\textsuperscript{2}  & 0.10   & 0.10   & 0.09 & 0.10  & 0.10  & 0.04  & 0.0011 & 0.003 & 0.007  & 0.0007 & 0.003  & 0.03   & 0.0009 & 0.002 & 0.007  & 0.0007 & 0.003  & 35   & 1.0   & 2     & 7     & 0.7   & 3     \\
                     &                         & 0.25          & 0.23\textsuperscript{1}  & 0.26   & 0.25   & 0.26 & 0.25  & 0.26  & 0.06  & 0.005  & 0.002 & 0.004  & 0.002  & 0.003  & 0.04   & 0.004  & 0.002 & 0.003  & 0.0018 & 0.002  & 15   & 2     & 0.8   & 1.4   & 0.7   & 1.0   \\
                     &                         & 0.28          & 0.26\textsuperscript{1}  & 0.29   & 0.28   & 0.28 & 0.28  & 0.28  & 0.03  & 0.002  & 0.002 & 0.0009 & 0.0012 & 0.0010 & 0.03   & 0.002  & 0.002 & 0.0007 & 0.0010 & 0.0010 & 10   & 0.8   & 0.7   & 0.3   & 0.4   & 0.3   \\
                     &                         & 0.31          & 0.31\textsuperscript{4}  & 0.31   & 0.31   & 0.31 & 0.31  & 0.31  & 0.009 & 0.003  & 0.004 & 0.003  & 0.002  & 0.0013 & 0.007  & 0.002  & 0.004 & 0.003  & 0.0015 & 0.0011 & 2    & 0.8   & 1.1   & 0.9   & 0.5   & 0.4   \\ \midrule
\multirow{7}{*}{\rotatebox{90}{Cl}}  & \multirow{7}{*}{\rotatebox{90}{Cement}} & 0.23          & 0.22\textsuperscript{4}  & 0.29   & 0.27   & 0.27 & 0.26  & 0.27  & 0.03  & 0.10   & 0.07  & 0.08   & 0.08   & 0.08   & 0.03   & 0.07   & 0.05  & 0.06   & 0.06   & 0.06   & 11   & 31    & 23    & 26    & 27    & 26    \\
                     &                         & 0.41          & 0.44\textsuperscript{1}  & 0.56   & 0.52   & 0.54 & 0.49  & 0.42  & 0.08  & 0.18   & 0.17  & 0.14   & 0.09   & 0.06   & 0.07   & 0.15   & 0.13  & 0.13   & 0.08   & 0.05   & 17   & 36    & 31    & 32    & 18    & 12    \\
                     &                         & 0.50          & 0.53\textsuperscript{4}  & 0.52   & 0.54   & 0.54 & 0.52  & 0.52  & 0.09  & 0.08   & 0.12  & 0.08   & 0.08   & 0.09   & 0.07   & 0.07   & 0.08  & 0.06   & 0.06   & 0.07   & 14   & 13    & 16    & 12    & 12    & 14    \\
                     &                         & 0.59          & 0.60\textsuperscript{4}  & 0.40   & 0.49   & 0.41 & 0.41  & 0.44  & 0.09  & 0.2    & 0.14  & 0.19   & 0.2    & 0.17   & 0.07   & 0.2    & 0.12  & 0.2    & 0.2    & 0.2    & 14   & 33    & 21    & 31    & 31    & 26    \\
                     &                         & 0.87          & 1.13\textsuperscript{1}  & 1.35   & 1.29   & 1.20 & 1.25  & 1.31  & 0.3   & 0.5    & 0.5   & 0.3    & 0.4    & 0.4    & 0.3    & 0.5    & 0.4   & 0.3    & 0.4    & 0.4    & 30   & 55    & 48    & 38    & 43    & 50    \\
                     &                         & 1.02          & 1.02\textsuperscript{1}  & 1.04   & 1.02   & 0.99 & 0.99  & 1.01  & 0.08  & 0.09   & 0.07  & 0.07   & 0.09   & 0.09   & 0.07   & 0.08   & 0.07  & 0.06   & 0.07   & 0.08   & 7    & 8     & 6     & 6     & 7     & 8     \\
                     &                         & 1.51          & 1.49\textsuperscript{1}  & 1.55   & 1.57   & 1.58 & 1.57  & 1.60  & 0.10  & 0.14   & 0.18  & 0.12   & 0.10   & 0.12   & 0.08   & 0.10   & 0.12  & 0.09   & 0.08   & 0.11   & 5    & 7     & 8     & 6     & 6     & 7     \\ \bottomrule
\end{tabular}%
}
\caption{%
    \emph{Complete analysis of the results of the \mt model as function of the noise parameter.}
    Results of inference concern the independent test set.
    The number of synthetic samples is fixed to \num{1000}, for illustration purposes.
    Best results are selected from \Cref{tab:trad_an} according to the smallest \rmse per sample.
    The \mape has been used to discriminate possible equal values.
    Legend: \textsuperscript{1}\lr, \textsuperscript{2}\mlr, \textsuperscript{3}\pls[1], \textsuperscript{4}\fcnn (*negative prediction, set to zero).
}
\label{tab:mt_noise}
\end{table}
\end{landscape}

\restoregeometry

\newgeometry{margin=1.5cm}

\begin{landscape}
\begin{table}[h]
\centering
\resizebox{\columnwidth}{!}{%
\begin{tabular}{@{}cc|ccc|cccccccc|cccccccc@{}}
\toprule
                    &                         &                    &               &                   & \multicolumn{8}{c|}{\textbf{MAPE (\%)}}                                                                                                                               & \multicolumn{8}{c}{\textbf{t-value}}                                                                                                                                  \\ \cmidrule(l){6-21} 
\textbf{Matrix}     & \textbf{[Fe] wt \%} & \multicolumn{3}{c|}{\textbf{Predicted [Fe] wt \%}} & \textbf{358.12 nm} & \textbf{373.49 nm} & \textbf{373.71 nm} & \textbf{374.56 nm} & \textbf{382.04 nm} & \textbf{385.99 nm} & \textbf{404.58 nm} & \textbf{438.35 nm} & \textbf{358.12 nm} & \textbf{373.49 nm} & \textbf{373.71 nm} & \textbf{374.56 nm} & \textbf{382.04 nm} & \textbf{385.99 nm} & \textbf{404.58 nm} & \textbf{438.35 nm} \\ \midrule
\multirow{5}{*}{Al} & 0.11                    & 0.11               & \textpm             & 0.02              & 76                 & 45                 & 65                 & 28                 & 29                 & 28                 & 56                 & 66                 & 0.97               & 1.45               & 1.49               & 0.93               & 1.08               & 0.65               & 1.84               & 1.59               \\
                    & 0.36                    & 0.32               & \textpm             & 0.09              & 32                 & 29                 & 61                 & 29                 & 25                 & 24                 & 15                 & 35                 & 0.88               & 0.86               & 1.10               & 0.98               & 0.70               & 0.61               & 0.66               & 0.84               \\
                    & 0.41                    & 0.42               & \textpm             & 0.02              & 7                  & 12                 & 40                 & 10                 & 13                 & 7                  & 19                 & 17                 & 0.70               & 0.95               & 1.22               & 0.66               & 1.14               & 0.57               & 0.62               & 1.04               \\
                    & 0.56                    & 0.53               & \textpm             & 0.04              & 8                  & 12                 & 40                 & 10                 & 18                 & 4                  & 16                 & 7                  & 0.66               & 0.84               & 1.56               & 0.81               & 1.04               & 0.81               & 0.77               & 0.90               \\
                    & 0.81                    & 0.77               & \textpm             & 0.05              & 10                 & 11                 & 21                 & 15                 & 6                  & 8                  & 10                 & 13                 & 0.49               & 0.46               & 0.60               & 0.78               & 0.44               & 0.47               & 0.62               & 0.48               \\ \midrule
\multirow{5}{*}{Co} & 0.10                    & 0.08               & \textpm             & 0.03              & 32                 & 10                 & 16                 & 9                  & 20                 & 20                 & 8                  & 12                 & 0.91               & 0.81               & 0.69               & 0.91               & 1.02               & 0.85               & 0.69               & 0.51               \\
                    & 0.33                    & 0.31               & \textpm             & 0.05              & 16                 & 8                  & 18                 & 20                 & 17                 & 14                 & 23                 & 15                 & 0.70               & 0.56               & 0.65               & 0.65               & 0.71               & 0.44               & 0.98               & 0.50               \\
                    & 0.61                    & 0.64               & \textpm             & 0.04              & 22                 & 13                 & 12                 & 7                  & 9                  & 14                 & 11                 & 19                 & 1.13               & 1.36               & 0.84               & 0.63               & 0.50               & 0.75               & 0.83               & 1.17               \\
                    & 1.07                    & 1.07               & \textpm             & 0.02              & 11                 & 10                 & 12                 & 11                 & 13                 & 18                 & 13                 & 10                 & 1.02               & 0.82               & 0.81               & 1.00               & 0.83               & 1.00               & 1.14               & 0.67               \\
                    & 1.14                    & 1.140              & \textpm             & 0.008             & 10                 & 8                  & 13                 & 7                  & 7                  & 10                 & 12                 & 15                 & 0.88               & 0.83               & 0.97               & 0.47               & 0.57               & 1.02               & 0.74               & 1.04               \\ \midrule
\multirow{5}{*}{Cu} & 0.02                    & 0.02               & \textpm             & 0.01              & 28                 & 109                & 79                 & 42                 & 74                 & 18                 & 12                 & 36                 & 0.70               & 1.68               & 0.90               & 0.89               & 1.09               & 0.59               & 0.49               & 0.93               \\
                    & 0.05                    & 0.045              & \textpm             & 0.009             & 37                 & 49                 & 138                & 38                 & 35                 & 22                 & 18                 & 24                 & 0.68               & 0.66               & 1.12               & 1.07               & 0.64               & 0.90               & 0.46               & 0.56               \\
                    & 0.08                    & 0.11               & \textpm             & 0.03              & 59                 & 30                 & 62                 & 40                 & 28                 & 39                 & 27                 & 31                 & 0.88               & 0.68               & 0.86               & 0.79               & 0.66               & 1.69               & 0.67               & 0.63               \\
                    & 0.10                    & 0.13               & \textpm             & 0.03              & 44                 & 49                 & 62                 & 28                 & 31                 & 20                 & 32                 & 31                 & 1.56               & 0.88               & 1.88               & 0.81               & 0.96               & 0.71               & 1.07               & 0.91               \\
                    & 0.17                    & 0.17               & \textpm             & 0.01              & 34                 & 28                 & 40                 & 22                 & 32                 & 20                 & 25                 & 26                 & 1.16               & 1.01               & 0.67               & 0.68               & 0.80               & 0.92               & 0.82               & 0.78               \\ \midrule
\multirow{4}{*}{Ni} & 0.06                    & 0.059              & \textpm             & 0.008             & 24                 & 15                 & 8                  & 12                 & 28                 & 24                 & 16                 & 27                 & 1.82               & 0.74               & 0.69               & 0.76               & 0.84               & 1.36               & 0.74               & 0.86               \\
                    & 0.25                    & 0.24               & \textpm             & 0.03              & 12                 & 12                 & 9                  & 9                  & 15                 & 14                 & 17                 & 19                 & 0.86               & 0.79               & 0.81               & 0.62               & 1.02               & 1.40               & 0.96               & 1.06               \\
                    & 0.34                    & 0.33               & \textpm             & 0.02              & 10                 & 12                 & 11                 & 9                  & 12                 & 14                 & 16                 & 14                 & 0.88               & 0.89               & 0.72               & 0.82               & 0.91               & 0.83               & 1.13               & 0.72               \\
                    & 0.97                    & 0.94               & \textpm             & 0.05              & 13                 & 16                 & 8                  & 14                 & 10                 & 8                  & 7                  & 22                 & 0.67               & 0.68               & 0.58               & 0.72               & 0.69               & 0.42               & 0.48               & 0.59               \\ \midrule
\multirow{6}{*}{Sn} & 0.01                    & 0.00               & \textpm             & 0.01              & 91                 & 269                & 58                 & 789                & 106                & 544                & 385                & 42                 & 1.05               & 1.29               & 0.61               & 0.70               & 0.83               & 1.33               & 0.75               & 0.93               \\
                    & 0.03                    & 0.02               & \textpm             & 0.01              & 40                 & 24                 & 30                 & 44                 & 14                 & 69                 & 38                 & 29                 & 0.65               & 0.57               & 0.60               & 0.85               & 0.59               & 0.75               & 0.58               & 0.78               \\
                    & 0.06                    & 0.052              & \textpm             & 0.009             & 50                 & 64                 & 32                 & 401                & 28                 & 30                 & 155                & 89                 & 0.92               & 0.79               & 0.89               & 0.63               & 0.36               & 0.38               & 0.93               & 0.58               \\
                    & 0.09                    & 0.07               & \textpm             & 0.02              & 25                 & 16                 & 36                 & 84                 & 18                 & 41                 & 29                 & 54                 & 0.83               & 0.50               & 0.53               & 0.84               & 0.63               & 0.63               & 0.71               & 0.75               \\
                    & 0.17                    & 0.13               & \textpm             & 0.05              & 24                 & 23                 & 31                 & 32                 & 61                 & 24                 & 24                 & 22                 & 0.52               & 0.73               & 0.60               & 0.74               & 0.53               & 0.49               & 0.44               & 0.48               \\
                    & 0.18                    & 0.14               & \textpm             & 0.05              & 23                 & 16                 & 43                 & 40                 & 19                 & 31                 & 22                 & 34                 & 0.73               & 0.50               & 0.58               & 0.52               & 0.43               & 0.66               & 0.43               & 0.61               \\ \midrule
\multirow{5}{*}{Ti} & 0.01                    & 0.05               & \textpm             & 0.04              & 13                 & 15                 & 19                 & 10                 & 12                 & 5                  & 13                 & 11                 & 0.70               & 0.69               & 1.05               & 1.26               & 0.73               & 0.60               & 0.84               & 1.84               \\
                    & 0.06                    & 0.10               & \textpm             & 0.06              & 11                 & 19                 & 7                  & 11                 & 9                  & 6                  & 7                  & 11                 & 0.74               & 0.73               & 0.51               & 0.86               & 0.59               & 0.68               & 0.61               & 0.58               \\
                    & 0.11                    & 0.15               & \textpm             & 0.05              & 9                  & 9                  & 8                  & 10                 & 7                  & 5                  & 6                  & 11                 & 1.01               & 0.55               & 0.73               & 0.72               & 0.62               & 0.51               & 0.51               & 0.69               \\
                    & 0.22                    & 0.24               & \textpm             & 0.04              & 5                  & 8                  & 6                  & 8                  & 12                 & 4                  & 9                  & 10                 & 0.35               & 0.62               & 0.53               & 0.58               & 0.72               & 0.40               & 0.48               & 0.66               \\
                    & 1.19                    & 1.15               & \textpm             & 0.05              & 8                  & 12                 & 7                  & 7                  & 7                  & 9                  & 8                  & 4                  & 0.87               & 0.73               & 0.72               & 0.78               & 0.97               & 0.85               & 0.78               & 0.45               \\ \midrule
\multirow{5}{*}{Zn} & 0.02                    & 0.02               & \textpm             & 0.01              & 123                & 361                & 174                & 42                 & 135                & 65                 & 338                & 155                & 1.33               & 1.54               & 1.06               & 0.90               & 1.49               & 0.99               & 1.57               & 1.86               \\
                    & 0.05                    & 0.04               & \textpm             & 0.01              & 75                 & 127                & 136                & 80                 & 95                 & 70                 & 125                & 128                & 0.56               & 1.03               & 0.96               & 1.13               & 0.96               & 0.80               & 1.18               & 0.96               \\
                    & 0.06                    & 0.059              & \textpm             & 0.008             & 119                & 219                & 123                & 45                 & 159                & 110                & 301                & 76                 & 0.52               & 0.55               & 0.52               & 0.52               & 0.56               & 0.55               & 0.55               & 0.52               \\
                    & 0.07                    & 0.07               & \textpm             & 0.01              & 93                 & 67                 & 110                & 76                 & 48                 & 118                & 75                 & 79                 & 0.59               & 0.59               & 0.58               & 0.68               & 0.59               & 0.62               & 0.56               & 0.54               \\
                    & 0.10                    & 0.094              & \textpm             & 0.008             & 281                & 292                & 191                & 173                & 131                & 176                & 194                & 111                & 0.71               & 0.86               & 0.74               & 0.75               & 0.73               & 0.77               & 0.81               & 0.59               \\ \midrule
\multirow{5}{*}{Zr} & 0.02                    & 0.028              & \textpm             & 0.010             & 8                  & 6                  & 13                 & 7                  & 5                  & 9                  & 5                  & 7                  & 0.79               & 0.80               & 0.76               & 0.70               & 0.61               & 0.84               & 0.87               & 0.75               \\
                    & 0.10                    & 0.099              & \textpm             & 0.003             & 9                  & 7                  & 6                  & 5                  & 8                  & 12                 & 5                  & 7                  & 0.65               & 0.50               & 0.63               & 0.51               & 0.62               & 0.70               & 0.48               & 0.53               \\
                    & 0.25                    & 0.255              & \textpm             & 0.003             & 9                  & 9                  & 7                  & 8                  & 7                  & 10                 & 7                  & 5                  & 0.70               & 0.63               & 0.68               & 0.61               & 0.77               & 0.75               & 0.55               & 0.57               \\
                    & 0.28                    & 0.282              & \textpm             & 0.001             & 5                  & 11                 & 6                  & 8                  & 9                  & 6                  & 5                  & 8                  & 0.48               & 0.68               & 0.75               & 1.06               & 1.48               & 0.56               & 0.63               & 1.53               \\
                    & 0.31                    & 0.3092             & \textpm             & 0.0013            & 10                 & 8                  & 6                  & 4                  & 10                 & 10                 & 4                  & 8                  & 0.86               & 0.51               & 0.61               & 0.45               & 0.79               & 0.74               & 0.44               & 0.79               \\ \bottomrule
\end{tabular}%
}
\caption{%
    \emph{Results of the validation analysis on the alloy matrices.}
    Results concern the independent test set.
    The number of synthetic samples is fixed to \num{1000} and the noise factor to $\beta = 0.10$.
}
\label{tab:mt_valid}
\end{table}
\end{landscape}

\restoregeometry

\clearpage

\section{Complementary Figures}\label{app:figures}

\begin{figure}[h]
    \centering
    \begin{tabular}{@{}c|c@{}}
        {\LARGE \textsc{Global Effects}}                                          & {\LARGE \textsc{Local Coverage}}
        \\
        \textsc{(whole spectra)}                                                  & \textsc{(Fe line at \SI{373.48}{\nano\meter})}
        \\
        \includegraphics[width=0.48\linewidth]{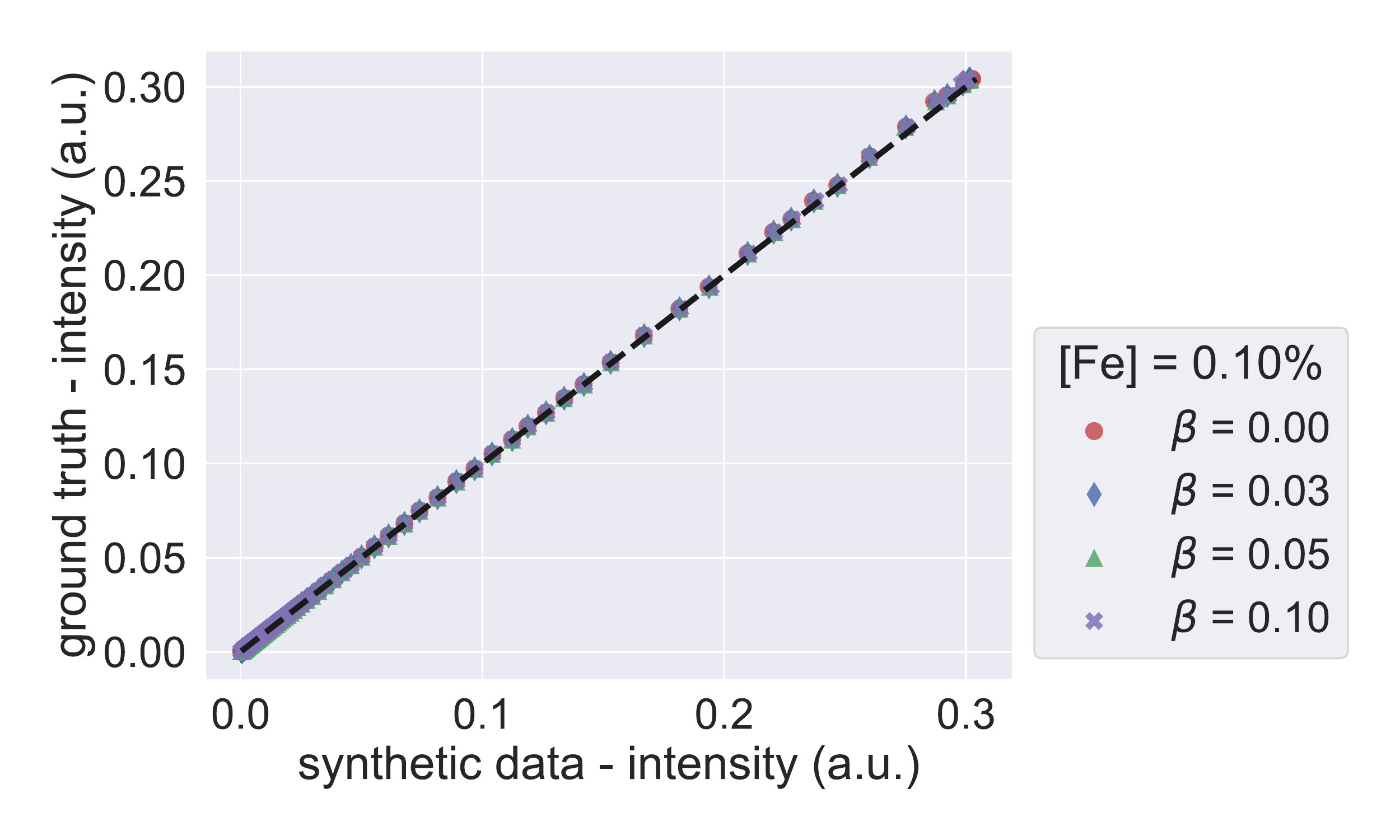}  & \includegraphics[width=0.48\linewidth]{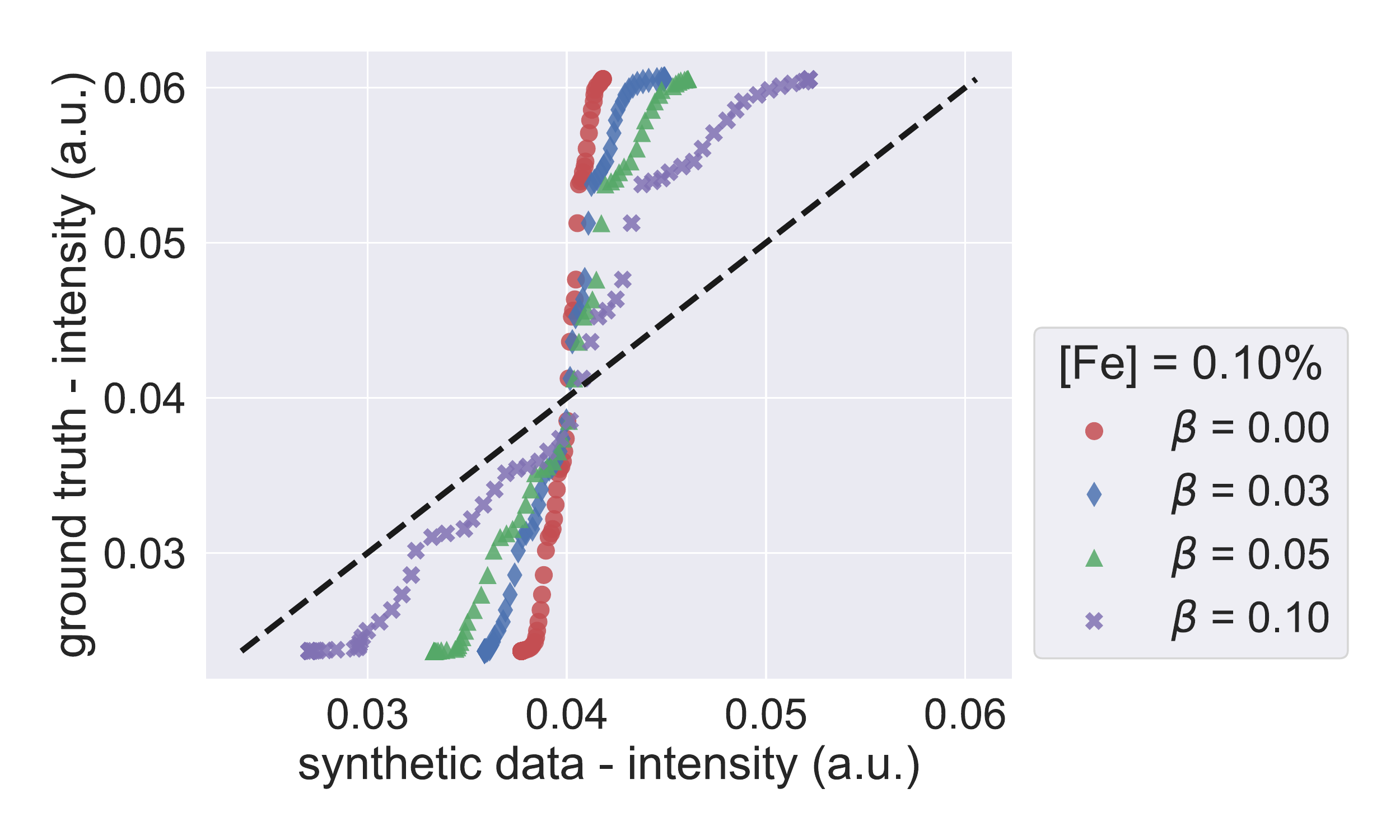}
        \\ \midrule
        \includegraphics[width=0.48\linewidth]{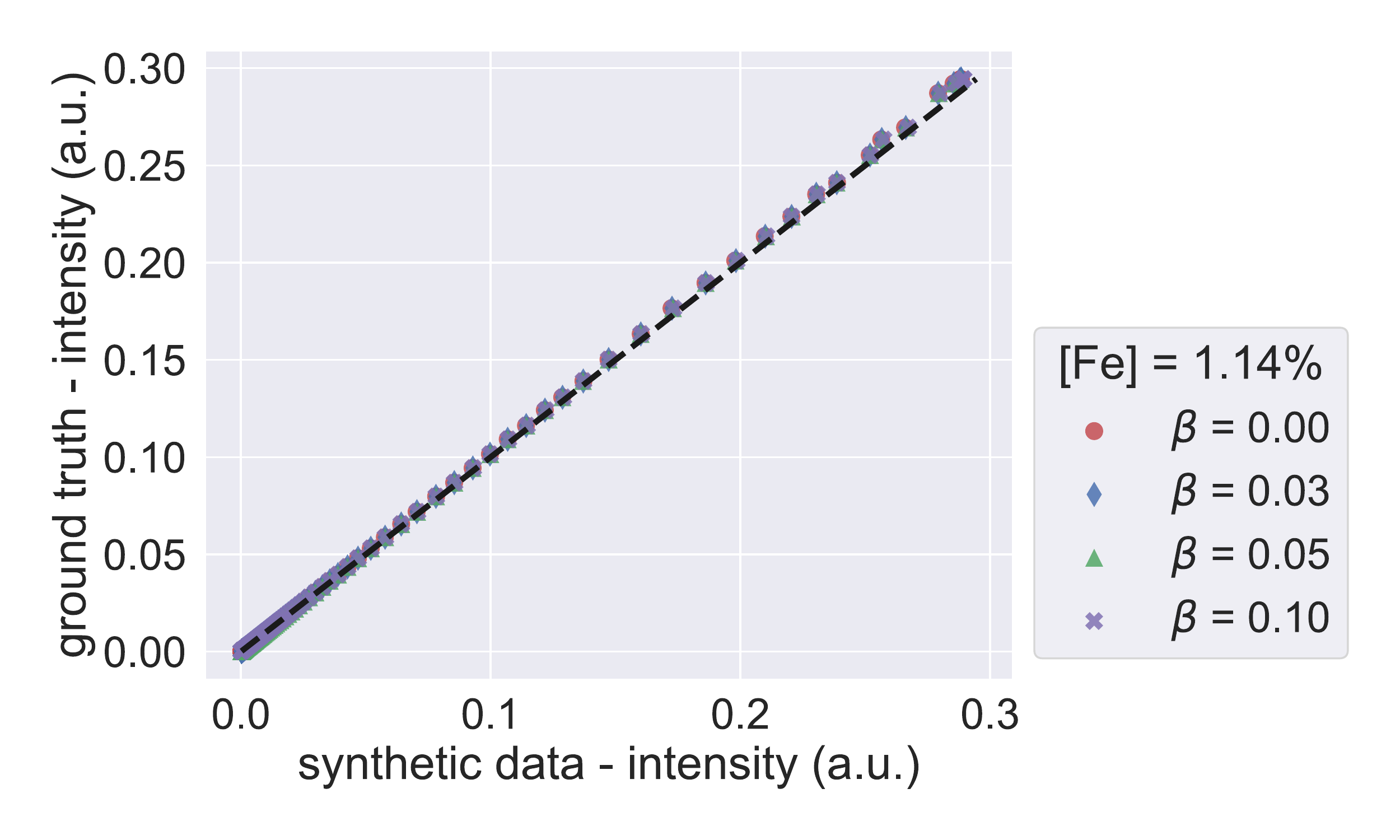} & \includegraphics[width=0.48\linewidth]{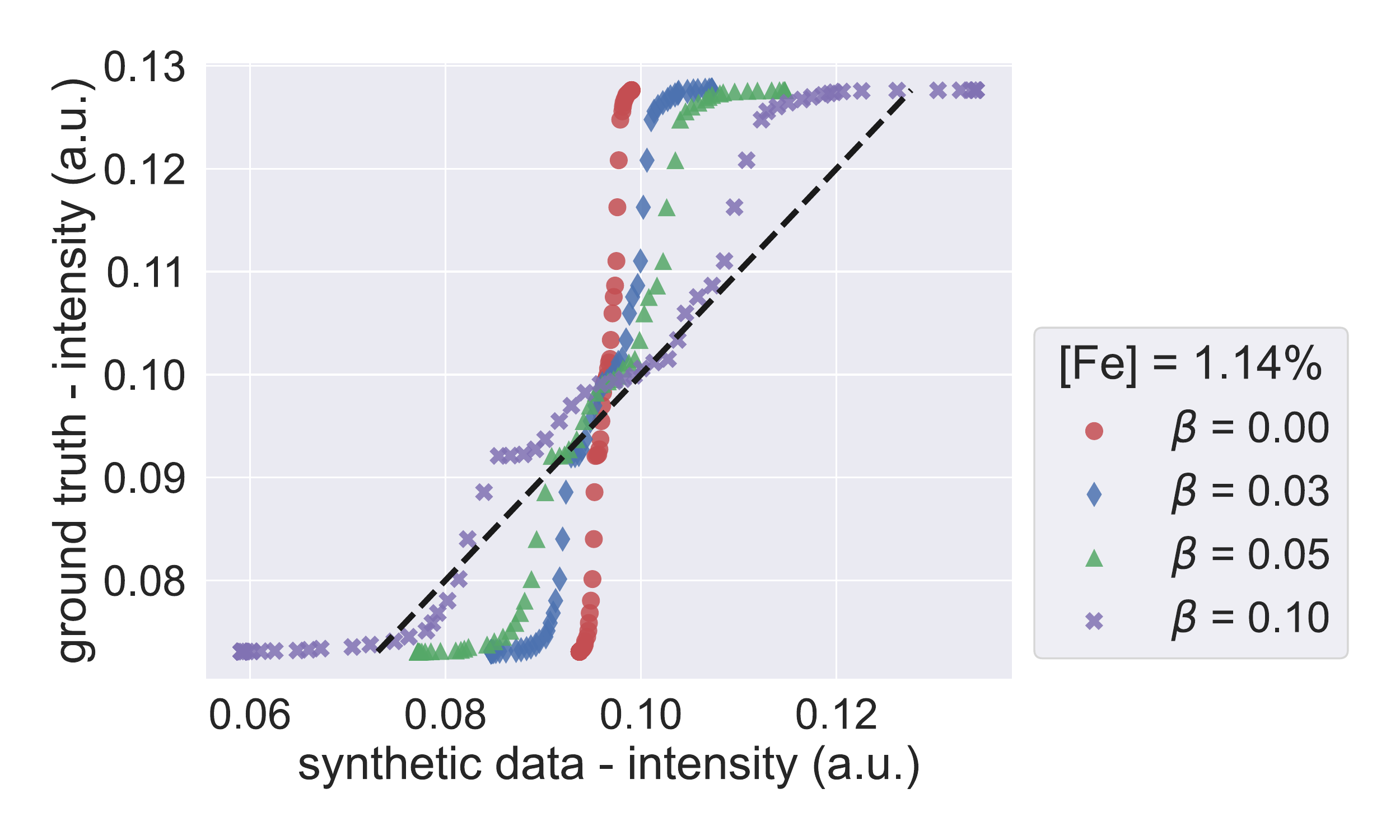}
    \end{tabular}
    \caption{%
        Global (average spectrum) and local (Fe line at \SI{373.48}{\nano\meter}) coverage of the synthetic Co matrix dataset (\num{1000} synthetic spectra per sample) at low and high Fe concentrations.
    }
    \label{fig:co_coverage}
\end{figure}

\begin{figure}[h]
    \centering
    \begin{tabular}{@{}c|c@{}}
        {\LARGE \textsc{Global Effects}}                                          & {\LARGE \textsc{Local Coverage}}
        \\
        \textsc{(whole spectra)}                                                  & \textsc{(Fe line at \SI{373.48}{\nano\meter})}
        \\
        \includegraphics[width=0.48\linewidth]{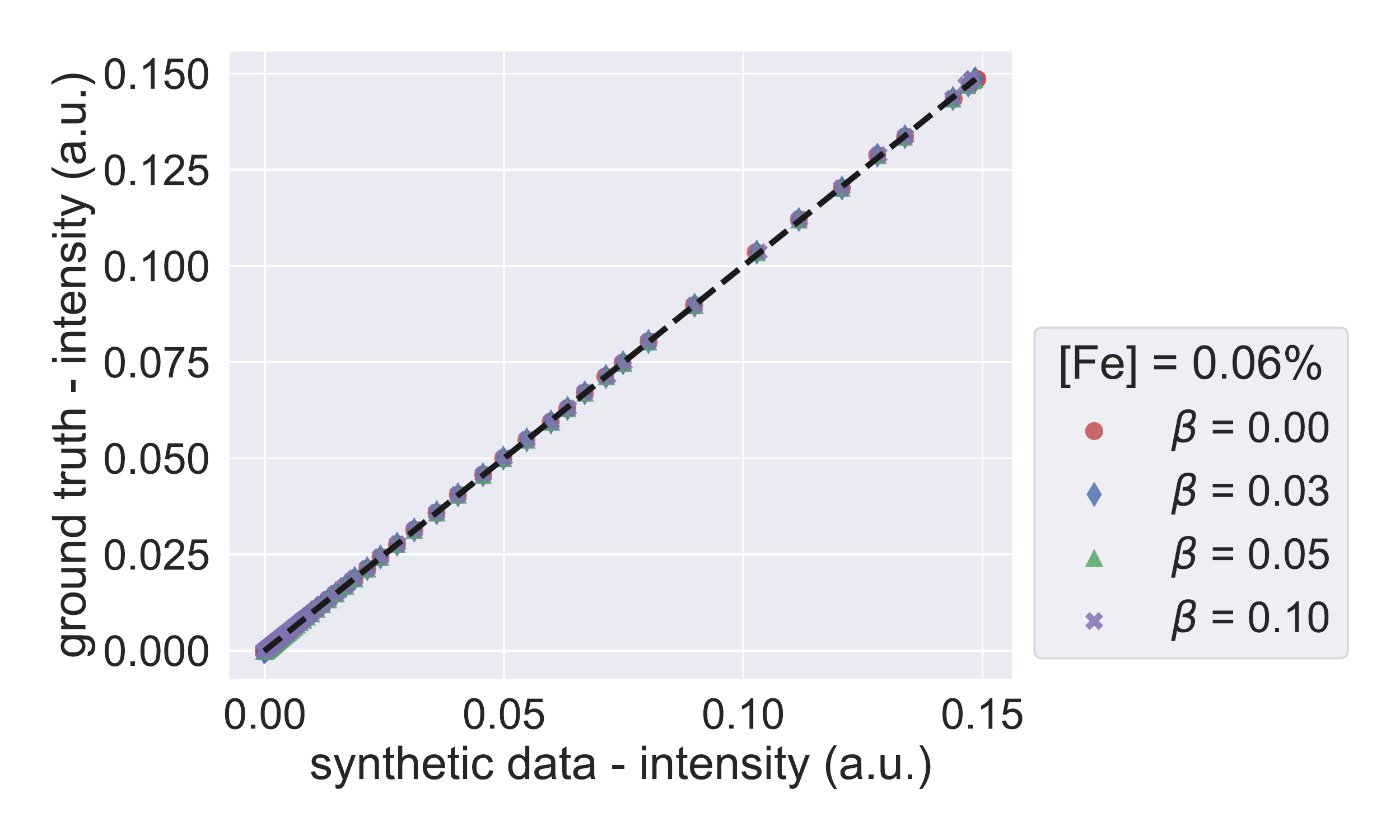}  & \includegraphics[width=0.48\linewidth]{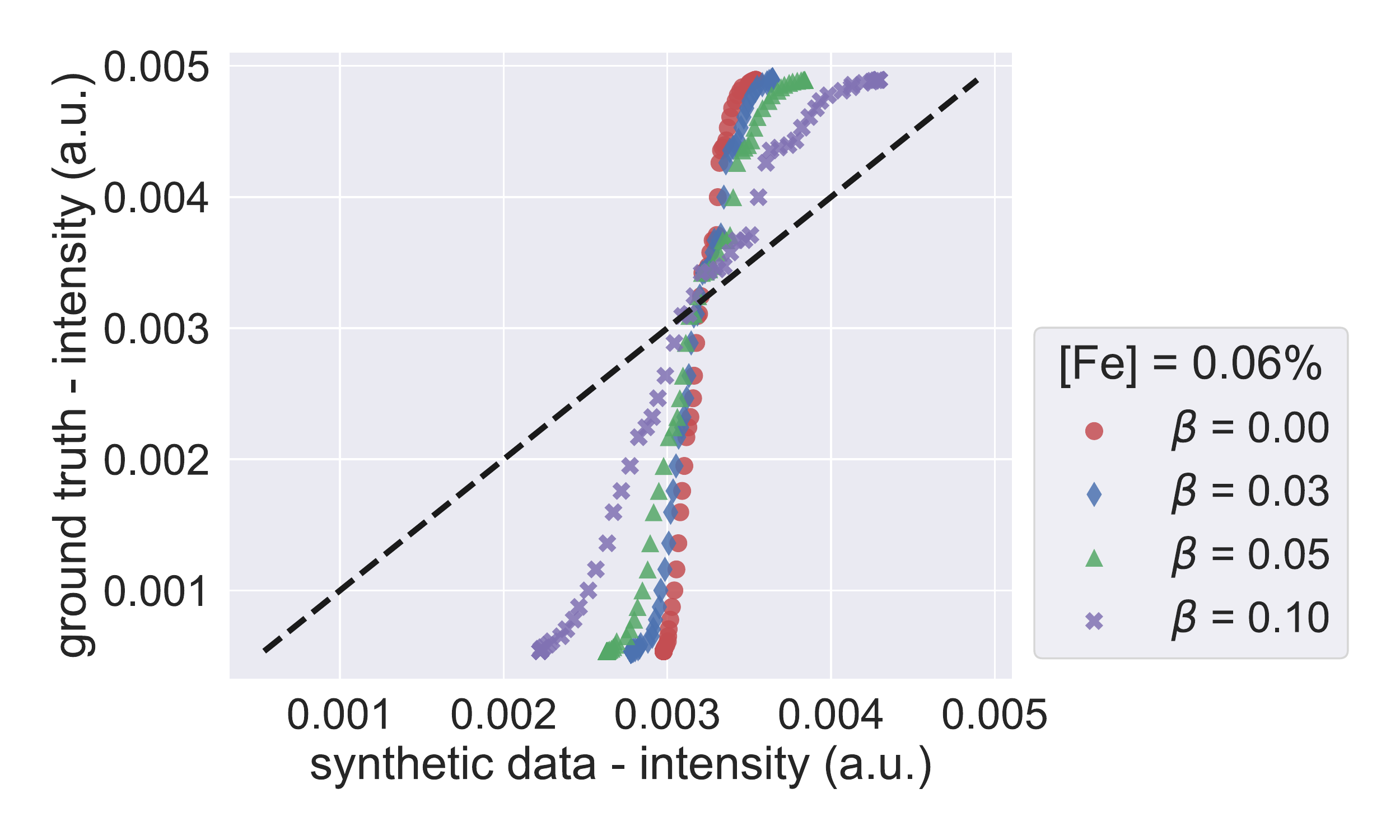}
        \\ \midrule
        \includegraphics[width=0.48\linewidth]{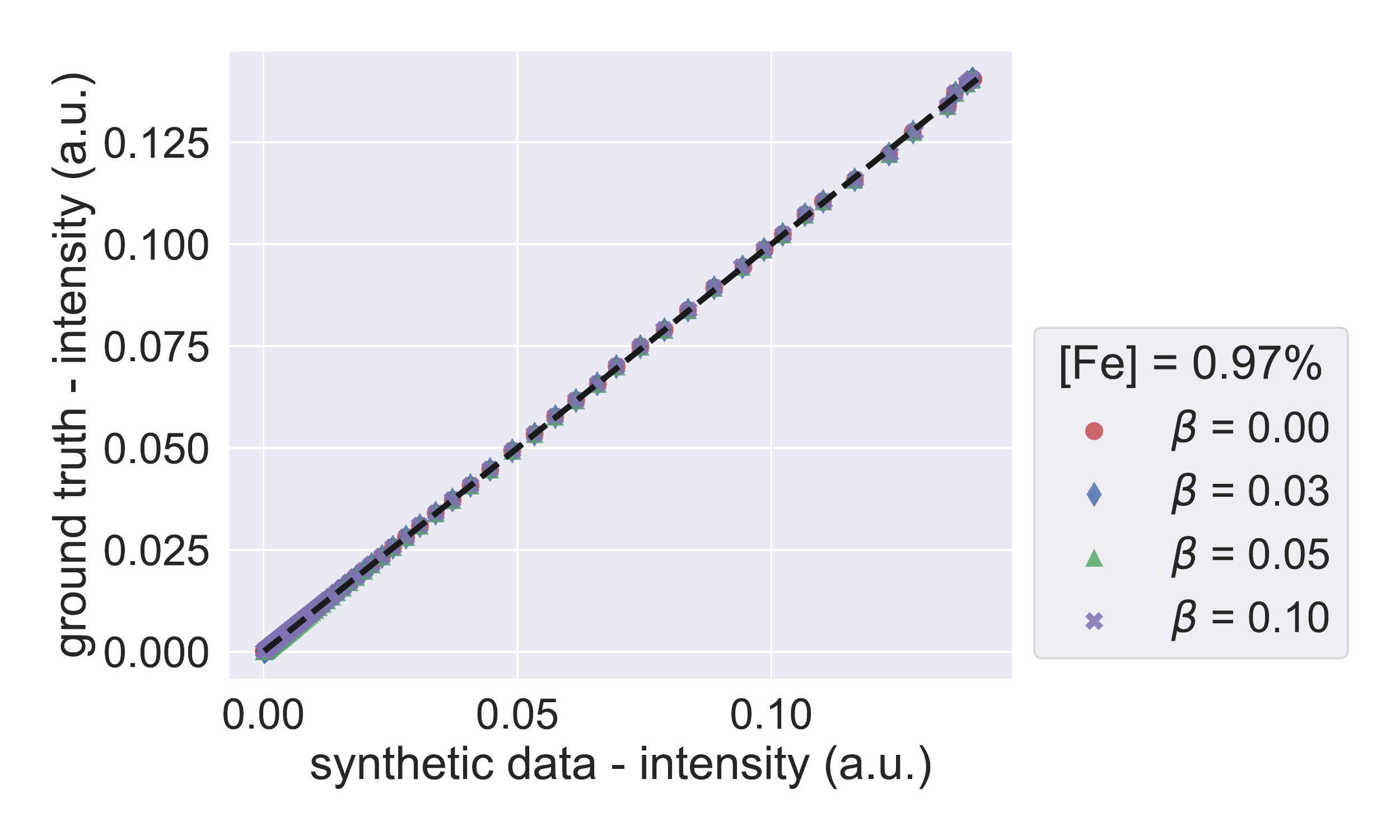} & \includegraphics[width=0.48\linewidth]{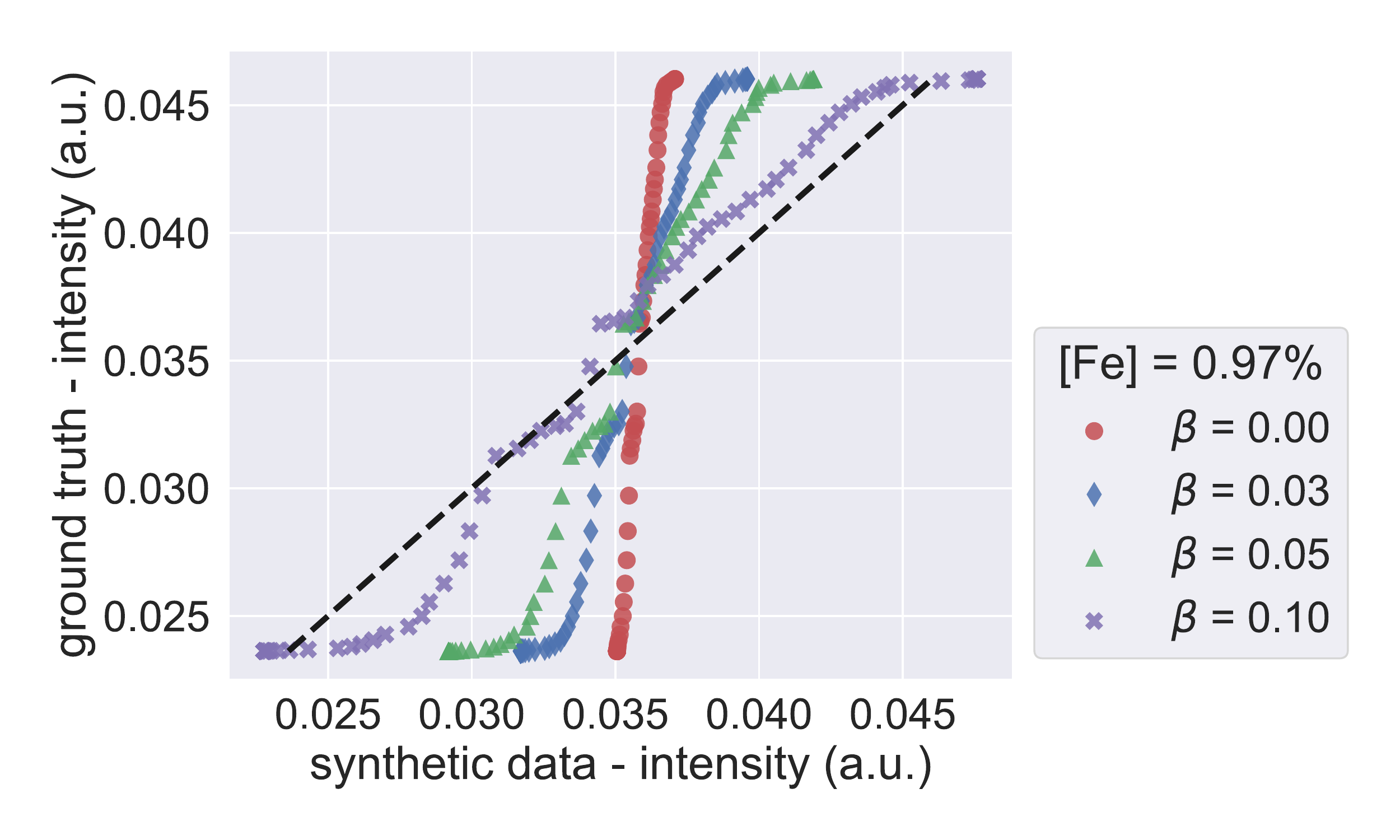}
    \end{tabular}
    \caption{%
        Global (average spectrum) and local (Fe line at \SI{373.48}{\nano\meter}) coverage of the synthetic Ni matrix dataset (\num{1000} synthetic spectra per sample) at low and high Fe concentrations.
    }
    \label{fig:ni_coverage}
\end{figure}

\begin{figure}[h]
    \centering
    \begin{tabular}{@{}c|c@{}}
        {\LARGE \textsc{Global Effects}}                                          & {\LARGE \textsc{Local Coverage}}
        \\
        \textsc{(whole spectra)}                                                  & \textsc{(Fe line at \SI{373.48}{\nano\meter})}
        \\
        \includegraphics[width=0.48\linewidth]{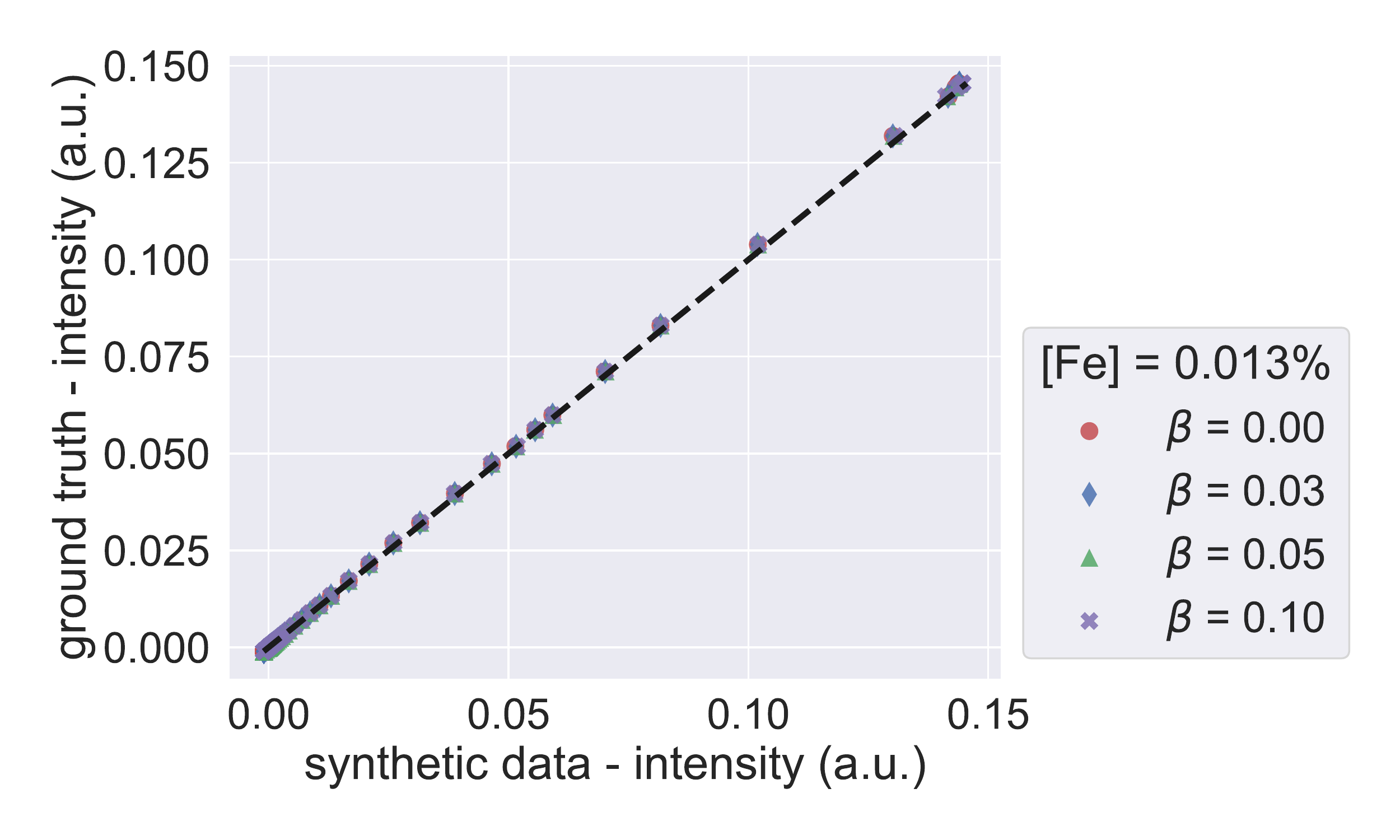}  & \includegraphics[width=0.48\linewidth]{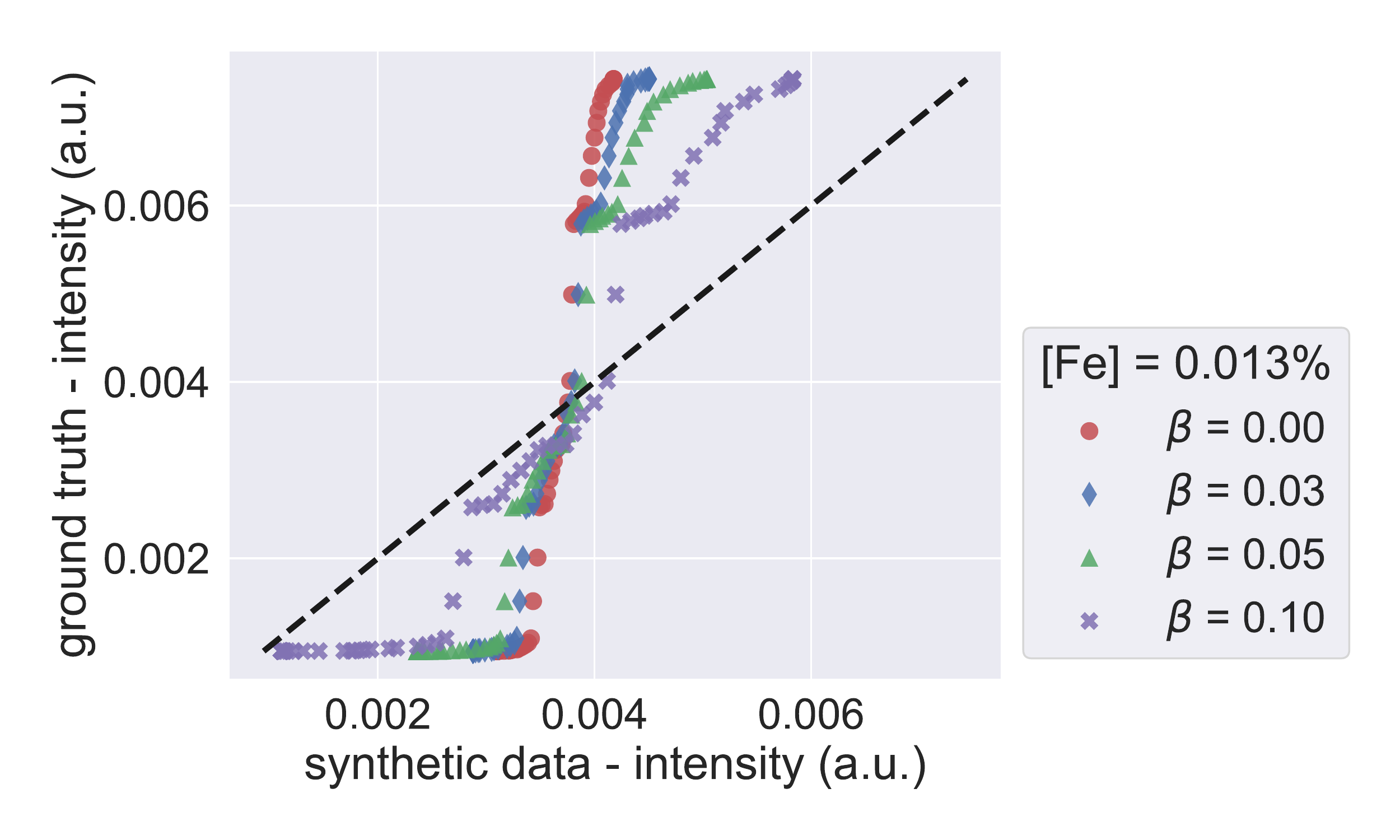}
        \\ \midrule
        \includegraphics[width=0.48\linewidth]{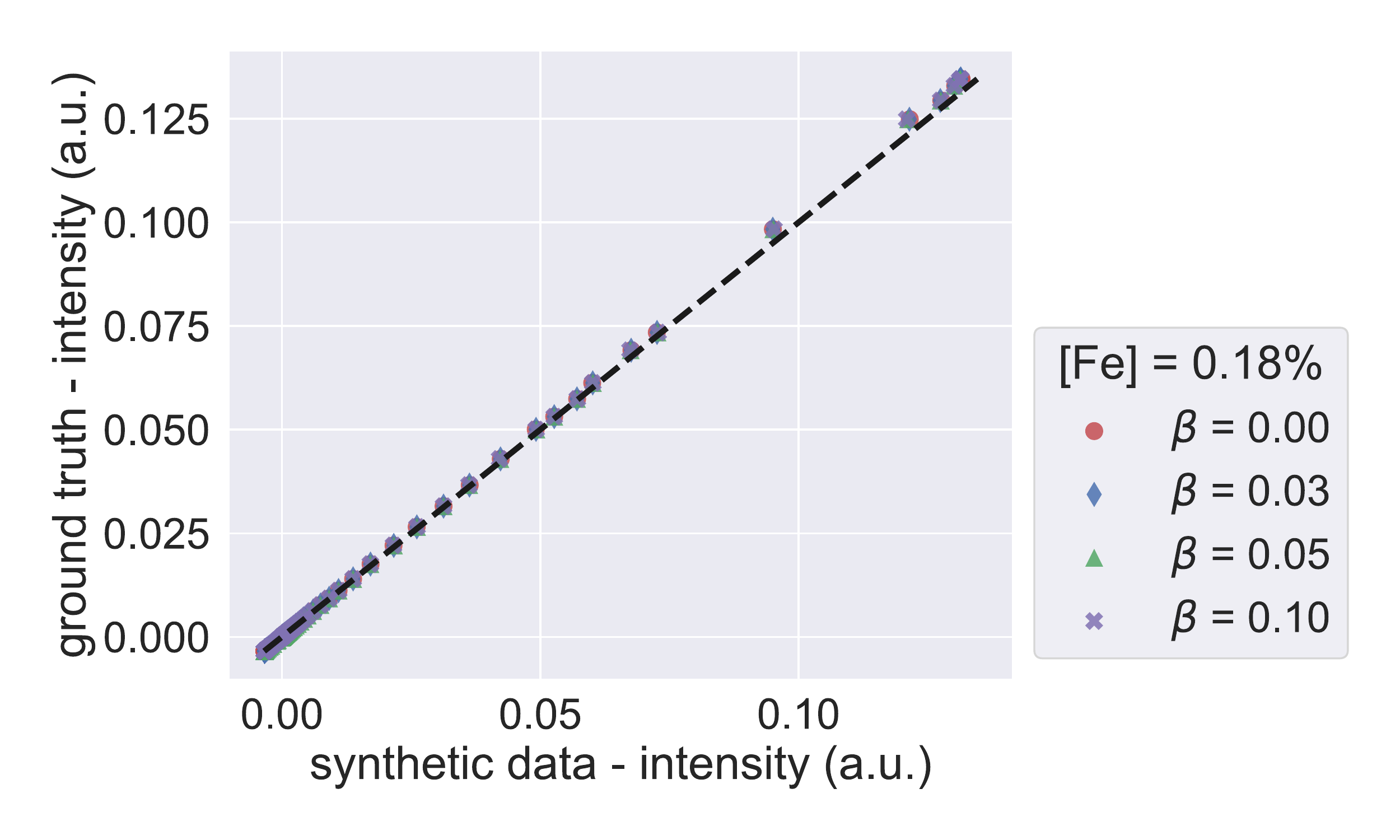} & \includegraphics[width=0.48\linewidth]{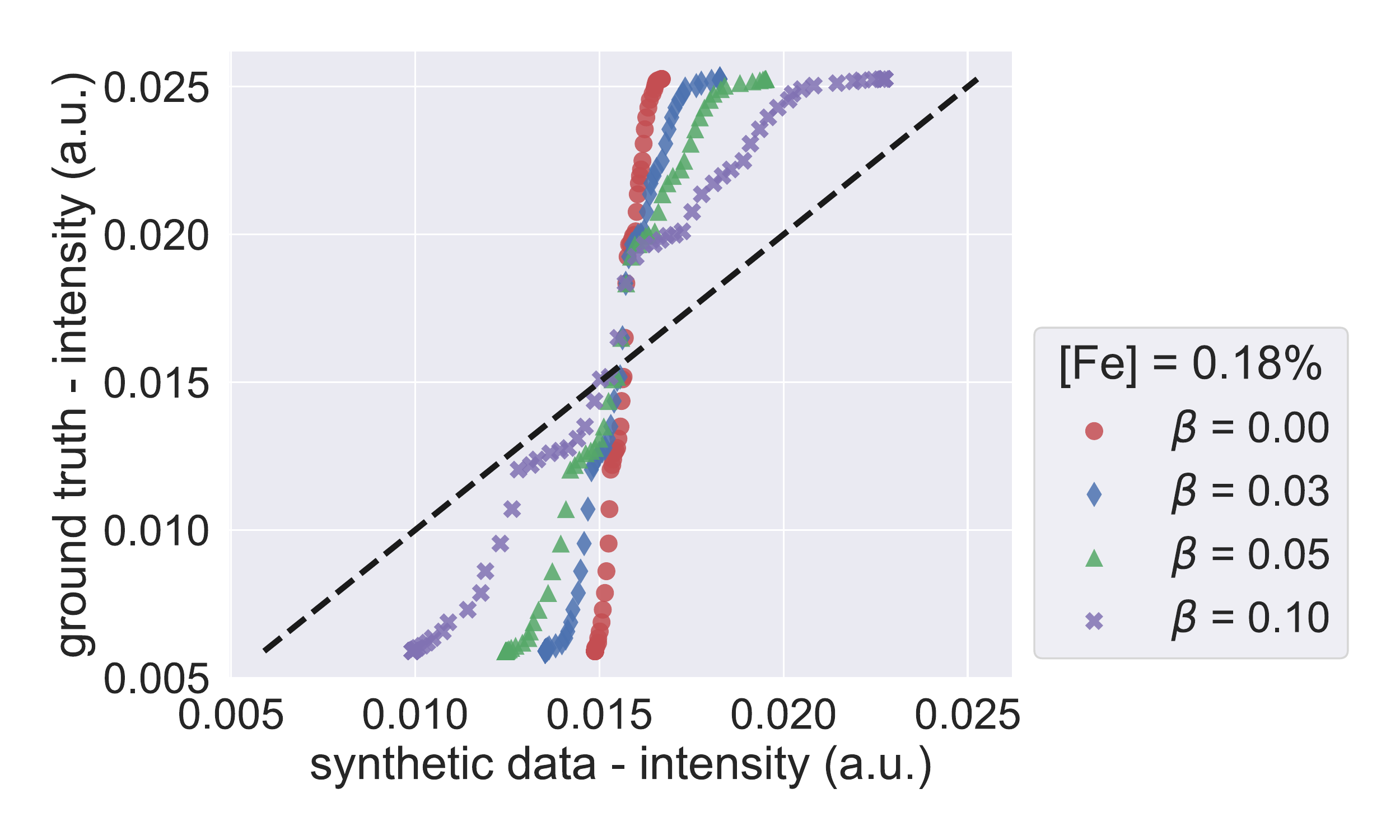}
    \end{tabular}
    \caption{%
        Global (average spectrum) and local (Fe line at \SI{373.48}{\nano\meter}) coverage of the synthetic Sn matrix dataset (\num{1000} synthetic spectra per sample) at low and high Fe concentrations.
    }
    \label{fig:sn_coverage}
\end{figure}

\begin{figure}[h]
    \centering
    \begin{tabular}{@{}c|c@{}}
        {\LARGE \textsc{Global Effects}}                                          & {\LARGE \textsc{Local Coverage}}
        \\
        \textsc{(whole spectra)}                                                  & \textsc{(Fe line at \SI{373.48}{\nano\meter})}
        \\
        \includegraphics[width=0.48\linewidth]{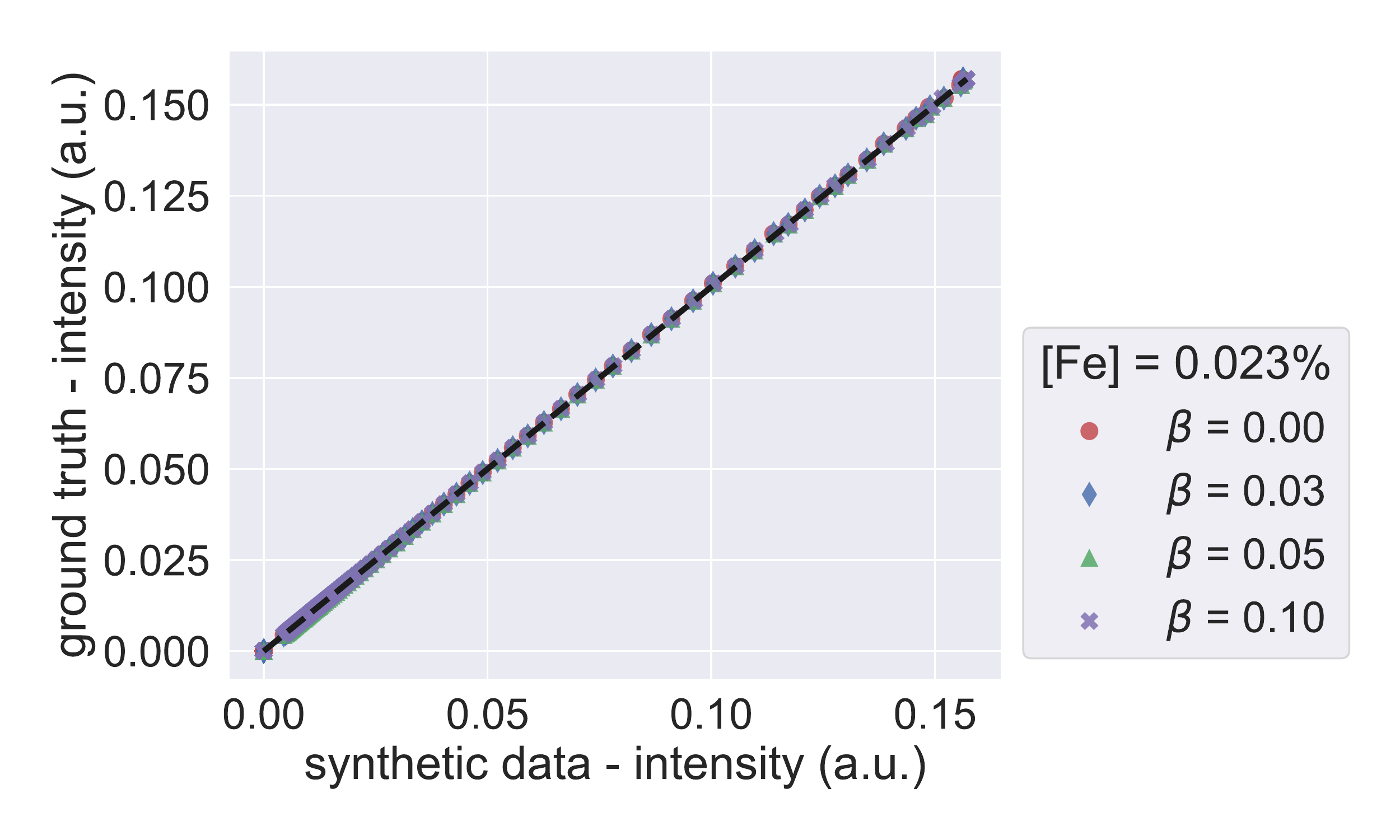}  & \includegraphics[width=0.48\linewidth]{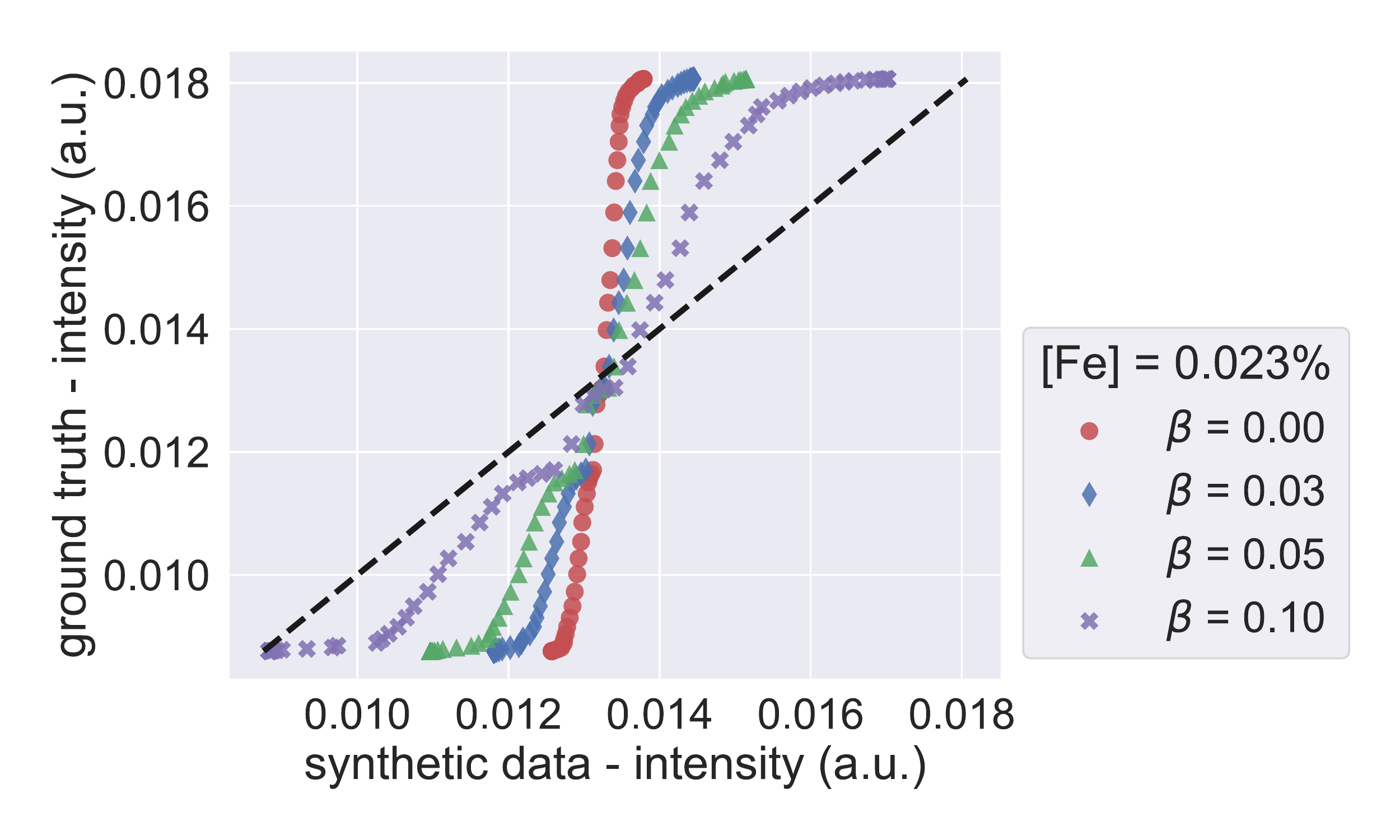}
        \\ \midrule
        \includegraphics[width=0.48\linewidth]{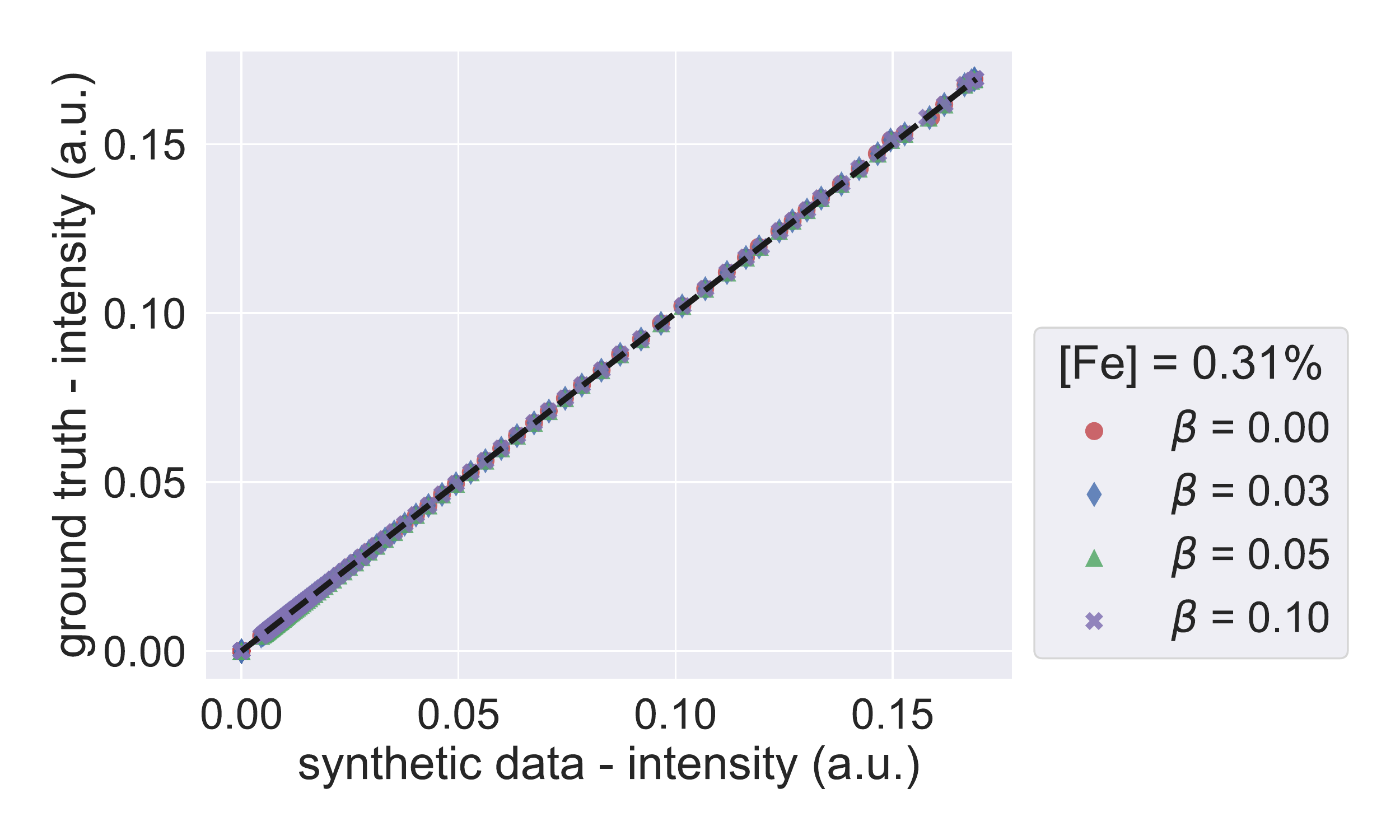} & \includegraphics[width=0.48\linewidth]{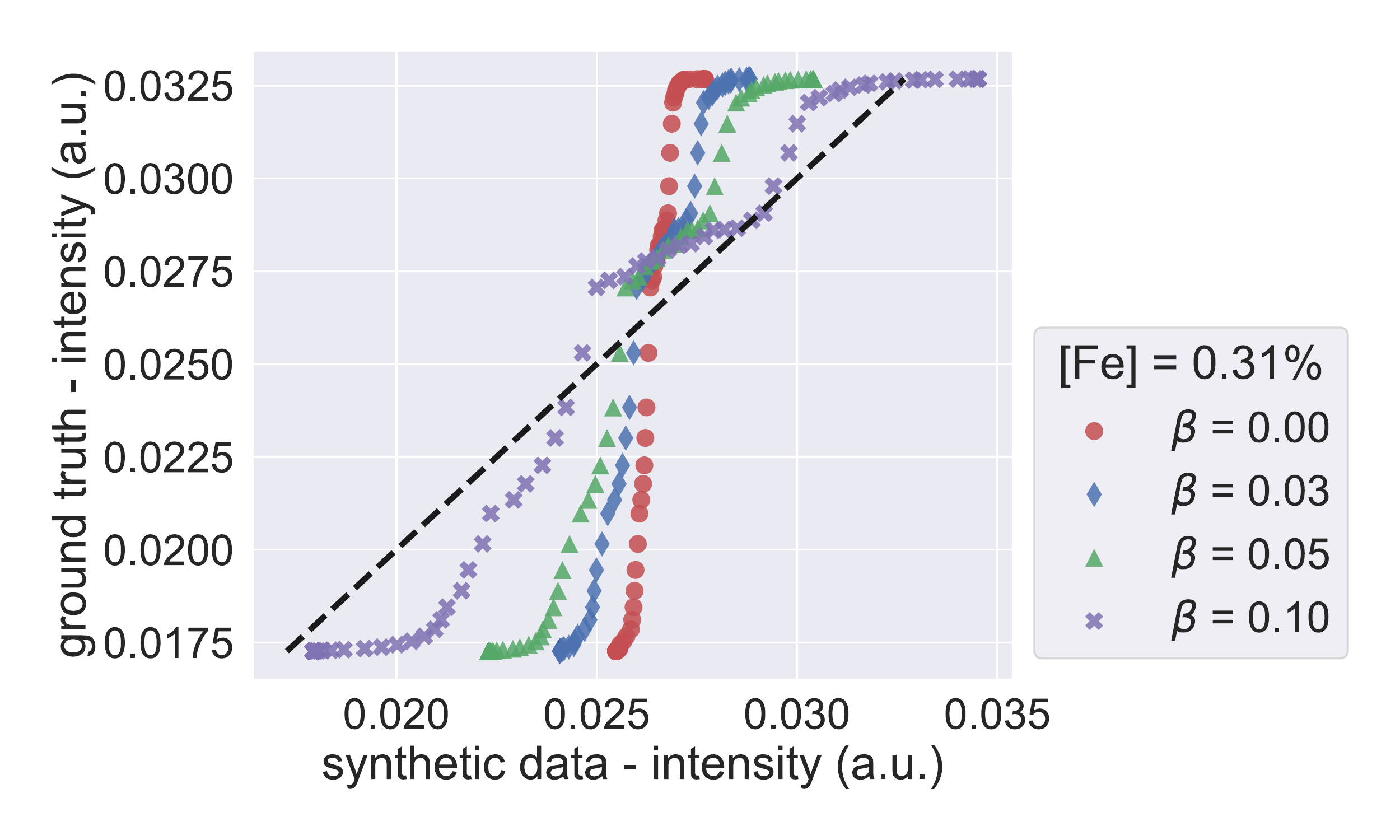}
    \end{tabular}
    \caption{%
        Global (average spectrum) and local (Fe line at \SI{373.48}{\nano\meter}) coverage of the synthetic Zr matrix dataset (\num{1000} synthetic spectra per sample) at low and high Fe concentrations.
    }
    \label{fig:zr_coverage}
\end{figure}

\begin{figure}[h]
    \centering
    \includegraphics[width=0.49\linewidth]{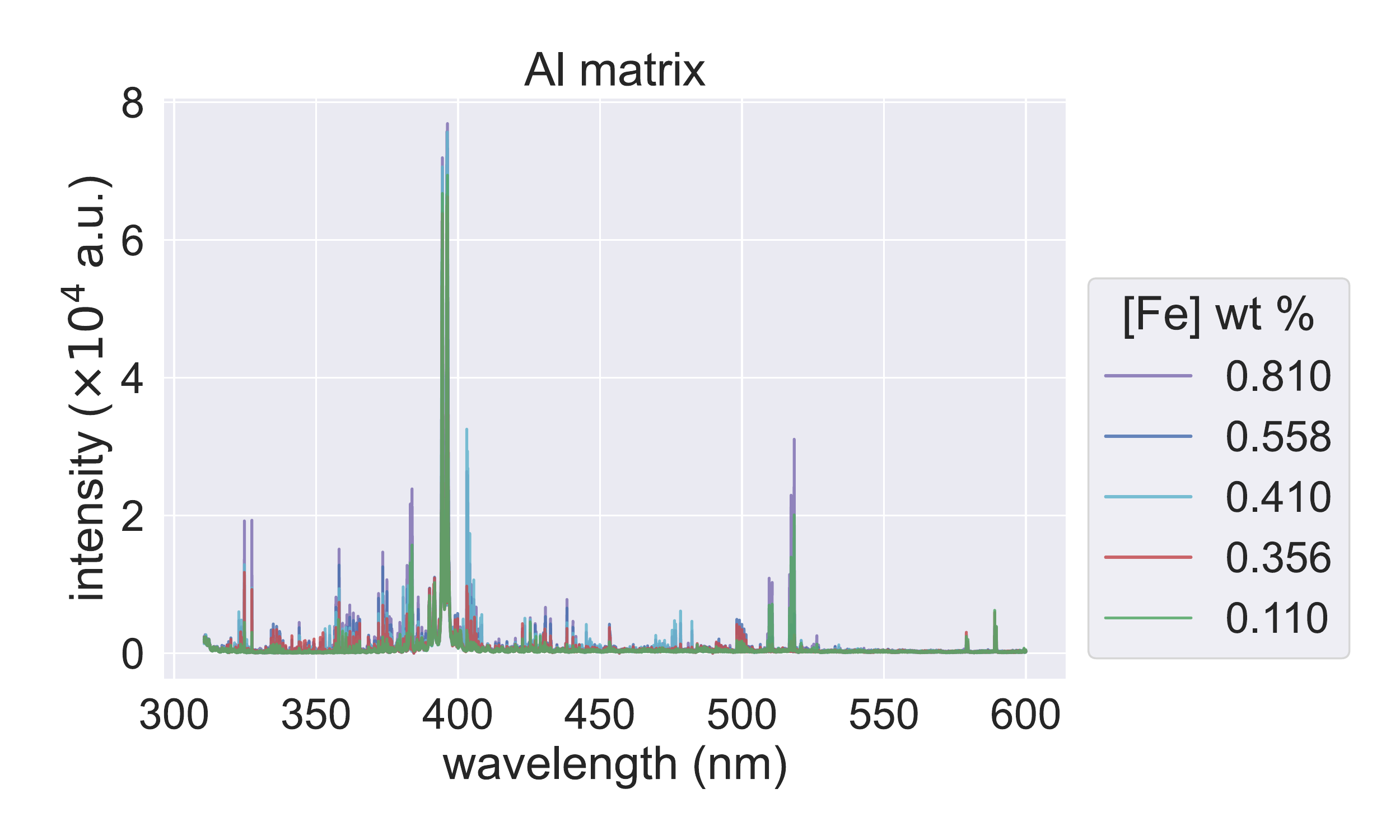}
    \hfill
    \includegraphics[width=0.49\linewidth]{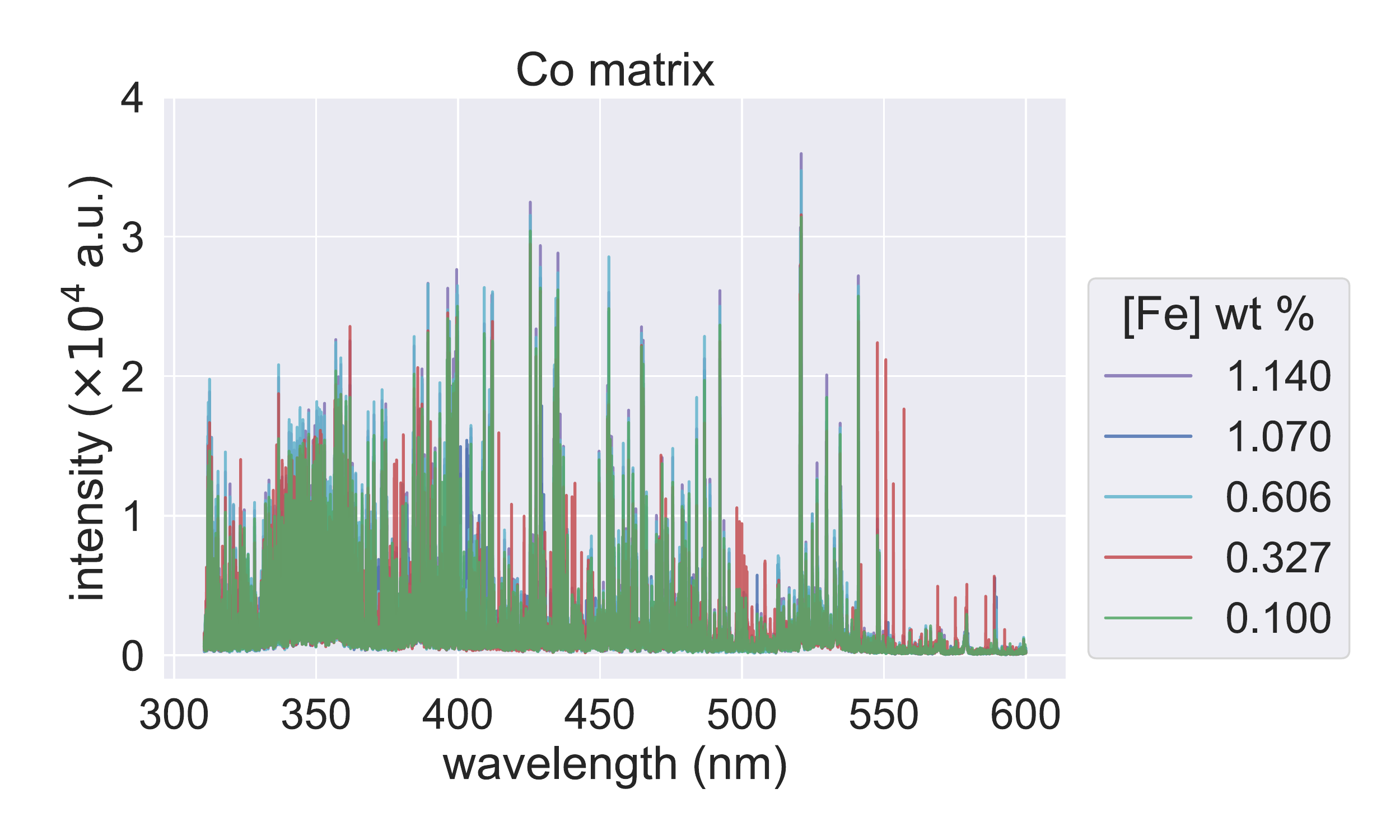}
    \\
    \includegraphics[width=0.49\linewidth]{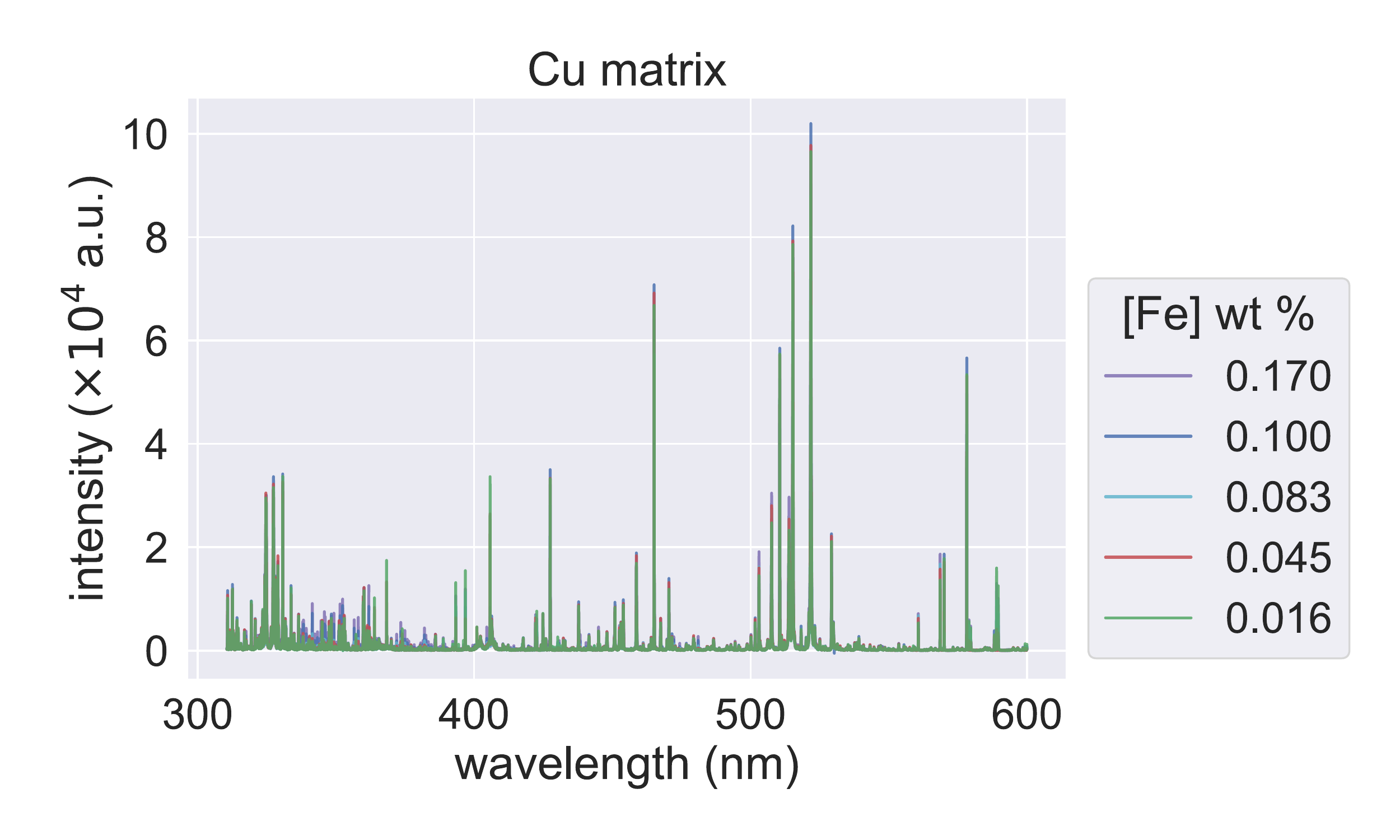}
    \hfill
    \includegraphics[width=0.49\linewidth]{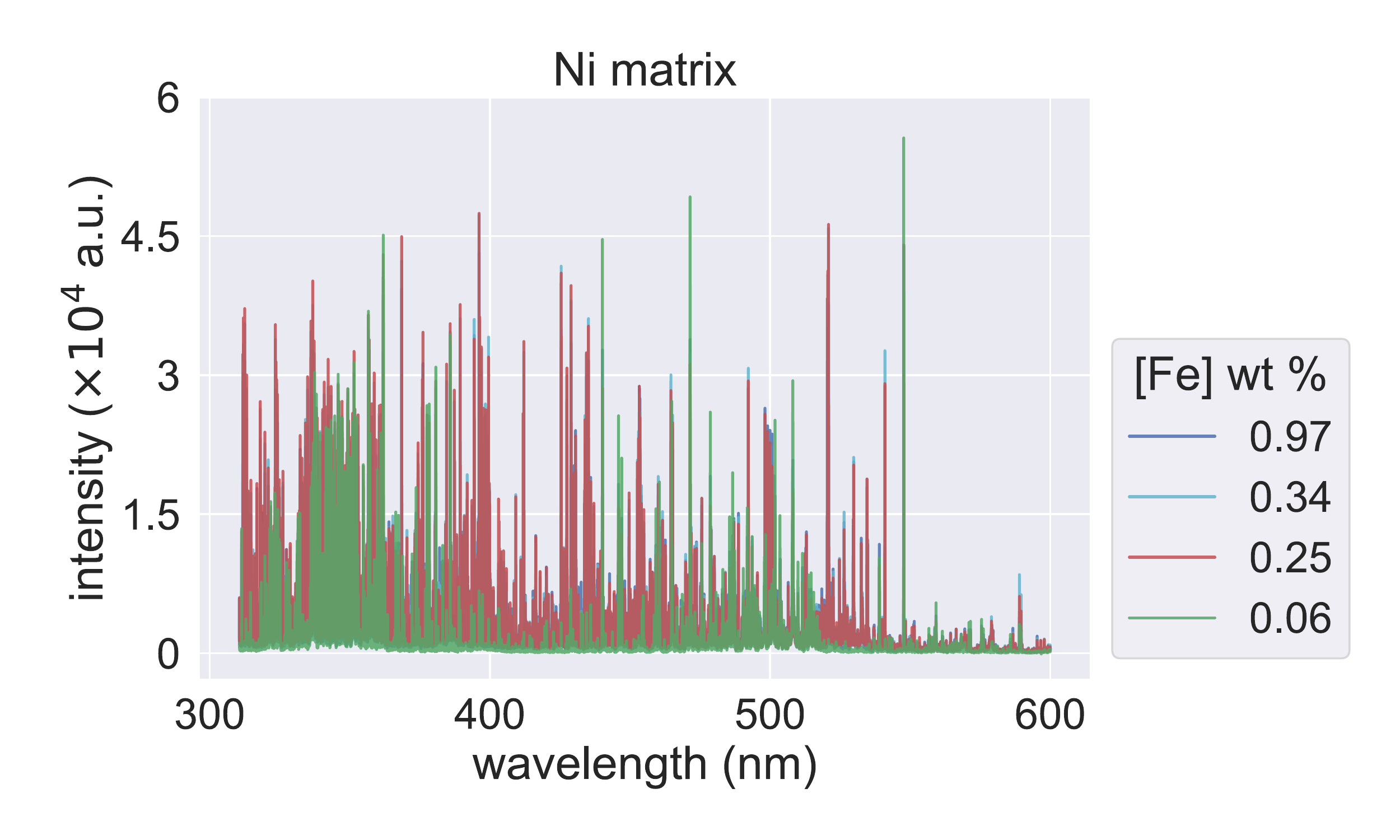}
    \\
    \includegraphics[width=0.49\linewidth]{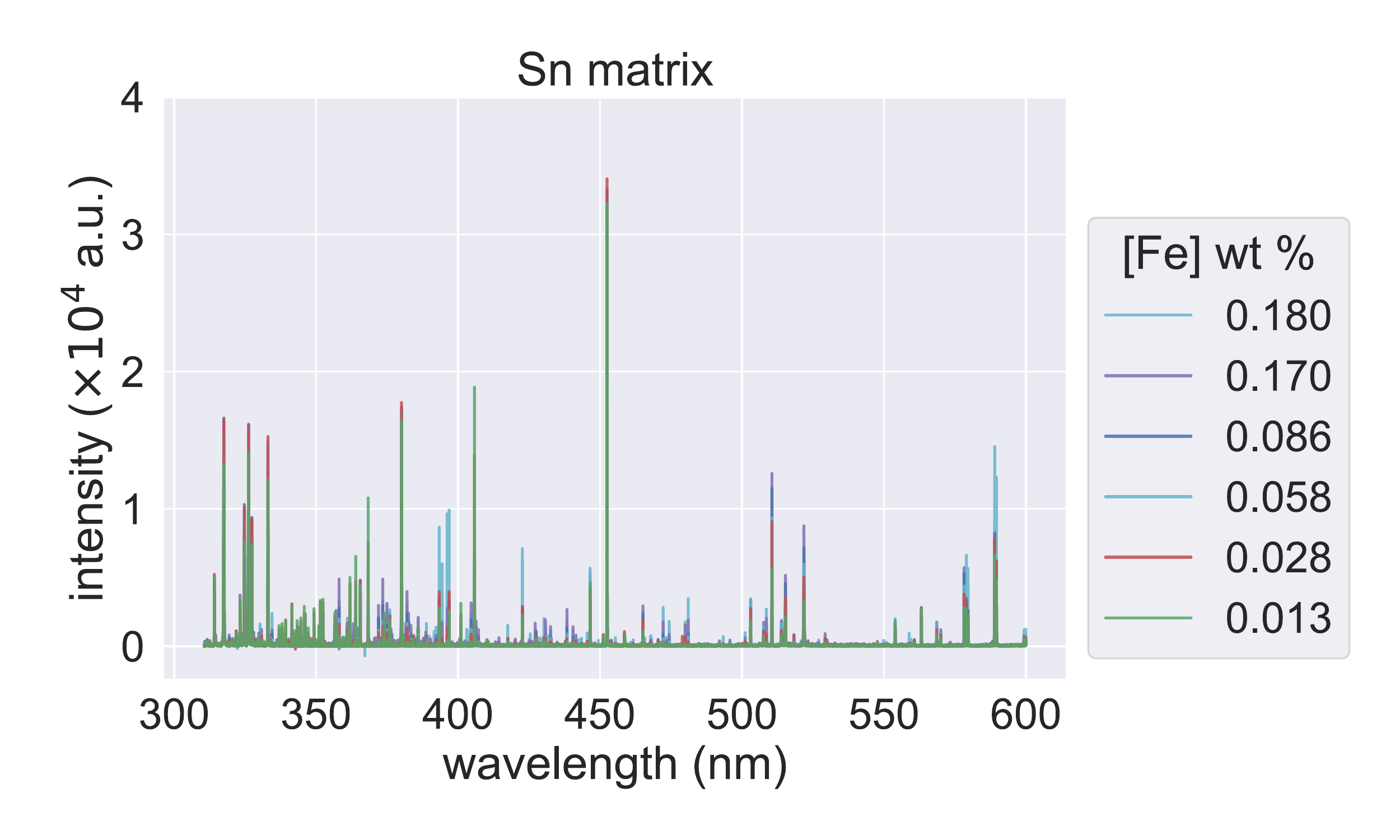}
    \hfill
    \includegraphics[width=0.49\linewidth]{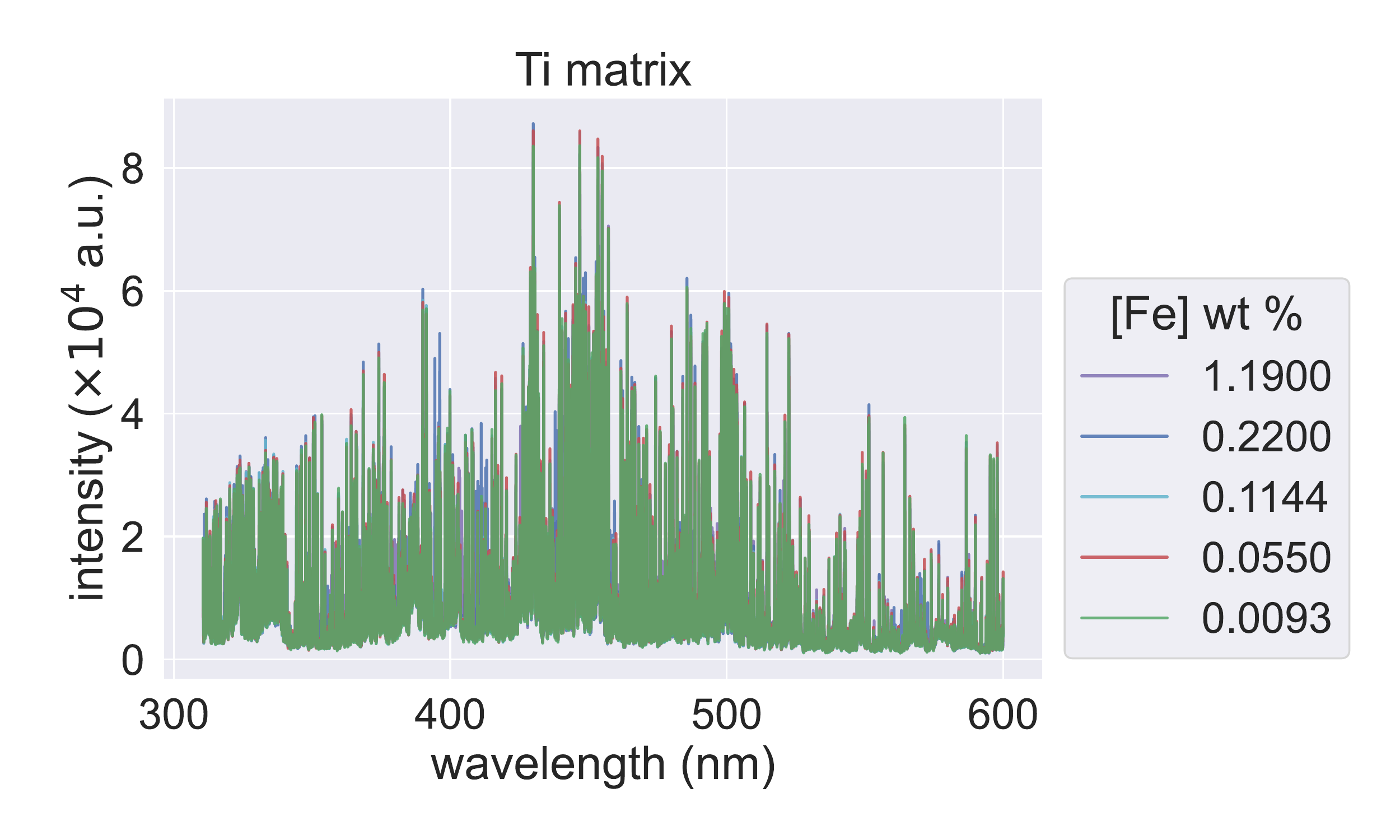}
    \\
    \includegraphics[width=0.49\linewidth]{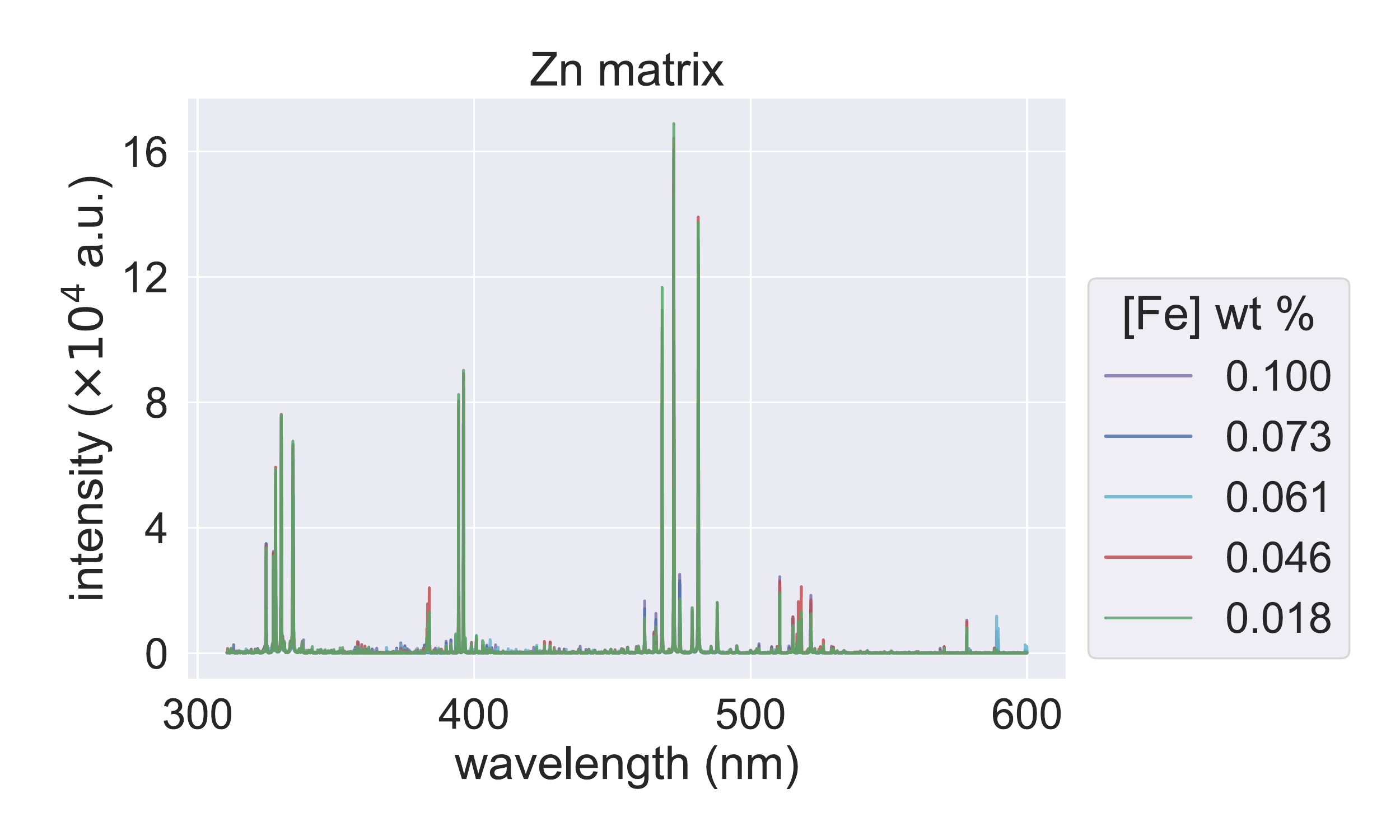}
    \hfill
    \includegraphics[width=0.49\linewidth]{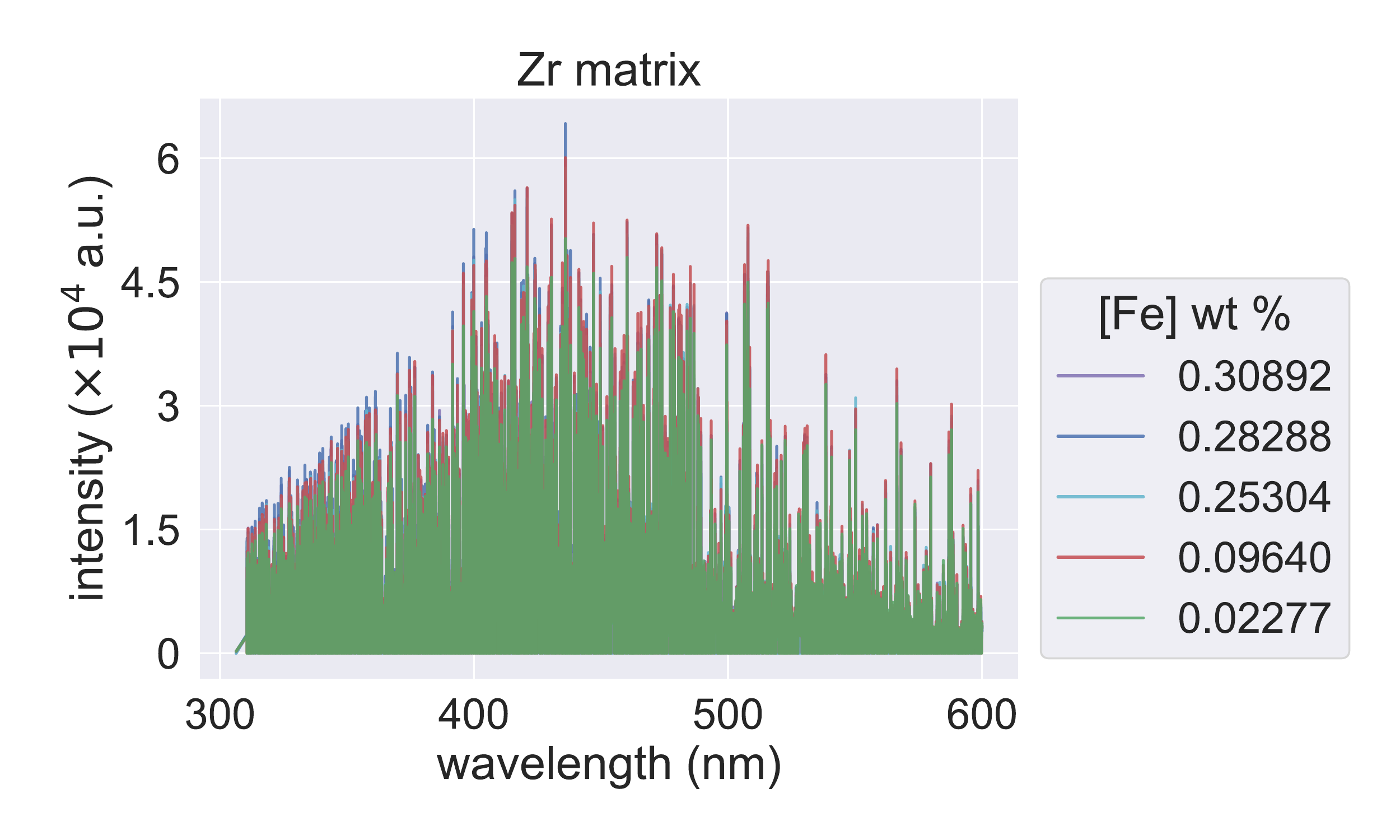}
    \caption{%
        Average spectra of the alloy matrices before preprocessing.
    }
    \label{fig:full_alloy_avg}
\end{figure}

\begin{figure}[h]
    \centering
    \includegraphics[height=0.38\linewidth]{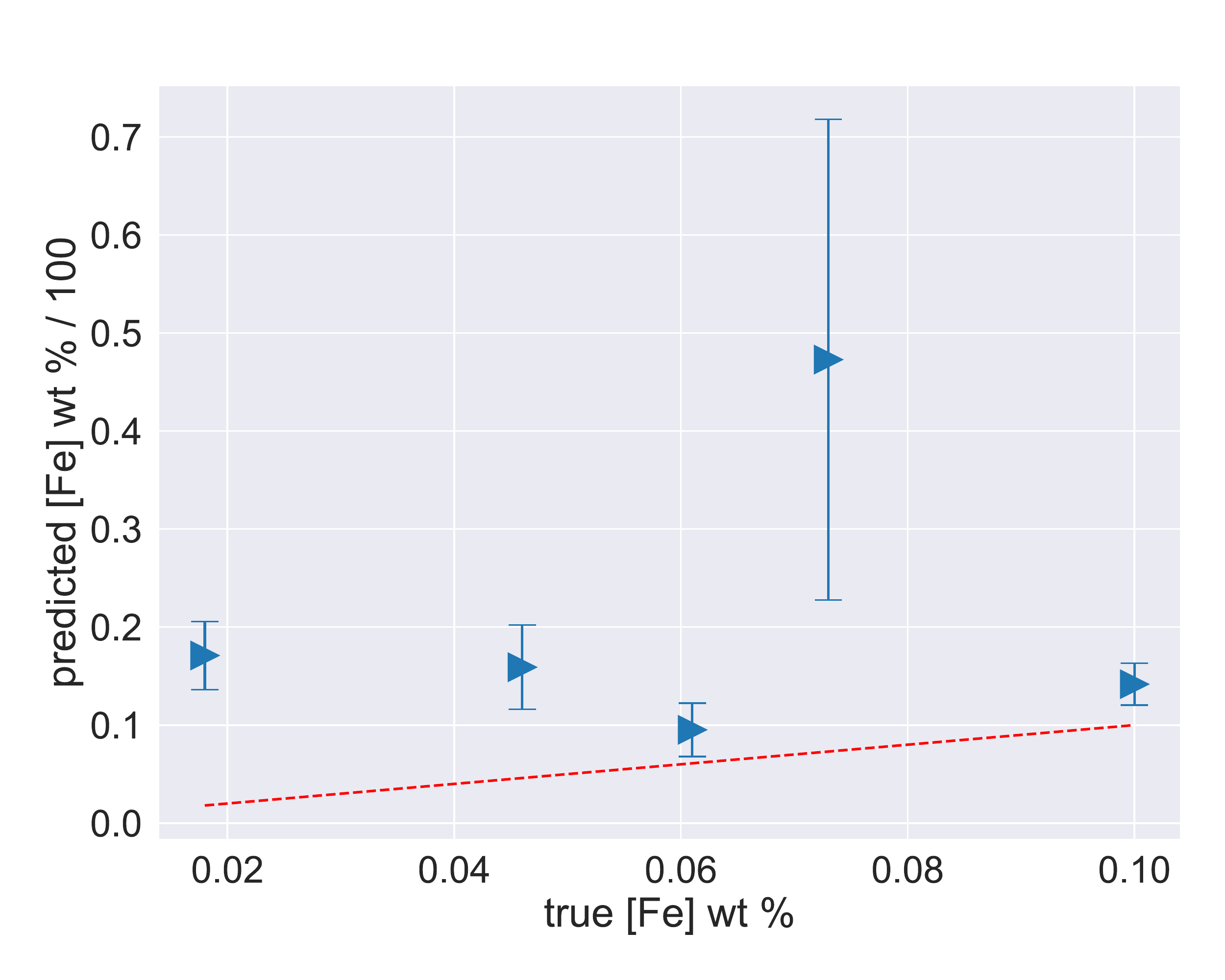}
    \hfill
    \includegraphics[height=0.38\linewidth]{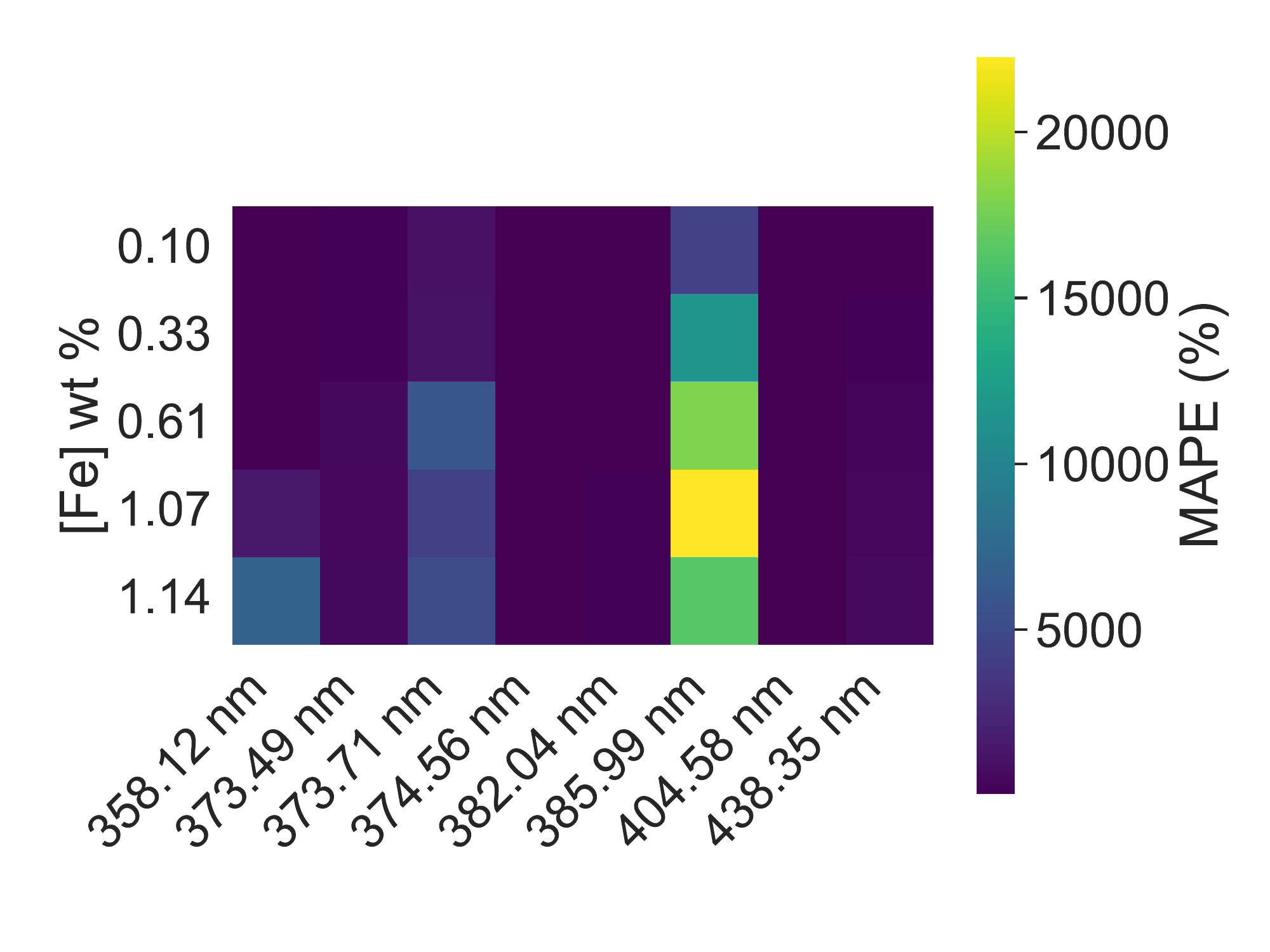}
    \caption{%
        Results of the calibration of the Zn matrix using the \mt model trained on the Co matrix.
    }
    \label{fig:transfer_2}
\end{figure}

\end{document}